\documentclass[aps,prb,twocolumn,showpacs,amsmath,amssymb,superscriptaddress,floatfix]{revtex4-2}

\usepackage[english]{babel}
\usepackage{blindtext}
\usepackage{latexsym}
\usepackage{amssymb}
\usepackage{physics}
\usepackage{amsmath}
\usepackage{bm}
\usepackage{amsfonts}
\usepackage{relsize}
\usepackage{xcolor}
\usepackage{verbatim}
\usepackage{bbold}
\usepackage{slashed}
\usepackage{appendix}
\usepackage{graphicx}
\usepackage{color}
\usepackage[normalem]{ulem}
\usepackage[colorlinks = true,
            linkcolor = blue,
            urlcolor  = blue,
            citecolor = blue,
            anchorcolor = blue]{hyperref}
\usepackage{graphicx} % Include figure files
\usepackage{dcolumn} % Align table columns on decimal point
\newcommand{\approptoinn}[2]{\mathrel{\vcenter{
  \offinterlineskip\halign{\hfil$##$\cr
    #1\propto\cr\noalign{\kern2pt}#1\sim\cr\noalign{\kern-2pt}}}}}

\newcommand{\up}{\uparrow}
\newcommand{\down}{\downarrow}

\definecolor{ao(english)}{rgb}{0.0, 0.5, 0.0}
\definecolor{amaranth}{rgb}{0.9, 0.17, 0.31}
\definecolor{green(html/cssgreen)}{rgb}{0.0, 0.5, 0.0}

\newcommand\greensout{\bgroup\markoverwith{\textcolor{green(html/cssgreen)}{\rule[0.5ex]{2pt}{1.0pt}}}\ULon}

\begin{document}

\title{Optically active Higgs and Leggett modes in multiband pair-density-wave superconductors with Lifshitz invariant}
%\thanks{A footnote to the article title}%

\author{Raigo Nagashima}
\affiliation{Department of Physics, The University of Tokyo, Hongo, Tokyo, 113-8656, Japan}

\author{Titouan Mouilleron}
\affiliation{Department of Physics, The University of Tokyo, Hongo, Tokyo, 113-8656, Japan}
\affiliation{Department of Physics, ENS-PSL, Paris, 75005, France}

\author{Naoto Tsuji}
\affiliation{Department of Physics, The University of Tokyo, Hongo, Tokyo, 113-8656, Japan}
\affiliation{RIKEN Center for Emergent Matter Science (CEMS), Wako 351-0198, Japan}

\date{\today}% It is always \today, today,
             %  but any date may be explicitly specified

\begin{abstract}
Lifshitz invariant is a symmetry invariant composed of multiple order parameters that contain a single spatial derivative in a Ginzburg-Landau (GL) free energy, which may induce a nonuniform configuration of the order parameters.
In multiband superconductors, we find phase transitions from a uniform superconducting state to qualitatively distinct two pair-density-wave (PDW) states with small and large momenta $\bm{q}$ based on the GL theory. The former is induced by the Lifshitz invariant, while the latter originates from the drag effect.
In the PDW states, the Higgs and Leggett modes (i.e., collective amplitude and relative phase oscillations of the order parameters) are shown to couple to electromagnetic fields linearly. 
We construct microscopic models of multiband superconductors with Lifshitz invariant that exhibit PDW states,
and calculate the linear optical conductivity using the diagrammatic approach.
We find optically active Higgs and Leggett modes in the small $\bm q$ PDW state, 
indicating that the PDW state is a suitable platform to explore collective modes of multiband superconductors in the linear response regime.

%\begin{description}
%\item[Usage]
%Secondary publications and information retrieval purposes.
%\item[Structure]
%You may use the \texttt{description} environment to structure your abstract;
%use the optional argument of the \verb+\item+ command to give the category of each item. 
%\end{description}
\end{abstract}

%\keywords{Suggested keywords}%Use showkeys class option if keyword
                              %display desired
\maketitle

\section{\label{sec:level1}Introduction}
Revealing dynamical aspects of various superconductors is essential in understanding the underlying microscopic nature of superconducting phases.
Among them, collective modes contain various information on pairing symmetries, fluctuations, spatial patterns of superconducting orders, and so on.
According to the BCS theory \cite{BCS_original}, single-band superconductors typically show two collective modes: the Higgs mode \cite{Anderson1958, Schmid1968, Littlewood_Varma1981, Littlewood_Varma1982, Pekker_Varma2015, Higgs_mode_review2020, Tsuji-Ency2024}, which corresponds to a massive excitation of an amplitude fluctuation of the order parameter, and the Nambu--Goldstone (NG) mode \cite{Goldstone1961, Nambu_JonaLasinio1961}, which would appear as a massless mode arising from the overall phase fluctuation.
The latter mode, however, couples to gauge fields, being lifted to the plasma frequency due to the Anderson--Higgs mechanism \cite{Anderson1963, Englert1964, Higgs1964, Guralnik1964}.
%Because of this mechanism, 
As a result, only the Higgs mode survives as a collective mode in the low-energy region on top of the continuum of quasiparticle excitations.

Unlike the NG mode, the Higgs mode does not linearly couple to gauge fields in ordinary superconductors, except for some special situations such as those in the presence of a supercurrent \cite{Moor2017, Nakamura2019} or with a coupling to an LC circuit \cite{Lu2023}.
This is the reason that previous studies have focused on nonlinear optical responses of superconductors \cite{Tsuji2015, Kemper2015, Cea2016, Tsuji2016, Jujo2018, Silaev2019, Schwarz2020, Tsuji_and_Nomura2020, Haenel2021, Udina2022}:
There have been performed Raman spectroscopy for superconductors with a charge density wave order \cite{Sooryakumar1980, Sooryakumar1981, Measson2014, Grasset2018, Majumdar2020}, THz pump-probe spectroscopy and third harmonic generations for superconductors \cite{Kemper2015, Matsunaga2013, Matsunaga2014, Matsunaga2017, Katsumi2018, Chu2020} 
to observe the Higgs mode experimentally.

\begin{figure}
    \centering
    \includegraphics[scale=0.2]{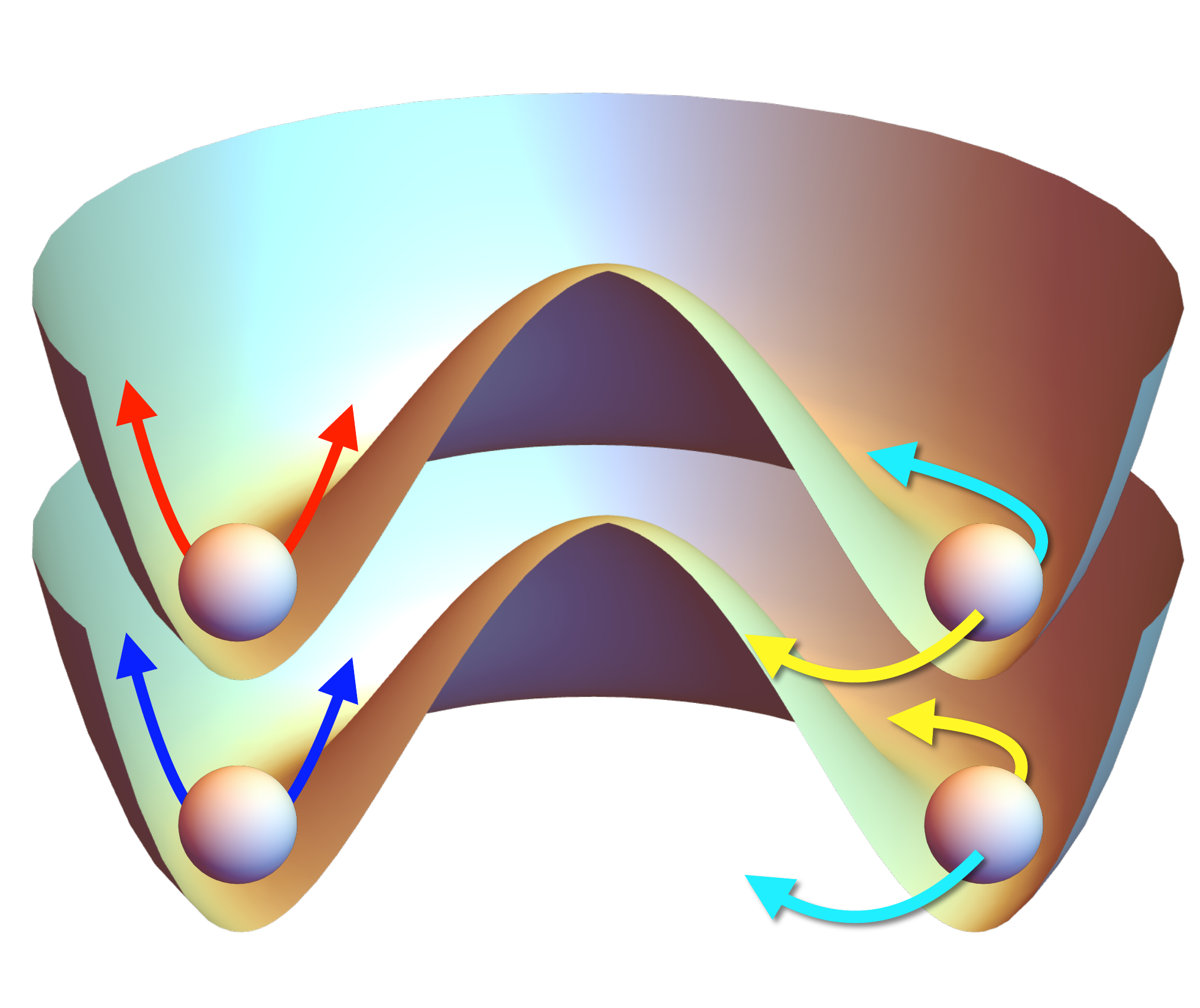}
    \caption{
    Schematic picture of the free energy for a two-band superconductor.
    In multiband superconductors, relative phase fluctuations between two order parameters generally comprise a massive mode (Leggett mode).
    Two-band superconductors have two Higgs modes (oscillations against the wall of the free energy) and one Leggett mode (anti-phase oscillations along the circular direction). The other phase degree of freedom corresponding to the massless Nambu--Goldstone mode is lifted to the plasma frequency via the Anderson--Higgs mechanism.
    }
    \label{Pic_TwoBand}
\end{figure}

When the system has multiple superconducting order parameters, even richer physics may be expected compared to the single-band cases.
A representative example of such a system is provided by multiband superconductors, e.g., iron-based superconductors \cite{Iron_SC_review}, niobium-based superconductors \cite{Niobium_SC_review}, $\text{MgB}_{2}$ \cite{MgB2_review}, and recently discovered Kagome superconductors \cite{Kagome_SC_review}.
Due to the multiband nature, there appears a relative phase mode between two different order parameters called the Leggett mode \cite{Leggett1966},
in addition to the Higgs and NG modes.
The schematic picture of the collective motion of the order parameters (here Higgs and Leggett modes) on the free energy surface of a two-band superconductor in Fig.~\ref{Pic_TwoBand}.
Previously, the Leggett mode has been suggested to be observed by nonlinear responses in multiband superconductors \cite{Balatsuky2000, Burnell2010, Ota2011, Lin2012, Marciani2013, Bittner2015, Cea2016, Krull2016, Murotani2017, Murotani2019, Giorgianni2019, Fiore2022, Seibold2021}. In fact, Raman spectroscopy has been used to experimentally observe the Leggett mode in a typical multiband superconductor $\text{MgB}_{2}$ \cite{Blumberg2007, Giorgianni2019}. On the other hand, it has been unclear whether the Leggett mode is optically visible in the linear response regime.

Recently, the Leggett mode is predicted to appear in the linear response regime \cite{Kamatani2022} when the system has a Lifshitz invariant \cite{LandauLifshitz}, which is a linear derivative term in the Ginzburg--Landau (GL) free energy in the antisymmetric form, $\bm d\cdot (\psi_i^\dagger \nabla \psi_j-\psi_j^\dagger \nabla \psi_i)$ ($\psi_i$ is the superconducting order parameters and $\bm d$ is a constant vector) \cite{Nagashima2024}.
Detecting a collective mode in the linear response would drastically simplify the situation and make an experimental observation more accessible.
Kagome superconductors are one of the candidate materials for observing the Leggett mode in the linear response since the charge-density-wave order allows the existence of the Lifshitz invariant from a symmetry point of view (for the group theoretical classification of the Lifshitz invariant, we refer to Ref.~\cite{Nagashima2024}).
While the Lifshitz invariant has been often discussed in the context of, e.g., noncentrosymmetric superconductors \cite{Mineev1994, Mineev2008, Samokhin2013, Kochan2022, Kochan2023}, parity- and time-reversal symmetry broken superconductors \cite{Kanasugi2022, Kitamura2023}, commensurate-incommensurate transitions \cite{Kopsky1977, Ishibashi1978}, and liquid crystals \cite{Sparavigna2009}, the Lifshitz invariant may exist even in inversion symmetric systems.
There is an experimental report for a possible observation of the Leggett mode by scanning tunneling microscopy in a Kagome superconductor $\text{CsV}_{3-x}\text{Ta}_{x}\text{Sb}_{5}$ \cite{Hu2024}.
There is also a study to detect collective modes, including the Leggett mode, in the linear response regime via gate electrodes \cite{Levitan2024}.

Since the Lifshitz invariant is linear in the spatial derivative of the order parameter \cite{LandauLifshitz}, one might expect that a uniform superconducting state becomes unstable, resulting in a nonuniform state with a finite center-of-mass momentum $\bm{q}$ of the superconducting order parameter.
This kind of nonuniform states is referred to as a pair-density-wave (PDW) state (see, e.g., Ref.~\cite{PDW_review}).
%in the language of the microscopic theory \cite{PDW_review}.
There have been experimental reports on PDW states in cuprate superconductors \cite{Cuprate_expr_Fujita2012, Cuprate_expr_Edkins2019, Cuprate_expr_Du2020}, iron-based superconductors \cite{Iron_expr_Liu2023, Iron_expr_Zhao2023}, $\text{NbSe}_{2}$ \cite{NbSe2_expr_Liu2021}, $\text{UTe}_{2}$ \cite{UTe2_expr_Aishwarya2023, UTe2_expr_Gu2023}, and Kagome superconductors \cite{Kagome_expr_Chen2021, Kagome_expr_Zhao2021, Kagome_expr_Deng2024}. The equilibrium properties of PDW states (and the stripe phase in the context of cuprates) have extensively been studied \cite{Himeda2002, Berg2007, Agterberg2008, Berg2009, Berg_2009_NJP, Berg2009_NatPhys, Loder2010, Berg2010, Loder2011, Jaefari2012, PALee2014, Corboz2014, RSotoGarrido2014, RSotoGarrido2015, Wang2018, Wardh2018, Choubey2020, Huang2022, Setty2023, Shaffer2023, Ponsionen2023, Ticea2024}.
Although collective modes in PDW states of a single-band model have been discussed \cite{Boyack2017, Soto-Garrido2017}, it has not been well understood how the collective modes in PDW states respond to light, particularly in relation to the Lifshitz invariant.

Moreover, it has not been known whether the Higgs and Leggett modes become optically active, i.e., whether these modes appear in the linear optical response. If so, the question is, what is the condition for the Higgs and Leggett modes to be optically active in the PDW state?

In this paper, we study phase transitions from a uniform superconducting state to PDW states in multiband superconductors with the Lifshitz invariant based on the GL and microscopic theories.
We find two qualitatively distinct PDW states with different finite center-of-mass momenta $\boldsymbol{q}$: small and large $\bm{q}$ states.
The Lifshitz invariant prefers to ``twist" the phase between two different order parameters. 
The small $\bm{q}$ state can appear when the effect of the Lifshitz invariant exceeds that of the other GL terms (e.g., the Josephson coupling term) that tend to ``align" the phase.
On the other hand, the large $\bm{q}$ state is found to appear when the drag-effect term becomes dominant.
We emphasize that an infinitesimal Lifshitz term does not necessarily lead to the instability of the uniform solution: There is a threshold for the strength of the Lifshitz invariant beyond which a nonuniform solution starts to appear.

We then study, using the GL theory, how the collective modes appear in the PDW state and how they respond to external electromagnetic fields.
We find that both the Higgs and Leggett modes appear in the linear response regime because of the finite $\bm{q}$, which is unique to the PDW state.
This might look similar to the situation where the Higgs mode can appear in the linear response regime when a supercurrent is injected to induce a finite momentum of Cooper pairs \cite{Moor2017}. The difference is that in the present case, the momentum of Cooper pairs is generated from an intrinsic effect (originating from the Lifshitz invariant) and not due to an external operation.
Moreover, in the case of supercurrent injection, the generated momentum $\bm q$ of Cooper pairs is usually much smaller than that of PDW states, which results in a small change in the optical conductivity \cite{Nakamura2019}. In contrast, PDW states may have large momentum $\bm q$ in the order of the inverse of the lattice constant. Hence we can expect a significant change in the optical conductivity spectrum due to the resonance with collective modes in PDW states.
Although the collective modes in the FFLO state, which is also the superconducting state with a finite center-of-mass momentum $\bm{q}$, have been recently discussed \cite{Miwa2024}, its $\bm q$ is usually much smaller than that of the PDW states, and much larger signals are expected in the PDW states.

To microscopically investigate the linear optical responses of the two distinct PDW states, we construct a one-dimensional (1D) two-band model with a mean-field approximation that exhibits the $\pi/2$- and the $\pi$-PDW state, each of which roughly corresponds to the small and large $\bm{q}$ states in the GL theory, respectively.
The model has two hopping strengths $-(t\pm\delta t)$ and a pair-hopping interaction $V$ in addition to the attractive on-site interaction $U$.
The competition between the effect of the difference of the single hopping $\delta t$ and the pair-hopping $V$ generates the two distinct PDW states of $q=\pi/2$ and $\pi$.
Based on this model, we obtain the equilibrium phase diagrams and compare them to the GL theory.
We then calculate the linear optical conductivity of the two PDW states using the diagrammatic technique.
In the $\pi/2$-PDW state, the Higgs and Leggett modes are hybridized with almost equal weights, while in the $\pi$-PDW state, the Higgs mode contribution vanishes, and the phase modes dominantly contribute to the linear optical spectrum.

We also demonstrate results for a two-dimensional (2D) two-band model to see that the results of the 1D model are robust, i.e., they are not sensitive to the system's dimension.
We confirm that two PDW states with $\bm{q}=(\pi/2,0)$ and $(\pi,\pi)$ can have the lowest free energy when the parameters are appropriately chosen, which is qualitatively consistent with the GL theory.
We then calculate the linear optical conductivity of these two PDW states using the diagrammatic approach.
In the $(\pi/2,0)$-PDW state, the Higgs and Leggett modes and their hybridized modes appear in the optical conductivity, while in the $(\pi,\pi)$-PDW state, phase modes are observed instead.
These results indicate that the PDW state is a suitable platform for investigating multiband superconductors' collective modes in the linear response regime.

This paper is organized as follows.
In Sec.~\ref{sec:level2}, we study the GL theory for multiband superconductors with the Lifshitz invariant to macroscopically understand how the small and large $\bm{q}$ PDW states emerge and how the collective Higgs and Leggett modes appear in a linear response regime.
In Sec.~\ref{sec:level3}, we analyze a microscopic theory in the real-space formulation to investigate the PDW state in the 1D two-band model.
We then perform a similar analysis for the PDW states based on the momentum-space formulation in Sec.~\ref{sec:level4}.
This section also provides the phase diagrams and the linear optical conductivity that is calculated microscopically.
Sec.~\ref{sec:level5} demonstrates the results for the 2D $(\pi/2,0)$- and $(\pi,\pi)$-PDW states including the linear optical conductivity.
Finally, we summarize the paper and discuss future issues in Sec.~\ref{sec:level6}.
Throughout this paper, we set $\hbar=k_{\text{B}}=1$.

\section{\label{sec:level2}Phenomenological theory of the Pair-Density-Wave state}
Let us first study the PDW state in multiband superconductors with the Lifshitz invariant based on the phenomenological GL theory.
It is helpful to begin with the macroscopic GL theory to understand how the two distinct PDW states appear and how the collective Higgs and Leggett modes dynamically respond to external electromagnetic fields.
The essence of the physics of the PDW states can be extracted by considering two-band systems.
One can extend the argument of the two-band case to arbitrary $n$-band cases straightforwardly.
We refer to the review article \cite{Higgs_mode_review2020} for the single-band case.

\subsection{\label{sec:level2-1}Two-band superconductor}
We consider the following form of the GL free-energy density $\mathcal{F}$ for a two-band superconductor \cite{Kamatani2022, Nagashima2024}:
\begin{align}
    \mathcal{F} &= \sum_{i=1,2}\qty[ a_{i}|\psi_{i}|^{2} + \frac{b_{i}}{2}|\psi_{i}|^{4} + \frac{1}{2m^{*}}|\bm{D}\psi_{i}|^{2}] \notag \\
    &\quad +\qty[\epsilon\psi_{1}^{*}\psi_{2} + \eta\qty(\bm{D}^{*}\psi_{1}^{*})\cdot\qty(\bm{D}\psi_{2}) + \text{c.c.}] \notag \\
    &\quad +\qty[\mathrm{i}\bm{d}_{\text{I}}\cdot\qty(\psi_{1}^{*}\bm{D}\psi_{2} - \psi_{2}^{*}\bm{D}\psi_{1}) + \text{c.c}],
    \label{GLFreeEnergy}
\end{align}
where $\psi_{i}$ is a complex order parameter of a band $i$, $a_{i} = a_{0,i}\qty(T-T_{\text{c}})$, $a_{0,i}$ and $b_{i}$ are positive constants, $T$ is the system's temperature, $T_{\text{c}}$ is the transition temperature, $m^{*}_{i}$ is the effective mass of the condensates, and $\bm{D}=-\mathrm{i}\nabla-e^{*}\bm{A}$ is the covariant derivative with an electric charge $e^{*}=2e$ and an electromagnetic vector potential $\bm{A}$.
The first line in Eq.~(\ref{GLFreeEnergy}) constitutes the free-energy density of two individual single-band superconductors,
each describing a Mexican hat potential below the transition temperature $T_{\text{c}}$.

The other terms correspond to the couplings between the two different order parameters with coefficients $\epsilon$, $\eta$, and $\bm{d}_{\text{I}}$ ($\bm{d}_{\text{I}}$ should be real due to the hermiticity and particle-hole symmetry \cite{Nagashima2024}).
The $\epsilon$ term represents the Josephson coupling between the two order parameters ($\epsilon$ is taken to be real), the $\eta$ term describes the drag effect \cite{Drag_effect_Doh1999, Drag_effect_Grogorishin2016, Yerin2022, Yerin2023} ($\eta$ must be real due to the time-reversal symmetry),
and the $\bm{d}_{\text{I}}$ term, the so-called Lifshitz invariant \cite{LandauLifshitz}, plays a role similar to the Dzyaloshinskii-Moriya interaction in magnets \cite{DMI_Dzyaloshinsky1958, DMI_Moriya1960}.
The Lifshitz invariant term is known to appear in multiband systems even when the system has an inversion symmetry \cite{Nagashima2024}.

To discuss the instability of the uniform superconducting state without external fields, let us neglect the vector potential $\bm{A}$ for the moment. 
The following argument is similar to the Appendix B of Ref.~\cite{Nagashima2024}.
However, we give a physical interpretation to realize the instability of the uniform superconductor and discuss the instability at large $\bm{q}$ limit, which was not well understood in Ref.~\cite{Nagashima2024}.
We assume that the order parameter $\psi_i$ has a spatial dependence in the form of $e^{\mathrm{i}\bm{q}_{i}\cdot\bm{r}}$ (not $\cos\qty(\bm{q}_i\cdot\bm{r})$, see the end of this section).
We substitute
\begin{equation}
    \bm{D} = -\mathrm{i}\nabla, \quad \psi_{i} = \psi_{i,0}e^{\mathrm{i}\theta_{i,0}}e^{\mathrm{i}\bm{q}_{i}\cdot\bm{r}}
    \quad (\psi_{i,0}\in\mathbb R)
\end{equation} 
into Eq.~(\ref{GLFreeEnergy}) and obtain
\begin{align}
    \mathcal{F} &= \sum_{i=1,2}\qty[ a_{i}\psi_{i,0}^{2} + \frac{b_{i}}{2}\psi_{i,0}^{4} + \frac{q_{i}^{2}}{2m^{*}}\psi_{i,0}^{2}] \notag \\
    &\quad + 2\epsilon\psi_{1,0}\psi_{2,0}\cos{\qty[\theta_{1,0}-\theta_{2,0}  + \qty(\bm{q}_{1}-\bm{q}_{2})\cdot\bm{r}]} \notag \\
    &\quad + 2\eta\qty(\bm{q}_{1}\cdot\bm{q}_{2})\psi_{1,0}\psi_{2,0}\cos{\qty[\theta_{1,0}-\theta_{2,0}  + \qty(\bm{q}_{1}-\bm{q}_{2})\cdot\bm{r}]} \notag \\
    &\quad + 2\psi_{1,0}\psi_{2,0}\qty[\bm{d}_{\text{I}}\cdot\qty(\bm{q}_{1}+\bm{q}_{2})] \notag \\
    &\quad \times \sin{\qty[\theta_{1,0}-\theta_{2,0} + \qty(\bm{q}_{1}-\bm{q}_{2})\cdot\bm{r}]}.
\end{align}
If $\bm{q}_{1}\neq\bm{q}_{2}$, the spatially oscillating terms disappear after integration, and the two bands are decoupled from each other,
which is not our focus.
We thus put $\bm{q}_{1}=\bm{q}_{2}=\bm{q}$. By setting $\phi=\theta_{1,0}-\theta_{2,0}$, we obtain
\begin{align}
    \mathcal{F} &= \sum_{i=1,2}\qty[ a_{i}\psi_{i,0}^{2} + \frac{b_{i}}{2}\psi_{i,0}^{4} + \frac{q^{2}}{2m^{*}}\psi_{i,0}^{2} ] \notag \\
    &\quad + 2\psi_{1,0}\psi_{2,0} \left[ \qty(\epsilon + \eta q^{2})\cos{\phi} + 2\qty(\bm{d}_{\text{I}}\cdot\bm{q})\sin{\phi}\right].
    \label{GL_phi}
\end{align}
We take a functional derivative in terms of $\phi$ to minimize the free energy:
\begin{align}
    \frac{\delta \mathcal{F}}{\delta \phi} &= -2\psi_{1,0}\psi_{2,0} \times\qty[ \qty(\epsilon + \eta q^{2}) \sin{\phi} - 2(\bm{d}_{\text{I}}\cdot \bm{q})\cos{\phi}] \notag \\
    &= 0.
\end{align}
It follows that
\begin{equation}
    \tan{\phi} = \frac{2\qty(\bm{d}_{\text{I}}\cdot\bm{q})}{\epsilon + \eta q^{2}}.
    \label{TanConstraint}
\end{equation}
The sign of $\phi$ is not uniquely determined from this condition, but the following choice gives the lower free energy:
\begin{align}
    \sin{\phi} = -\frac{2\qty(\bm{d}_{\text{I}}\cdot\bm{q})}{\sqrt{(\epsilon + \eta q^{2})^{2} + 4(\bm{d}_{\text{I}}\cdot\bm{q})^{2}}}, \label{GL_sine} \\
    \cos{\phi} = -\frac{\epsilon + \eta q^{2}}{\sqrt{(\epsilon + \eta q^{2})^{2} + 4(\bm{d}_{\text{I}}\cdot\bm{q})^{2}}}. \label{GL_cosine}
\end{align}
Note that this choice does not depend on the sign of $\epsilon$ and $\eta$.
These expressions give us the free energy minimized with respect to $\phi$:
\begin{align}
    \mathcal{F} &= \sum_{i=1,2}\qty[ a_{i}\psi_{i,0}^{2} + \frac{b_{i,0}}{2}\psi_{i,0}^{4} + \frac{q^{2}}{2m^{*}_{i}}\psi_{i,0}^{2}] \notag \\
    & \quad - 2\psi_{1,0}\psi_{2,0} \sqrt{\qty(\epsilon + \eta q^{2})^{2} + 4\qty(\bm{d}_{\text{I}}\cdot\bm{q})^{2}},
    \label{GL_q}
\end{align}
which is always valid for an arbitrary value of $q$.

To proceed further, let us assume that
\begin{equation}
    m^{*}_{1}=m^{*}_{2}=m^{*}, \quad \psi_{1,0}=\psi_{2,0}=\psi_{0}.
    \label{GL_parameter}
\end{equation}
At the minimum of the free energy, $\bm{d}_{\text{I}}$ and $\bm{q}$ are pointing in the same direction.
To examine the instability around $q=0$, we expand the free energy (\ref{GL_q}) in terms of small $q$:
\begin{align}
    \mathcal{F} &\approx 2\qty(a-|\epsilon|)\psi_{0}^{2} + b\psi_{0}^{4} \notag \\
    &\quad +  2\psi_{0}^{2}\qty[ \frac{1}{2m^{*}} - \frac{\epsilon \eta + 2d_{\text{I}}^{2}}{|\epsilon|} ]q^{2}.
    \label{GL_Smallq_minimized}
\end{align}
The sufficient condition for the nonzero $q$ state to appear is given by the negativity of the coefficient of $q^{2}$:
\begin{equation}
    d_{\text{I}}^{2} > \frac{1}{2}\qty( \frac{|\epsilon|}{2m^{*}} - \epsilon\eta).
    \label{SufCondition}
\end{equation}
The r.h.s. of Eq.~(\ref{SufCondition}) describes the effect of ``aligning" the phase between the two order parameters, as we can see this from the $\cos{\phi}$ term in Eq.~(\ref{GL_phi}).
On the flip side, the l.h.s. of Eq.~(\ref{SufCondition}) describes the effect of ``twisting" the phase between the two order parameters, as the $\sin{\phi}$ term in Eq.~(\ref{GL_phi}) suggests.
Thus, Eq.~(\ref{SufCondition}) says that when the effect of ``twisting" is larger than that of the ``aligning," the nonzero $q$ state (one of the PDW states, which we shall call a small $\bm{q}$ state because the state is realized by the instability at $\bm{q}=\bm{0}$) becomes more stable than the uniform ($\bm{q}=0$) state.
We emphasize that the existence of the Lifshitz invariant term does not necessarily lead to the PDW state.
To realize the PDW state, one needs a sufficiently large $d_{\text{I}}$ that should exceed the threshold value
given by Eq.~(\ref{SufCondition}).
The emergence of the PDW state is not merely determined by the presence of $d_{\text{I}}$ but by
the competition between the ``twisting" and ``aligning."

There is another condition for the PDW state that is given by the behavior of the free energy at large $q$, where we have
\begin{equation}
    \mathcal{F}\approx 2\psi_{0}^{2}\qty( \frac{1}{2m^{*}} - |\eta|)q^{2}.
\end{equation}
Hence, when the inverse of the condensate's effective mass (times $\frac{1}{2}$) becomes smaller than the coupling strength of the drag effect,
\begin{equation}
    \frac{1}{2m^{*}} < |\eta|,
    \label{StrongCondition}
\end{equation}
the coefficient of $q^{2}$ becomes negative.
A higher $q$ state may have a lower free energy, and the system becomes unstable within the GL theory based on Eq.~(\ref{GLFreeEnergy}).
The true minimum of the free energy can be found outside our treatment, which we call the large $\bm{q}$ state.
A condition similar to Eq.~(\ref{StrongCondition}) has been derived previously \cite{Yerin2008}. We emphasize that Eq.~(\ref{StrongCondition}) can be used as a criterion for the phase transition to the PDW state.

The condition (\ref{StrongCondition}) is stronger than that of Eq.~(\ref{SufCondition}): when Eq.~(\ref{StrongCondition}) is satisfied, the nonzero $\bm{q}$ state is more stable than the uniform one even when Eq.~(\ref{SufCondition}) is not satisfied.
The instability caused by Eq.~(\ref{StrongCondition}) is independent of $d_{\text{I}}$, meaning the large $\bm{q}$ state can happen even when the Lifshitz invariant is absent.
These two instabilities are qualitatively different because the former instability occurs in competition between the "twisting" and ``aligning" effects, while the latter is caused by the dominance of the drag effect.
This difference is reflected in the destination state: the former instability with the Lifshitz invariant realizes the small $\bm{q}$ PDW state, and the latter, with the huge electron effective mass, leads to the large $\bm{q}$ PDW state.

Note that the GL theory may not be suitable for describing the ground state of the large $\bm{q}$ state because when $q$ becomes quite large, the higher-order terms will come to have an effect.
Nevertheless, one can effectively discuss the instability for large $\bm{q}$ state within our treatment of the GL theory.

Another important note is that we expand the order parameter with $e^{\mathrm{i}\bm{q}\cdot\bm{r}}$.
Although the PDW state has a $\cos{(\bm{q}\cdot\bm{r})}$ dependence, the order parameter should be expanded by the exponential function, not by the cosine function, as we will see later when we microscopically derive the GL free energy.

\subsection{\label{sec:level2-2}Optically active Higgs and Leggett modes in PDW states}
Next, we demonstrate that the Higgs mode can appear together with the Leggett mode in the linear response regime by confirming the linear coupling between the amplitude or relative phase fluctuation of the order parameters and the vector potential.
We begin with Eq.~(\ref{GL_phi}), taking the fluctuations into account by $\psi_{i,0}\to\psi_{0}+H_{i}$ and $\phi\to\phi + \phi_{\text{L}}$, where $H_{i}$ and $\phi_{\text{L}}$ are amplitude and relative phase fluctuations, respectively.
We also include the effect of the external field via the vector potential $\bm{A}$ with $\bm{q}\to \bm{q}-e^{*}\bm{A}$.
After expanding the free energy $\mathcal{F}$ around the ground state and neglecting the spatial derivatives of the fluctuations, we obtain the terms related to the linear response of the fluctuations $\mathcal{F}_{\text{FL}}$ as
\begin{align}
    \mathcal{F}_{\text{FL}} &= \mathcal{F}_{\text{Higgs}} + \mathcal{F}_{\text{Leggett}}, \\
    \mathcal{F}_{\text{Higgs}} &= \qty[\frac{1}{m^{*}} + 2\eta(\cos{\phi})] \notag \\
    &\quad\times\psi_{0}\qty( (\bm{q}-e^{*}\bm{A})^{2} - \bm{q}^{2})\qty(H_{1}+H_{2}) \notag \\
    &\quad -4\bm{d}_{\text{I}}\cdot\qty(e^{*}\bm{A})\psi_{0}\qty(\sin{\phi})\qty(H_{1}+H_{2}), 
    \label{eq: F_Higgs}
    \\
    \mathcal{F}_{\text{Leggett}} &= -4\eta\qty(\sin{\phi})\psi_{0}^{2}\qty( (\bm{q}-e^{*}\bm{A})^{2} - \bm{q}^{2})\phi_{\text{L}} \notag \\
    &\quad    -8\bm{d}_{\text{I}}\cdot\qty(e^{*}\bm{A})\psi_{0}^{2}\qty(\cos{\phi})\phi_{\text{L}},
\end{align}
where $\sin\phi$ and $\cos\phi$ are defined in Eqs.~(\ref{GL_sine}) and (\ref{GL_cosine}), respectively.
One can see that there is a linear coupling between the Higgs/Leggett mode and the vector potential ($\propto\bm A H_i, \bm A \phi_L$) in the case of $\bm q\neq \bm 0$.
When $\bm{q}=\bm{0}$, the linear coupling to the Higgs mode vanishes since $\sin{\phi}=0$ (see Eq.~(\ref{TanConstraint})), and the coupling between the amplitude fluctuation $H_{i}$ and the vector potential $\bm{A}$ appears in the third order ($\bm{A}^{2}H_{i}$).
In contrast, the linear Leggett mode response survives even for $\bm{q}=\bm{0}$ due to the presence of the second term in $\mathcal{F}_{\text{Leggett}}$, which is consistent with the previous results ~\cite{Nagashima2024, Kamatani2022}.
The remaining terms in the free energy $\mathcal F$ that exist for $\bm{A}=\bm{0}$ are shown in Appendix~\ref{Apdx.B} (see Eq.~(\ref{appendix: F})).

Note that in the absence of the Lifshitz invariant the leading coupling between the Leggett mode and vector potential appears in the fourth order ($\bm A^2 \phi_L^2$) if the time-reversal symmetry exists (when the time-reversal symmetry is broken, the coefficient $\eta$ in the drag term can take a complex value, which then leads to the third-order coupling ($\bm A^2 \phi_L$)). 
This is in stark contrast to the situation of the Higgs mode, where the leading coupling term in the absence of the Lifshitz invariant (or with $\bm q=\bm 0$) is in the third order ($\bm{A}^{2}H_{i}$).
This difference implies that the coupling between the Leggett mode and light is much weaker than that for the Higgs mode.
Particularly, the contribution of the Leggett mode to the third harmonic generation will become much weaker than expected from the mean-field analysis \cite{Murotani2017} due to the screening effect.

A special case happens when $H_{1}=-H_{2}$ with $\bm{q}\neq \bm{0}$.
For instance, in the $\pi$-PDW state, the fluctuation that enhances one of the order parameters and simultaneously weakens the other one, which corresponds to the case in which $H_{1}$ and $H_{2}$ have the same sign, does not seem to occur because this kind of fluctuation does not fit with the existing PDW order.
Hence, $H_{1}=-H_{2}$ is preferred, and the linear coupling between the Higgs mode and the vector potential vanishes (see Eq.~(\ref{eq: F_Higgs})), and only the Leggett mode can appear in the linear response regime.
Although we consider the two-band case for simplicity, we can extend this argument to the $n$-band case: when $H_{1}=-H_{2}=H_{3}=\cdots$ holds, the number of the Higgs mode in the linear response regime will be zero (even number band case) or one (odd number band case).
Therefore, by checking the number of Higgs modes in the linear response regime, we can, in principle, distinguish the $\pi$ (and ($\pi,\pi$))-PDW state and other PDW states.

We also note that the terms should be invariant under  $(\bm{q},\bm{A},\phi_{\text{L}})$ $\to$ $(-\bm{q},-\bm{A},-\phi_{\text{L}})$ to satisfy the time-reversal symmetry, but the consequence does not change.

\subsection{\label{sec:level2-3}Particle-hole symmetry}
Let us briefly mention the particle-hole symmetry in the finite $\bm{q}$ state.
The particle-hole symmetry is a built-in symmetry in the BCS theory and should also be respected in the GL theory.
The particle-hole transformation changes the order parameter of the finite $q$ state as $\psi\to\psi^{\dagger}$,
which is similar to the uniform case.
However, because of the real-space modulation factor $e^{\mathrm{i}\bm{q}\cdot\bm{r}}$, the symmetry operation should be read as
\begin{align}
    H_{i}&\to H_{i}, \notag \\
    \theta_{i}&\to -\theta_{i}, \notag \\
    \bm{A}&\to -\bm{A}, \notag \\
    \bm{q} &\to -\bm{q}.
\end{align}
We can easily confirm that the Higgs and Leggett terms contributing to the linear response are invariant under the particle-hole transformation.
We can also see the particle-hole symmetry later in the BdG Hamiltonian within the mean-field approximation.

\section{\label{sec:level3}Microscopic theory of the PDW states in real space}
Here we introduce a simple 1D microscopic model that shows the PDW state, 
which is treated in the real-space formulation of the mean-field approximation.
We calculate the superconducting gap function in real space and find that the gap oscillates spatially in a certain parameter regime.
As we will see later, the oscillation period depends on the value of the parameters. 
The obtained values of the finite center-of-mass momentum $q$ are $\pi/2$ and $\pi$, corresponding to the small and large $q$ states.
We fix the lattice constant $a=1$ from now on.

\begin{figure}
    \centering
    \includegraphics[scale=1.0]{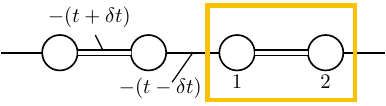}
    \caption{
    Schematic picture of the 1D model (\ref{Hamiltonian_1D real space}) that exhibits a PDW state.
    The model has two kinds of hopping amplitudes, $-(t-\delta t)$ (single lines) and $-(t+\delta t)$ (double lines).
    The yellow rectangle encloses one of the unit cells, including two lattice sites labeled by $1$ and $2$.
    The point group of this model is $C_{i}$, in which the Lifshitz invariant is allowed to exist in the free energy \cite{Kamatani2022, Nagashima2024}.
    }
    \label{Pic_model}
\end{figure}

Let us consider a microscopic model with the Hamiltonian $\mathcal{H}_{\text{r}}$ given by
\begin{align}
    \mathcal{H}_{\text{r}} &= \sum_{j=1}^{N}\sum_{\sigma}\left[ -\left\{ t + \qty(-1)^{j}\qty(\delta t) \right\}c^{\dagger}_{j,\sigma}c_{j+1,\sigma} + \text{H.c.} \right] \notag \\
    &\quad - \mu\sum_{j=1}^{N}\sum_{\sigma}c^{\dagger}_{j,\sigma}c^{\dagger}_{j,\sigma} + U\sum_{j=1}^{N}c^{\dagger}_{j,\up}c_{j,\up}c^{\dagger}_{j,\down}c_{j,\down} \notag \\
    &\quad + V\sum_{j=1}^{N}\left[c^{\dagger}_{j,\up}c^{\dagger}_{j,\down}c_{j+1,\down}c_{j+1,\up} + \text{H.c.} \right],
    \label{Hamiltonian_1D real space}
\end{align}
where $c_{j,\sigma}^\dagger$ is the creation operator of fermions at site $j$ with spin $\sigma$,
$-(t + (-1)^{j}(\delta t))$ is the staggered hopping strength, $\mu$ is the chemical potential, $U$ is the on-site interaction, $V$ is the nearest-neighbor pair-hopping strength, and $N$ is the number of lattice sites (which is assumed to be an even integer).
We impose the periodic boundary condition.
Because of the two kinds of hoppings, $-(t\pm \delta t)$, the unit cell of this model contains two non-equivalent lattice sites.
The schematic picture of the kinetic part of the model is depicted in Fig.~\ref{Pic_model}.
We can verify that the model has a Lifshitz invariant arising from two distinguishable lattice sites according to the previous study \cite{Nagashima2024}.

To capture nonuniform superconducting states, we employ the mean-field approximation in real space to describe the $s$-wave superconductivity with a spatial structure.
The mean-field Hamiltonian $\mathcal{H}_{\text{rMF}}$ reads
\begin{align}
    \mathcal{H}_{\text{rMF}}% \notag \\
    &= \sum_{j=1}^{N}\sum_{\sigma}\left[ -\left\{ t + \qty(-1)^{j}\qty(\delta t) \right\}c^{\dagger}_{j,\sigma}c_{j+1,\sigma} + \text{H.c.} \right] \notag \\
    &\quad - \mu\sum_{j=1}^{N}\sum_{\sigma}c^{\dagger}_{j,\sigma}c_{j,\sigma} + \sum_{j=1}^{N}\qty[ \Delta_{j}c^{\dagger}_{j,\up}c^{\dagger}_{j,\down} + \text{H.c.} ],
\end{align}
where
\begin{equation}
    \Delta_{j} = U\ev*{c_{j,\down}c_{j,\up}} + V\ev*{c_{j+1,\down}c_{j+1,\up}} + V\ev*{c_{j-1,\down}c_{j-1,\up}}
    \label{Gap_realspace_MF}
\end{equation}
is the site-dependent gap function.
By defining the generalized Pauli matrices $\tau_{x(y)j}$ as
\begin{equation}
    \tau_{xj} = \mqty[ O & A_{j} \\ A_{j} & O] \text{ and } \tau_{yj} = \mqty[ O & -\mathrm{i}A_{j} \\ \mathrm{i}A_{j} & O],
    \label{PauliMatrix}
\end{equation}
with $[A_{j}]_{\gamma\gamma'} = \delta_{j\gamma}\delta_{\gamma\gamma'}$, the self-consistent equation to be solved is written as
\begin{equation}
    \ev*{c_{j,\down}c_{j,\up}} = \sum_{l}\frac{1}{2}\frac{\mel{\varphi_{l}}{\tau_{xj} - \mathrm{i}\tau_{yj}}{\varphi_{l}}}{e^{\beta E_{l}} + 1}.
\end{equation}
Here, $\beta$ is the inverse temperature, and we introduce the $l$-th eigenvector $\ket{\varphi_{l}}$ of $\mathcal{H}_{\text{rMF}}$ with an eigenvalue $E_{l}$.
To judge whether the self-consistent result gives the lowest free energy state or not, we calculate the free energy $\mathcal{F}_{\text{r}}$:
\begin{equation}
    \mathcal{F}_{\text{r}} = -\sum_{i,j}\Delta^{*}_{i}\tilde{U}^{-1}_{ij}\Delta_{j} - \sum_{j}\qty[E_{j} + \frac{2}{\beta}\ln{\qty(1 + e^{-\beta E_{j}})}],
\end{equation}
where $E_{j}$ for $j=1,2,\cdots,N$ are the $j$-th largest eigenvalues of the mean-field Hamiltonian $\mathcal{H}_{\text{rMF}}$ and $\tilde{U}_{ij} = U\delta_{i,j} + V\qty(\delta_{i,j-1} + \delta_{i,j+1})$.

\begin{figure}
    \centering
    \includegraphics[scale=0.4]{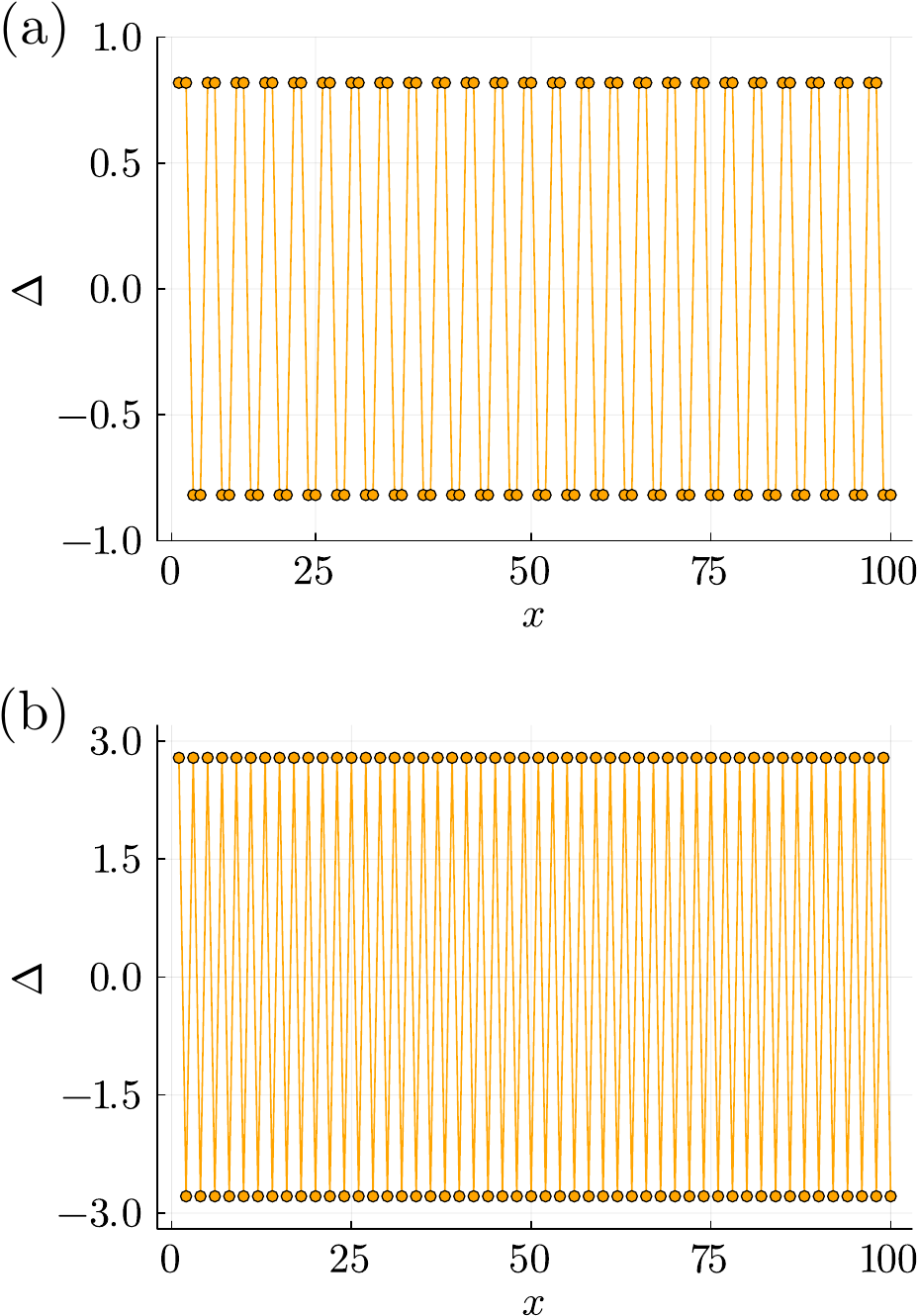}
    \caption{
    Real-space configurations of the gap function in the PDW states 
    for the 1D model (\ref{Hamiltonian_1D real space}) with different pair hoppings $V$.
    (a) $\pi/2$-PDW state with $V=0.5$.
    (b) $\pi$-PDW state with $V=1.4$.
    In both cases, the parameters $\delta t=0.35$, $\mu=-0.8$, $T=0.001$, and $U=-3.0$ are used.
    The number of lattice sites is $N=100$.
    }
    \label{RealSpaceGap}
\end{figure}

We now show two examples of the PDW gap function calculated in the above formalism.
We hereby fix $t=1$ as the unit of energy.
We choose $\delta t= 0.35$, $\mu=-0.8$, $T=0.001$, $U=-3.0$, and $V=0.5$ for the $\pi/2$-PDW state, 
and $V=1.4$ with the other parameters fixed for the $\pi$-PDW state.
The finite center-of-mass momentum $\bm{q}$ is measured on a site and not on a unit cell, meaning that if the gap function acquires the phase factor $e^{\mathrm{i}q}$ by moving from one site to another, the finite center-of-mass momentum is determined to be $q$.
We take the number of lattice sites $N=100$.
The computed results are shown in Fig.~\ref{RealSpaceGap}.
When $\delta t$ is large and $V$ is moderately large, electrons tend to localize within the bond whose hopping strength is $-(t+\delta t)$ but the sign of the gap functions should be changed to lower the energy due to the pair-hopping $V$, the every two site oscillation of the gap function is realized.
In contrast, when $V$ is sufficiently large, the electrons can lower the energy more when the gap function changes its sign at every site, and hence, every one-site oscillation of the gap function is realized.
It should be noted that the $\pi$-PDW state has an odd parity even though the Hamiltonian has an even parity.
This is reminiscent of the odd-parity superconductors \cite{Sato2010, Landaeta2022}.
In this paper, we call the phase the $\pi$-PDW state since the superconducting order parameter oscillates in real space, and the $\pi$-PDW state is continuously connected to the case of $\delta t=0$, where the unit cell only contains one lattice site and the $\pi$-PDW state has a doubled periodicity.

The PDW gap function is written as $\Delta(j)=\Delta_{0}\cos{\qty(q_{0}j + \theta)}$, where $j$ is the lattice site index, and $\theta$ is a phase shift, as discussed in previous research \cite{PDW_review}.
The gap function $\Delta(j)$  is decomposed into two parts,
\begin{align}
    \Delta(j) &= \frac{\Delta_{0}}{2}e^{\mathrm{i}\qty(q_{0}j + \theta)} + \frac{\Delta_{0}}{2}e^{-\mathrm{i}\qty(q_{0}j + \theta)} \notag \\
    &= \frac{\Delta_{q_{0}}}{2}e^{\mathrm{i}q_{0}j} + \frac{\Delta_{-q_{0}}}{2}e^{-\mathrm{i}q_{0}j},
\end{align}
with $\Delta_{q_{0}} = \Delta_{0}e^{\mathrm{i}\theta}$, $\Delta_{-q_{0}} = \Delta_{0}e^{-\mathrm{i}\theta}$, and $\Delta_{q_{0}} = \Delta^{*}_{-q_{0}}$. 
As we will see in the next section, the order parameters appearing in the GL free energy are $\Delta_{\pm q_{0}}$ associated with $e^{\pm\mathrm{i}q_{0}j}$, and not $\Delta(j)$ associated with $\cos{(q_{0}j+\theta)}$.
This is why the expansion of the order parameter of the GL theory in the form of $e^{\mathrm{i}q_{0}j}$ in the previous section (Sec.~\ref{sec:level2}) is justified.

We would like to note that the translation is equivalent to shifting the global phase of the gap function.
In the case of $\pi/2$-PDW state, the two-site translation is equivalent to shifting the phase of the gap function by $\pi$, or namely applying $e^{\mathrm{i}\pi}$ to the gap function.
Hence, the system's unit cell does not need to enlarge, and as in the normal state, we can treat it as having two non-equivalent lattice sites.
The situation is the same in the $\pi$-PDW state; the one-site translation is equivalent to shifting the phase of the gap function by $\pi$, and the unit cell stays unchanged.

\section{\label{sec:level4}Microscopic theory of the PDW states in momentum space}
We now move on to the momentum-space formulation. 
In momentum space, it is possible to compute various physical quantities by taking advantage of the periodicity of the lattice and PDW states, reducing the numerical cost of the calculations.
At the end of this section, we calculate the linear optical conductivity of the two distinct $\pi/2$- and $\pi$-PDW states.
Still, it will be essential to scrutinize these two PDW states' equilibrium properties first and confirm the consistency with the macroscopic GL theory.

To this end, we start by deriving the GL parameters (up to an overall factor), free energy at the saddle point, gap equation, and linear optical conductivity in the momentum space of the PDW state for the model in Fig.~\ref{Pic_model}.
Based on the GL parameters, we can see that the condition for the small $q$ PDW state [Eq.~(\ref{SufCondition})] is satisfied when the difference of the hopping strengths $\delta t$ is large and the pair hopping $V$ is also large. 
The condition for large $q$ PDW state [Eq.~(\ref{StrongCondition})] is satisfied when the pair hopping $V$ is sufficiently large.
In addition, we can validate the expansion of the ``order parameter" by the exponential function $e^{\mathrm{i}qx}$, not by the cosine function because the free energy is a function of $\Delta_{\pm q_{0}}$, and not of $\Delta(x)$.

After the formulation, we will see the normal state phase diagram, $\delta t$-$|\mu|$ and $\delta t$-$V$ phase diagrams based on the GL parameters with our microscopic model using Eqs.~(\ref{SufCondition}) and (\ref{StrongCondition}), $q$-dependence of the free energy, $\delta t$-$|\mu|$ and $\delta t$-$V$ phase diagrams by the microscopic calculations, and the temperature dependence of the $\pi/2$-PDW gap function.
Depending on the parameter values, we can see that the uniform, $\pi/2$-PDW, and the $\pi$-PDW states appear in the phase diagrams.
In particular, the $\pi/2$-PDW state occurs when $|V/U|$ is moderately large with large $|\mu|$, and the $\pi$-PDW state occurs when $|V/U|$ is sufficiently large, which roughly agrees with the GL theory.
Finally, we calculate the linear optical conductivity to determine how the Higgs and Leggett modes appear in the linear response regime in the PDW states.
We find that the hybridization occurs between the Higgs and Leggett modes in the $\pi/2$-PDW state, while only the phase mode appears in the $\pi$-PDW state, indicating the substantial difference between the $\pi/2$- and $\pi$-PDW states. 
For the derivation details, we refer to Appendix~\ref{Apdx.A}.

\subsection{\label{sec:level4-1}Formalism}
The total Hamiltonian $\mathcal H$ of the model in Fig.~\ref{Pic_model} is written in the momentum space as
\begin{align}
    \mathcal{H} &= \mathcal{H}_{\text{kin}} + \mathcal{H}_{\text{int}}, \notag \\
    \mathcal{H}_{\text{kin}} &= \sum_{k\alpha\alpha'\sigma}\xi_{\alpha\alpha'}(k)c^{\dagger}_{k\alpha\sigma}c_{k\alpha'\sigma}, \notag \\
    \mathcal{H}_{\text{int}} &= \sum_{kk'\alpha\alpha'}\sum_{q=\pm q_{0}}\frac{1}{2}\tilde{U}_{\alpha\alpha'}(q) \notag \\
    &\quad\times c^{\dagger}_{k+q,\alpha\up}c^{\dagger}_{-k+q,\alpha\down}c_{-k'+q,\alpha'\down}c_{k'+q,\alpha'\up},
    \label{Hamiltonian_1D}
\end{align}
where the kinetic part $\xi(k)$ is
\begin{equation}
    \xi(k) = \mqty[ -\mu & f(k) \\ f^{\dagger}(k) & -\mu ]
    \label{Kinetic_N}
\end{equation}
with $f(k) = -2t\cos{(k/2)}-\mathrm{i}\cdot2(\delta t)\sin{(k/2)}$, and the interaction $\tilde{U}(q)$ is
\begin{equation}
    \tilde{U}(q) = \mqty[ U & 2V\cos{(q)} \\ 2V\cos{(q)} & U ].
\end{equation}
Here we assume $U<0$, $V>0$, and $|V/U|<1/2$.
The index $\alpha=1,\cdots, n_{0}$ labels the bands (in our model $n_{0}=2$).
We will use the path-integral approach in imaginary time $\tau$.
We first arrange the momentum $k$ of the creation/annihilation operators using the Brillouin zone periodicity.
We shift the momentum $k$ in $\mathcal{H}_{\text{kin}}$ by $\pm q_{0}$, and decompose it into two parts, $k\pm q_{0}$, as below:
\begin{equation}
    \mathcal{H}_{\text{kin}} = \sum_{k\alpha\alpha'\sigma}\sum_{q=\pm q_{0}}\frac{1}{2}\xi_{\alpha\alpha'}(k+q)c^{\dagger}_{k+q,\alpha\sigma}c_{k+q,\alpha'\sigma}.
\end{equation}
The partition function of the whole system $\mathcal{Z}$ is expressed as $\mathcal{Z}=\int\mathcal{D}\qty(c^{\dagger}c)e^{-S[c^{\dagger},c]}$ with the Euclidean action,
\begin{align}
    S[c^{\dagger},c] &= \int_{0}^{\beta}d\tau\qty( \sum_{k\alpha\sigma}c^{\dagger}_{k\alpha\sigma}\partial_{\tau}c_{k\alpha\sigma} + \mathcal{H} ) \notag \\
    &= \int_{0}^{\beta}d\tau\qty(\sum_{k\alpha\sigma}\sum_{q=\pm q_{0}}\frac{1}{2}c^{\dagger}_{k+q,\alpha\sigma}\partial_{\tau}c_{k+q,\alpha\sigma} + \mathcal{H} ).
\end{align}
For simplicity, we set the volume size of the whole system to be one.
We perform the Hubbard--Stratonovich transformation to split the interaction part of the Hamiltonian $\mathcal{H}_{\text{int}}$ into auxiliary bosonic fields, $\Delta_{\pm q_{0}\alpha}$, $\Delta^{*}_{\pm q_{0}\alpha}$, and fermions.
The resulting expression is
\begin{align}
    \mathcal Z
    &=
    \prod_{q=\pm q_{0}}\int\mathcal{D}\qty(\Delta^{*}_{q},\Delta_{q})\text{exp}\left(-\int_{0}^{\beta}d\tau\left[ 2\frac{\Delta^{*}_{q\alpha}}{2}\tilde{U}^{-1}_{\alpha\alpha'}(q)\frac{\Delta_{q\alpha'}}{2} \right.\right. \notag \\
    &+ \left.\left.\sum_{\alpha k}\qty(\frac{\Delta^{*}_{q\alpha}}{2}c_{k+q,\alpha\up}c_{-k+q,\alpha\down} + \frac{\Delta_{q\alpha}}{2}c^{\dagger}_{-k+q,\alpha\down}c_{k+q,\alpha\up} ) \right]\right).
\end{align}
To simplify the expression, we take a basis,
\begin{equation}
    \Psi_{q,k} = \qty[C^{\dagger}_{q\up}(k)\ C_{q\down}(-k)\ C_{-q\down}(-k)\ C^{\dagger}_{-q\up}(k)]^{\text{T}},
\end{equation}
where $C^{\dagger}_{q\sigma}(k) = \qty[c^{\dagger}_{k+q,1\sigma}\ \cdots \ c^{\dagger}_{k+q,n_{0}\sigma}]$.
On this basis, we move on to the frequency space and obtain the action $S$ with the fluctuation $\delta\Delta_{\pm q_{0}\alpha}(\mathrm{i}\omega_{n})$ as
\begin{align}
    &S[c^{\dagger},c,\Delta^{*}_{q_{0}},\Delta_{q_{0}},\Delta^{*}_{-q_{0}},\Delta_{-q_{0}}] \notag \\
    =&-\frac{\beta}{2}\sum_{q=\pm q_{0}}\sum_{\alpha\alpha'}\Delta^{*}_{q,0\alpha}\tilde{U}^{-1}_{\alpha\alpha'}(q)\Delta_{q,0\alpha'} \notag \\
    &-\frac{1}{2}\sum_{q=\pm q_{0}}\sum_{\alpha\alpha'}\sum_{n}\delta\Delta^{*}_{q\alpha}(\mathrm{i}\omega_{n})\tilde{U}^{-1}_{\alpha\alpha'}(q)\delta\Delta_{q\alpha'}(\mathrm{i}\omega_{n}) \notag \\
    &-\frac{1}{2\beta^{2}}\sum_{k}\sum_{m,n}\Psi^{\dagger}_{q_{0},k}(\mathrm{i}\omega_{m})\tilde{G}_{q_{0}}\qty(\mathrm{i}\omega_{m},\mathrm{i}\omega_{n};q)\Psi_{q_{0},k}(\mathrm{i}\omega_{n}).
\end{align}
Here, the fluctuation $\delta\Delta_{\pm q_{0}\alpha}(\mathrm{i}\omega_{n})$ is introduced by $\Delta_{\pm q_{0}\alpha} = \Delta_{\pm q_{0},0\alpha} + \delta\Delta_{\pm q_{0}\alpha}$ with the saddle point contribution $\Delta_{\pm q_{0},0\alpha}$ and $\delta\Delta_{\pm q_{0}\alpha}= \Delta_{\pm q_{0},x\alpha} - \mathrm{i}\Delta_{\pm q_{0},y\alpha}$ ($\delta\Delta_{\pm q_{0},\mu\alpha}$ ($\mu=x,y$) is real).
We note that the fluctuation of the momentum around $\pm q_{0}$ is not included for simplicity.
The inverse of Green's function $\tilde{G}^{-1}_{q_{0}}(\mathrm{i}\omega_{m},\mathrm{i}\omega_{n};k)$ is defined as
\begin{align}
    &\tilde{G}^{-1}_{q_{0}}\qty(\mathrm{i}\omega_{m},\mathrm{i}\omega_{n};k) \notag \\
    =&\frac{\beta\delta_{m,n}}{2}\mqty[ G^{-1}_{+,0}\qty(\mathrm{i}\omega_{n};k) & O \\ O & G^{-1}_{-,0}(\mathrm{i}\omega_{n};k)] \notag \\
    &-\frac{\beta}{2}\mqty[ \delta\Delta\qty(\mathrm{i}\omega_{m}-\mathrm{i}\omega_{n}) & O \\ O & \delta\Delta(\mathrm{i}\omega_{m}-\mathrm{i}\omega_{n})],
\end{align}
where
\begin{equation}
    G^{-1}_{\pm,0}(\mathrm{i}\omega_{n};k) = \mqty[ \mathrm{i}\omega_{n} \mp \xi(k+q_{0}) & -\Delta_{q_{0},0} \\ -\Delta_{-q_{0},0} & \mathrm{i}\omega_{n}\pm\xi(k-q_{0})],
    \label{Green_pm}
\end{equation}
and
\begin{equation}
    \delta\Delta(\mathrm{i}\omega_{m}-\mathrm{i}\omega_{n}) = \mqty[ O & \delta\Delta_{q_{0}}(\mathrm{i}\omega_{m}-\mathrm{i}\omega_{n}) \\ \delta\Delta_{-q_{0}}(\mathrm{i}\omega_{m}-\mathrm{i}\omega_{n}) & O],
\end{equation}
with $\delta\Delta_{\pm q_{0}}(\mathrm{i}\omega_{m}-\mathrm{i}\omega_{n})$ being an $n_{0}\times n_{0}$ diagonal matrix.
The fermionic path integral with the constraint $\Delta^{*}_{q_{0}}=\Delta_{-q_{0}}$ provides us with the expression below:
\begin{align}
    &S[\Delta_{q_{0}},\Delta_{-q_{0}}] \notag \\
    =&-\beta\sum_{\alpha\alpha'}\Delta_{-q_{0},0\alpha}\tilde{U}^{-1}_{\alpha\alpha'}(q_{0})\Delta_{q_{0},0\alpha'} \notag \\
    &-\sum_{\alpha\alpha'}\sum_{n}\delta\Delta_{-q_{0},0\alpha}(\mathrm{i}\omega_{n})\tilde{U}^{-1}_{\alpha\alpha'}(q_{0})\delta\Delta_{q_{0},0\alpha'}(\mathrm{i}\omega_{n}) \notag \\
    &-\frac{1}{2\beta^{2}}\sum_{k}\sum_{m,n}\text{Tr}\text{ln} \qty[ -\tilde{G}^{-1}_{q_{0}}\qty(\mathrm{i}\omega_{m},\mathrm{i}\omega_{n};k)].
\end{align}
Here we used the property $\tilde{U}^{-1}_{\alpha\alpha'}(q_{0})=\tilde{U}^{-1}_{\alpha'\alpha}(q_{0})=\tilde{U}^{-1}_{\alpha\alpha'}(-q_{0})$, and the coefficient $1/2$ in front of the trace term appears from the fact that we doubled the basis in response to the splitting $k$ into $k\pm q_{0}$.
By choosing the reference state and extracting the corresponding Green's function $\tilde{G}_{q_{0},0}$ with the self-energy $\Sigma$, we obtain
\begin{align}
    &S[\Delta_{q_{0}},\Delta_{-q_{0}}] \notag \\
    =&-\beta\sum_{\alpha\alpha'}\Delta_{-q_{0},0\alpha}\tilde{U}^{-1}_{\alpha\alpha'}(q_{0})\Delta_{q_{0},0\alpha'} \notag \\
    &-\sum_{\alpha\alpha'}\sum_{n}\delta\Delta_{-q_{0}\alpha}(\mathrm{i}\omega_{n})\tilde{U}^{-1}_{\alpha\alpha'}(q_{0})\delta\Delta_{q_{0}\alpha'}(\mathrm{i}\omega_{n}) \notag \\
    &-\frac{1}{2\beta^{2}}\sum_{\bm{k}}\sum_{m,n}\text{Tr}\text{ln}\qty[-\tilde{G}^{-1}_{q_{0},0}] \notag \\
    &-\frac{1}{2\beta^{2}}\sum_{k}\sum_{m,n}\text{Tr}\qty[ \sum_{L=1}^{\infty}\frac{\qty(\tilde{G}_{q_{0},0}\Sigma)^{L}}{L}].
    \label{Action}
\end{align}
One can see that the ``order parameters" that will appear in the free energy (or in the action here) are $\Delta_{\pm q_{0},0}$, and not $\Delta(x)$.
This indicates that the expansion of the order parameter as a cosine function is not allowed because each ``order parameter" $\Delta_{\pm q_{0}}$ is connected with $e^{\mathrm{i}q_{0}x}$.

Note that the saddle point inverse Green's function $G^{-1}_{\pm,0}$ in Eq.~(\ref{Green_pm}) manifests the particle-hole symmetry with a finite momentum $q_{0}$.
We need to exchange the electron and hole diagonal blocks and reverse the sign of the finite momentum $q_{0}$.
This feature of the particle-hole symmetry in the PDW state also appears in the microscopic GL formalism, as mentioned in Sec.~\ref{sec:level2-3}.

The free energy $\mathcal F$ without fluctuation is given by
\begin{align}
    \mathcal{F} &= -\sum_{\alpha\alpha'}\Delta_{-q_{0},0\alpha}\tilde{U}^{-1}_{\alpha\alpha'}(q_{0})\Delta_{q_{0},0\alpha'} \notag \\
    &\quad -\sum_{k}\sum_{\alpha=1}^{n_{0}}\qty[ E_{\alpha}(k) + \frac{2}{\beta}\ln{\qty(1+e^{-\beta E_{\alpha}(k)} )} ],
    \label{FE_w/_fluctuation}
\end{align}
where $E_{\alpha}(k)$ is the $\alpha$-th largest eigenvalues of the matrix,
\begin{equation}
    \mqty[ \xi(k+q_{0}) & \Delta_{q_{0},0} \\ \Delta_{-q_{0},0} & -\xi(k-q_{0})].
    \label{BdG}
\end{equation}

The self-consistent gap equation of $\Delta_{q_{0},0\alpha}$ is written as
\begin{align}
    &\Delta_{q_{0},0\alpha} \notag \\
    =& \frac{1}{\beta}\sum_{k}\sum_{n}\sum_{\alpha'}\tilde{U}_{\alpha\alpha'}(q_{0})\text{Tr}\qty[ \frac{\tau_{x\alpha}-\mathrm{i}\tau_{y\alpha}}{2}G_{+,0}(\mathrm{i}\omega_{n},k)].
    \label{Gap_equation}
\end{align}
The Green's function $G_{+,0}(\mathrm{i}\omega_{n},k)$ and generalized Pauli matrices $\tau_{x(y)\alpha}$ are defined in Eqs.~(\ref{Green_pm}) and (\ref{PauliMatrix}), respectively.

\subsubsection{GL parameters}
Let us suppose that the system's temperature is sufficiently close to the transition temperature and the magnitude of the gap function $|\Delta_{\pm q_{0}}|$ is small enough.
To consider the GL parameters and discuss the instability of the uniform state as we did in Sec.~\ref{sec:level2-1}, we expand the functions around $q_{0}=0$ and assume $q_{0}$ to be sufficiently small.
We then neglect the fluctuations and obtain the GL parameters as below:
\begin{align}
    \epsilon &= \frac{2V}{U^{2}-4V^{2}} + \epsilon_{\text{kin}}, \\
    d_{\text{I}} &= d_{\text{I},\text{kin}}, \\
    \frac{1}{2m^{*}} &= \frac{4UV^{2}}{(U^{2}-4V^{2})^{2}} + \frac{1}{2m^{*}_{\text{kin}}}, \\
    \eta &= -\frac{U^{2}+4V^{2}}{(U^{2}-4V^{2})^{2}}V + \eta_{\text{kin}},
    \label{GLparams_micro}
\end{align}
where $\epsilon_{\text{kin}}$, $d_{\text{I},\text{kin}}$, $1/2m^{*}_{\text{kin}}$, and $\eta_{\text{kin}}$ are the GL parameters without the contribution from the supercurrent term $S_{\text{SC}}=-\beta\sum_{\alpha\alpha'}\Delta_{-q_{0},0\alpha}\tilde{U}_{\alpha\alpha'}^{-1}(q_{0})\Delta_{q_{0},0\alpha'}$.
Here we assume $|V/U|<1/2$. The relations $\epsilon_{\text{kin}}<0$, $1/2m^{*}_{\text{kin}}>0$, and $\eta_{\text{kin}}>0$ with $1/2m^{*}_{\text{kin}}>|\eta_{\text{kin}}|$ hold since in the limit of $V\to0$ the system returns to the uniform state with $q_{0}=0$.
When $|V/U|$ is sufficiently small, the modification of the GL parameters by the interaction parts is small, and the kinetic parts of the GL parameters are dominant.
This is usually the case for uniform superconductors, and the conditions (\ref{SufCondition}) and (\ref{StrongCondition}) are not satisfied.
However, as $|V/U|$ becomes larger, the magnitude of the GL parameters except for $d_{\text{I}}$ becomes smaller, which leads to the condition (\ref{SufCondition}) being satisfied, and the small $q$ PDW state is realized.
When $|V/U|$ is sufficiently large, the condition (\ref{StrongCondition}) is satisfied at a certain point, inducing the large $q$ PDW state.

\subsubsection{\label{LinearOpticalConductivity}Linear optical conductivity}
To consider the electromagnetic contribution, we need to include the electromagnetic term arising from the kinetic part of the Hamiltonian, 
where we assume that the vector potential $A$ is small enough,
\begin{equation}
    \mathcal{H}_{\text{EM}}(k) = \sum_{k\alpha\alpha'\sigma}\partial_{k}\xi_{\alpha\alpha'}(k)(-eA)c^{\dagger}_{k\alpha\sigma}c_{k\alpha'\sigma}.
\end{equation}
There is also the diamagnetic term ($\propto\bm{A}^{2}$), which gives the delta-function contribution at zero frequency in the real part of the optical conductivity.
Since we are interested in the linear response of the collective modes at finite frequencies, which is insensitive to the diamagnetic term, we do not consider the diamagnetic term in the following.

There is also the electromagnetic contribution from the interaction part of the Hamiltonian through the pair-hopping term.
The modified interaction $\tilde{U}(q,A)$ is
\begin{equation}
    \tilde{U}(q,A) = \mqty[ U & 2V\cos{(q-2eA)} \\ 2V\cos{(q-2eA)} & U ],
\end{equation}
where the factor $2$ in front of $eA$ comes from the pair-hopping term having two creation/annihilation operators on adjacent sites.
Since we take a sum in terms of $q=\pm q_{0}$ in the Hamiltonian $\mathcal{H}_{\text{int}}$, the interaction is rewritten as
\begin{equation}
    \tilde{U}(q,A) = \tilde{U}(q) - (2eA)^{2} \mqty[ O & V\cos{(q)} \\ V\cos{(q)} & O] + O(A^{4}).
    \label{ModifiedInteraction}
\end{equation}

In our model with the pair hopping, there are three components in the optical conductivity, $\sigma_{\rm SC}$, $\sigma_{\rm QP}$, and $\sigma_{\rm CM}$, each corresponding to the contributions from the supercurrent, quasiparticles, and collective modes, respectively.
Only the first term $\tilde{U}(q)$ in the above contributes to $\sigma_{\text{CM}}$, and the terms of $A^{2}$ and higher ones give the $A^{4}$ and higher terms corresponding to the higher-order response, which is not our focus.
On the flip side, the second term can come about from the supercurrent term of the action in Eq.~(\ref{Action}):
$S_{\text{SC}}=-\beta\sum_{\alpha\alpha'}\Delta_{-q_{0},0\alpha}\tilde{U}^{-1}_{\alpha\alpha'}(q_{0})\Delta_{q_{0},0\alpha'}$, where the optical conductivity from the supercurrent $\sigma_{\text{SC}}$ is defined by $-\delta S_{\text{SC}}/(\beta\delta A) = \sigma_{\text{SC}}(\omega)E(\omega)$ with $E(\omega) = \mathrm{i}\omega A(\omega)$.
Although we can basically give the general form of $\sigma_{\text{SC}}$, we focus on the three cases: $q=0$ (uniform), $q=\pi/2$ ($\pi/2$-PDW), and $q=\pi$ ($\pi$-PDW).
For the uniform case,
\begin{equation}
    \sigma_{\text{SC},q=0}(\omega) = \frac{(2e)^{2}}{\mathrm{i}\omega}\frac{4V}{(U+2V)^{2}}\Delta_{0}^{2},
\end{equation}
where $\Delta_{0}$ is the magnitude of the gap function: $\Delta_{0} = |\Delta_{\pm q_{0},01}|=|\Delta_{\pm q_{0},02}|$.
For the $\pi/2$-PDW case,
\begin{equation}
    \sigma_{\text{SC},q=\pi/2}(\omega) = \frac{(2e^{2})}{\mathrm{i}\omega}\frac{16V^{2}}{U^{3}}\Delta_{0}^{2}.
\end{equation}
For the $\pi$-PDW case,
\begin{equation}
    \sigma_{\text{SC},q=\pi}(\omega) = -\frac{(2e)^{2}}{\mathrm{i}\omega}\frac{4V}{(U-2V)^{2}}\Delta_{0}^{2}.
\end{equation}
One can see that $\sigma_{\rm SC}(\omega)$ has only the imaginary part proportional to $1/\omega$.
For the details of the derivation of $\sigma_{\text{SC}}(\omega)$, we refer to Appendix~\ref{Apdx.A}. 
From now on, we focus on the linear optical conductivity $\sigma_{\text{QP}}$ and $\sigma_{\text{CM}}$ arising from the quasiparticles and collective modes.

\begin{figure}
    \centering
    \includegraphics[scale=1]{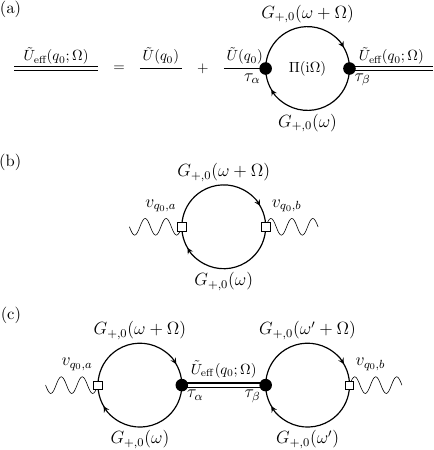}
    \caption{Diagrammatic representations of the (a) effective interaction $\tilde{U}_{\text{eff}}$ within the RPA approximation and [(b),(c)] optical conductivity arising from (b) quasiparticles and (c) collective modes.
    Here $\tilde{U}$ is the bare interaction, and $a$ and $b$ label the spatial directions.}
    \label{Diagrams}
\end{figure}

We shall calculate the linear optical conductivity of the quasiparticle and collective modes.
Because of the electromagnetic contribution of the kinetic part $\mathcal{H}_{\text{EM}}$, another term $S_{\text{FL}}$ (fluctuations due to the electromagnetic vector potential) appears in the action,
\begin{align}
    S_{\text{FL}} &= \frac{e^{2}}{2}\int\frac{d\omega}{2\pi}A(\omega)A(-\omega)\Phi(\omega) \notag \\
    &\quad +\frac{e^{2}}{4}\int\frac{d\omega}{(2\pi)}A(\omega)A(-\omega)Q^{\text{T}}(\omega)\tilde{U}_{\text{eff}}(q_{0};\omega)Q(-\omega).
\end{align}
Here we have introduced the current-current correlation function $\Phi(\mathrm{i}\Omega)$ by
\begin{align}
    &\Phi(\mathrm{i}\Omega) \notag \\
    =& \frac{1}{\beta}\sum_{n}\int\frac{dk}{2\pi}\text{Tr}\qty[v_{q_{0}}(k)G_{+,0}\qty(\mathrm{i}\omega+ \mathrm{i}\Omega;k)v_{q_{0}}(k)G_{+,0}\qty(\mathrm{i}\omega;k)] \notag \\
    =&\int\frac{dk}{2\pi}\sum_{j,l}\frac{f_{jl}v_{q_{0},jl}v_{q_{0},lj}}{\mathrm{i}\Omega - E_{lj}},
\end{align}
where $v_{q_{0}}(k)$ is the velocity operator
\begin{equation}
    v_{q_{0}}(k) = \mqty[ \partial_{k}\xi(k+q_{0}) & O \\ O & \partial_{k}\xi(k-q_{0}) ].
\end{equation}
To untangle the notation, the band representation of the generalized Pauli matrices $\tau_{\mu\alpha,jl} = \mel*{\varphi_{j}}{\tau_{\mu\alpha}}{\varphi_{l}}$ and $v_{q_{0},jl}=\mel*{\varphi_{j}}{v_{q_{0}}}{\varphi_{l}}$ are used, where $\ket{\varphi_{j}}$ is the $j$-th eigenvector $\ket{\varphi_{j}}$, $E_{lj}:=E_{l}-E_{j}$ with the $j$-th largest eigenvalue $E_{j}$, and $f_{jl}:=f(E_{j})-f(E_{l})$ with $f(E_{j})$ being the Fermi distribution: $f(E_{j}) = 1/(e^{\beta E_{j}}+1)$.
We have also introduced the effective interaction within the random phase approximation (RPA) $\tilde{U}_{\text{eff}}$:
\begin{equation}
    \tilde{U}_{\text{eff}} = \tilde{U} + \tilde{U}\Pi\tilde{U} + \cdots = (1-\tilde{U}\Pi)^{-1}\tilde{U},
\end{equation}
or, namely,
\begin{align}
    &\qty[\tilde{U}_{\text{eff}}\qty(q_{0};\mathrm{i}\Omega)]_{\mu\alpha,\mu'\alpha'} \notag \\
    =& \qty( \delta_{\mu\mu'}\delta_{\alpha\alpha'} - \tilde{U}_{\alpha\alpha'}(q_{0})\qty[\Pi\qty(\mathrm{i}\Omega)]_{\mu\alpha,\mu'\alpha'} )^{-1}\tilde{U}_{\alpha\alpha'}(q_{0}),
\end{align}
the polarization bubble $\Pi(\mathrm{i}\Omega)$,
\begin{align}
    &\qty[\Pi(\mathrm{i}\Omega)]_{\mu\alpha,\mu'\alpha'} \notag \\
    =& \frac{1}{2\beta}\sum_{n}\int\frac{dk}{2\pi}\text{Tr}\qty[\tau_{\mu\alpha}G_{+,0}(\mathrm{i}\omega_{n}+\mathrm{i}\Omega;k)\tau_{\mu\alpha'}G_{+,0}(\mathrm{i}\omega_{n};k)] \notag \\
    =&\frac{1}{2}\int\frac{dk}{2\pi}\sum_{j,l}\frac{f_{jl}\tau_{\mu\alpha,jl}\tau_{\mu'\alpha',lj}}{\mathrm{i}\Omega - E_{lj}},
\end{align}
and the vector $Q(\mathrm{i}\Omega)$:
\begin{align}
    &\qty[Q(\mathrm{i}\Omega)]_{\mu\alpha} \notag \\
    =&\frac{1}{\beta}\sum_{n}\int\frac{dk}{2\pi}\text{Tr}\qty[v_{q_{0}}(k)G_{+,0}(\mathrm{i}\omega_{n}+\mathrm{i}\Omega;k)\tau_{\mu\alpha}G_{+,0}(\mathrm{i}\omega_{n};k)] \notag \\
    =&\int\frac{dk}{2\pi}\sum_{j,l}\frac{f_{jl}v_{q_{0},jl}\tau_{\mu\alpha,lj}}{\mathrm{i}\Omega - E_{lj}}.
\end{align}
By performing the analytic continuation $\mathrm{i}\Omega \to \omega + \mathrm{i}0^{+}$, we can obtain the real-frequency form of $\tilde{U}_{\text{eff}}$ and $Q$.
We take the functional derivative of $S_{\text{FL}}$ with respect to $A(-\omega)$ to obtain the linear optical conductivity arising from the kinetic part, 
\begin{equation}
    j(\omega) = -\frac{\delta S_{\text{FL}}}{\delta A(-\omega)} = \qty(\sigma_{\text{QP}}\qty(\omega)+\sigma_{\text{CM}}(\omega))E(\omega),
\end{equation}
where
\begin{align}
    \sigma_{\text{QP}}(\omega) &= \frac{\mathrm{i}e^{2}}{\omega}\Phi(\omega), \\
    \sigma_{\text{CM}}(\omega) &= \frac{\mathrm{i}e^{2}}{2\omega}Q^{\text{T}}(\omega)\tilde{U}_{\text{eff}}(q_{0};\omega)Q(-\omega).
\end{align}
Here, $\sigma_{\text{QP}}$ corresponds to the quasiparticle response, and $\sigma_{\text{CM}}$ to the collective mode response.
The collective mode response originates from the poles of $\tilde{U}_{\text{eff}}(q_{0};\omega)$, which satisfies $1-\tilde{U}(q_{0})\Pi(\omega)=0$.
As pointed out in \cite{Kamatani2022}, $Q(\omega)$ is off-resonant and does not have a singularity at the collective-mode frequencies.
%for this $\omega$.
The diagrammatic representations of the effective interaction $\tilde{U}_{\text{eff}}$ within the RPA approximation and the optical conductivity of the quasiparticles and collective modes are given in Fig.~\ref{Diagrams}.

\begin{figure}
    \centering
    \includegraphics[scale=0.25]{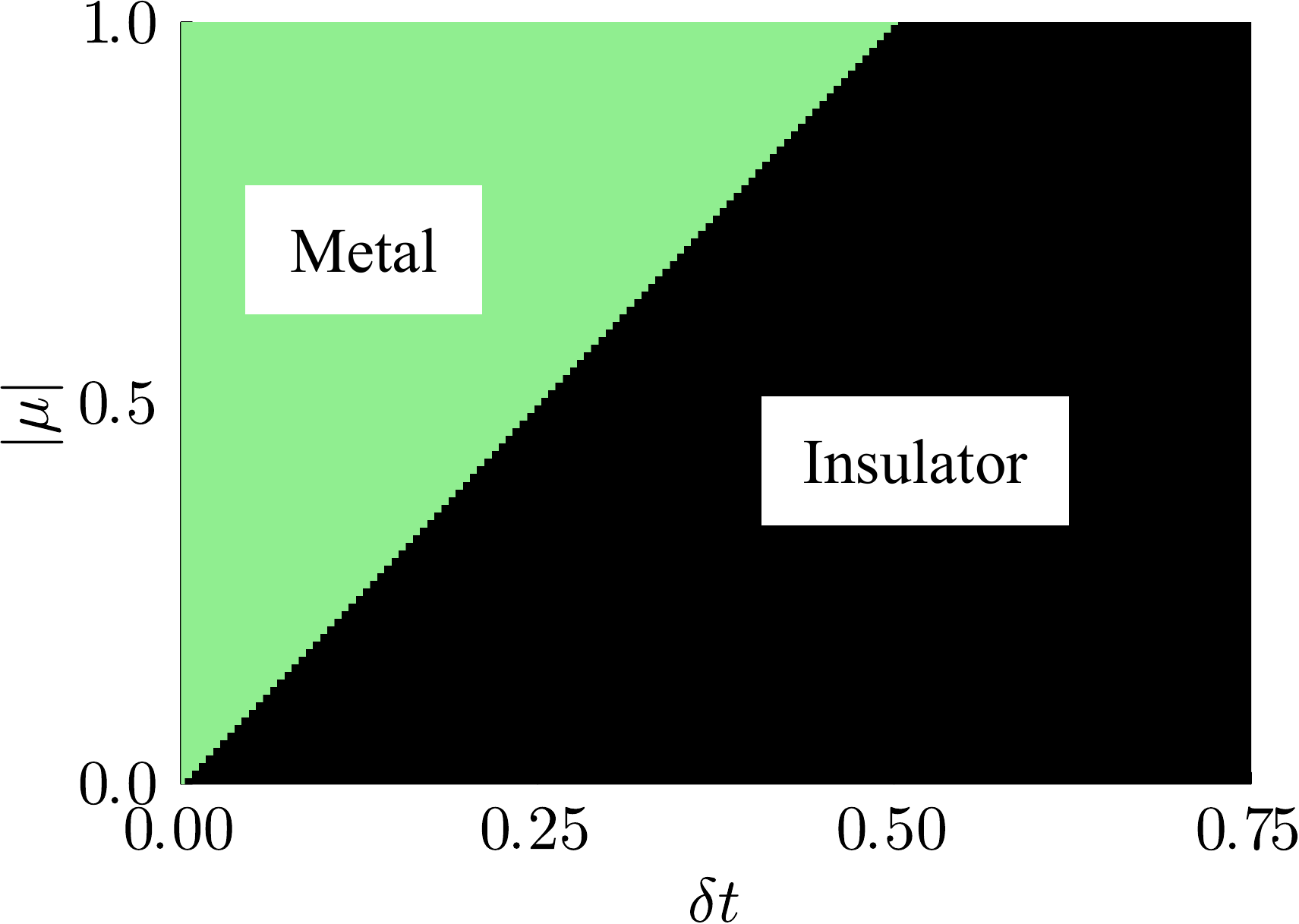}
    \caption{Phase diagram of the normal state without interactions ($U=V=0$) for the 1D model (\ref{Hamiltonian_1D}) in the space of $\delta t$ and $|\mu|$.
    The light green area corresponds to a metallic state, while the black area corresponds to an insulating state.
    The boundary between the two phases is given by $|\mu|=2\delta t$.}
    \label{PD_normal}
\end{figure}

\begin{figure*}[htbp]
    \centering
    \includegraphics[scale=0.4]{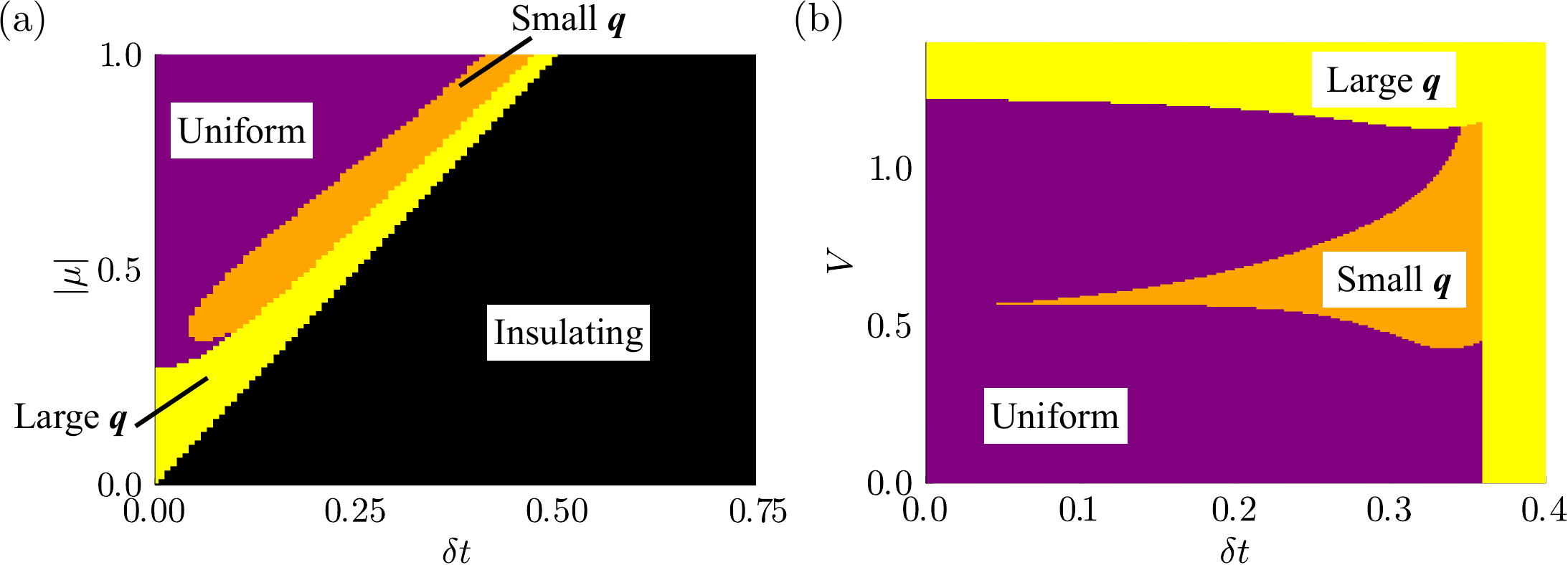}
    \caption{Phase diagrams based on the GL theory with the boundaries determined by
    (\ref{SufCondition}) and (\ref{StrongCondition})
    for the 1D model (\ref{Hamiltonian_1D}) with $t=1.0$, $U=-3.0$, and $T=0.05$:
    (a) $\delta t$-$|\mu|$ phase diagram with $V=0.5$,
    and (b) $\delta t$-$V$ phase diagram with $|\mu|=0.8$.
    The purple, orange, and yellow areas correspond to the uniform, small $\bm{q}$, and large $\bm{q}$ superconducting states, respectively.
    The black region shows the insulating phase.
    }
    \label{PD_GL_dtmu_dtV}
\end{figure*}

\subsection{\label{sec:level4-2}Normal state phase diagram}
Here, we provide a phase diagram of the normal state without interactions.
Since a superconducting state emerges in the region of a metallic phase in the BCS theory, it is necessary to identify the phase diagram of the normal state.
We judge whether the system is in the metallic or insulating phase by calculating the density of states using the kinetic part of the Hamiltonian (\ref{Kinetic_N}). 
We fix $t=1.0$ as the unit of energy. 
Figure~\ref{PD_normal} shows the phase diagram of the normal state of the model.
%In Figure.~\ref{PD_normal}, 
The light green area corresponds to the metallic state, while the black area corresponds to the insulating state.
The boundary between the two phases is $|\mu|=2\delta t$: When $\delta t$ is smaller than $|\mu|/2$, the system is metallic, and when $\delta t$ is larger than $|\mu|/2$, the system is insulating.
Later, we focus on the metallic phase, the region satisfying $|\mu|>2\delta t$, as a parent state of the superconducting phases.

\subsection{\label{sec:level4-3}Phase diagrams based on the GL theory}
We determine the phase diagrams within the GL theory by using the conditions derived in Sec.~\ref{sec:level2-2} [Eqs.~(\ref{SufCondition}) and (\ref{StrongCondition})] and the expressions of the GL parameters based on the microscopic model in Sec.~\ref{sec:level4-1},
where we assume the existence of the ordered state (here, the superconducting state) with the small order parameter (gap function).
We calculate the GL parameters at $\bm{q}=\bm{0}$ and consider the instability of the uniform state.
We fix $U=-3.0$ and $T=0.05$, and change other parameters $\delta t$, $|\mu|$, and $V$.
The computed phase diagrams are shown in Fig.~\ref{PD_GL_dtmu_dtV}.
In Fig.~\ref{PD_GL_dtmu_dtV}(a) and (b), the purple, orange, and yellow regions correspond to the uniform, small $\bm{q}$, and large $\bm{q}$ states, respectively.
When the GL parameters satisfy only Eq.~(\ref{SufCondition}), the state is determined to be the small $\bm{q}$ state.
When the GL parameters satisfy Eq.~(\ref{StrongCondition}), the state is judged to be the large $\bm{q}$ state even if the Eq.~(\ref{SufCondition}) is simultaneously fulfilled.
When both Eq.~(\ref{SufCondition}) and Eq.~(\ref{StrongCondition}) are not satisfied, the system is in the uniform superconducting phase.
The black area describing the insulating phase is not our focus.
We can see that the small $\bm{q}$ state is realized for large $|\mu|$, $\delta t$, and moderately large $|V/U|$, and the large $\bm{q}$ state is realized for sufficiently large $|V/U|$.

As $\delta t$ increases, electrons tend to localize in the stronger hopping bonds (the bond of the strength $-(t+\delta t)$).
This leads to the significant reduction of the kinetic energy, meaning the decrease of $1/2m^{*}$, or the l.h.s. of Eq.~(\ref{StrongCondition}).
When $\delta t$ is quite large and the state is very close to the insulating state, the magnitude of the l.h.s. of Eq.~(\ref{StrongCondition}) is sufficiently small, indicating that Eq.~(\ref{StrongCondition}) is always satisfied independent of the magnitude of $|\mu|$ and $V$.
This situation may explain the excessively stabilized large $\bm{q}$ state close to the metal-insulator boundary in Fig.~\ref{PD_GL_dtmu_dtV}.
Note that the GL theory does not tell the exact value of $\bm{q}$ to be realized. We only determined the phase boundaries based on the GL parameters at $\bm{q}=0$.

\subsection{\label{sec:level4-4}Phase diagrams based on the microscopic theory}
We microscopically calculate the phase diagrams for the model (\ref{Hamiltonian_1D}) based on the formalism in Sec.~\ref{sec:level4-1}.
First, we find that the possible values of $q_{0}$ are $q_0=0$ (uniform state),
$q_{0}=\pi/2$ ($\pi/2$-PDW state), and $q_0=\pi$ ($\pi$-PDW state), which are confirmed by calculating the free energy as a function of $q$ with the self-consistent calculation of the gap function according to Eq.~(\ref{Gap_equation}).
We also solve the self-consistent gap equation for each state with fixed $q_0$ by sweeping the parameter sets ($\delta t,|\mu|$) and ($\delta t, V)$.
After calculating the gap function, we evaluate the free energy of each state by Eq.~(\ref{FE_w/_fluctuation}) and compare them to see which state has the lowest free energy.
For comparison, we also show the free energy of the metallic state.
We choose $t=1.0$, $U=-3.0$ and $T=0.001$.
When looking at the temperature dependence of the $\pi/2$-PDW state, we fix $\delta t$, $|\mu|$ and $V$ and change the temperature $T$.

\subsubsection{$q$-dependence of the free energy}
\begin{figure}
    \centering
    \includegraphics[scale=0.55]{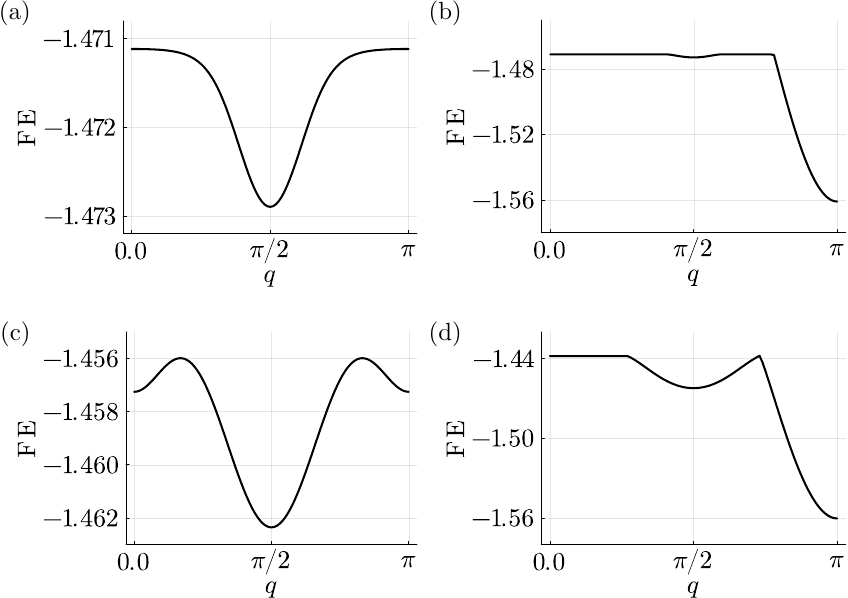}
    \caption{Free energy of the 1D model (\ref{Hamiltonian_1D}) as a function of $q$ for
    (a) $V=0.5$, $T=0.2$, (b) $V=1.4$, $T=0.2$, (c) $V=0.5$, $T=0.01$, and (d) $V=1.4$, $T=0.01$.}
    \label{FE_qdep}
\end{figure}

We evaluate the $q$ dependence of the free energy for several values of $V$ and $T$
with $U=-3.0$, $\delta t=0.35$, and $|\mu|=0.8$ being fixed.
The result for $V=0.5$ and $T=0.2$ is shown in Fig.~\ref{FE_qdep}(a).
The free energy $\mathcal F$ for this parameter takes a minimum at $q=\pi/2$. The curvature at $q=0$ becomes negative ($\lim_{q\to0}\partial_{q}^{2}\mathcal{F}<0$), indicating that the condition (\ref{SufCondition}) for nonzero $q$ states in the GL theory is satisfied.
On the other hand, the result for $V=1.4$ and $T=0.2$ shown in Fig.~\ref{FE_qdep}(b) does not obey the same condition: instead, the free energy has the minimum at $q=\pi$, suggesting that the system is in the regime of (\ref{StrongCondition}).
We note that there is a singularity at $q\sim 3\pi/4$ in Fig.~\ref{FE_qdep}(b), which may be an artifact of the calculations since the free energy with fixed $q$ away from the minimum is not necessarily related to physically realized states.

When the temperature decreases, the free energy takes a different shape.
We show the results for $V=0.5$ and $T=0.01$ in Fig.~\ref{FE_qdep}(c) and $V=1.4$ and $T=0.01$ in Fig.~\ref{FE_qdep}(d).
In the former case, the free energy takes a global minimum at $q=\pi$ and a local minimum at $q=0$, which differs from what we find in the GL theory.
This may be because the temperature ($T=0.01$) is far from the region around the critical temperature where the GL theory becomes valid.

\subsubsection{$\delta t$-$|\mu|$ phase diagram}
\begin{figure*}
    \centering
    \includegraphics[scale=0.4]{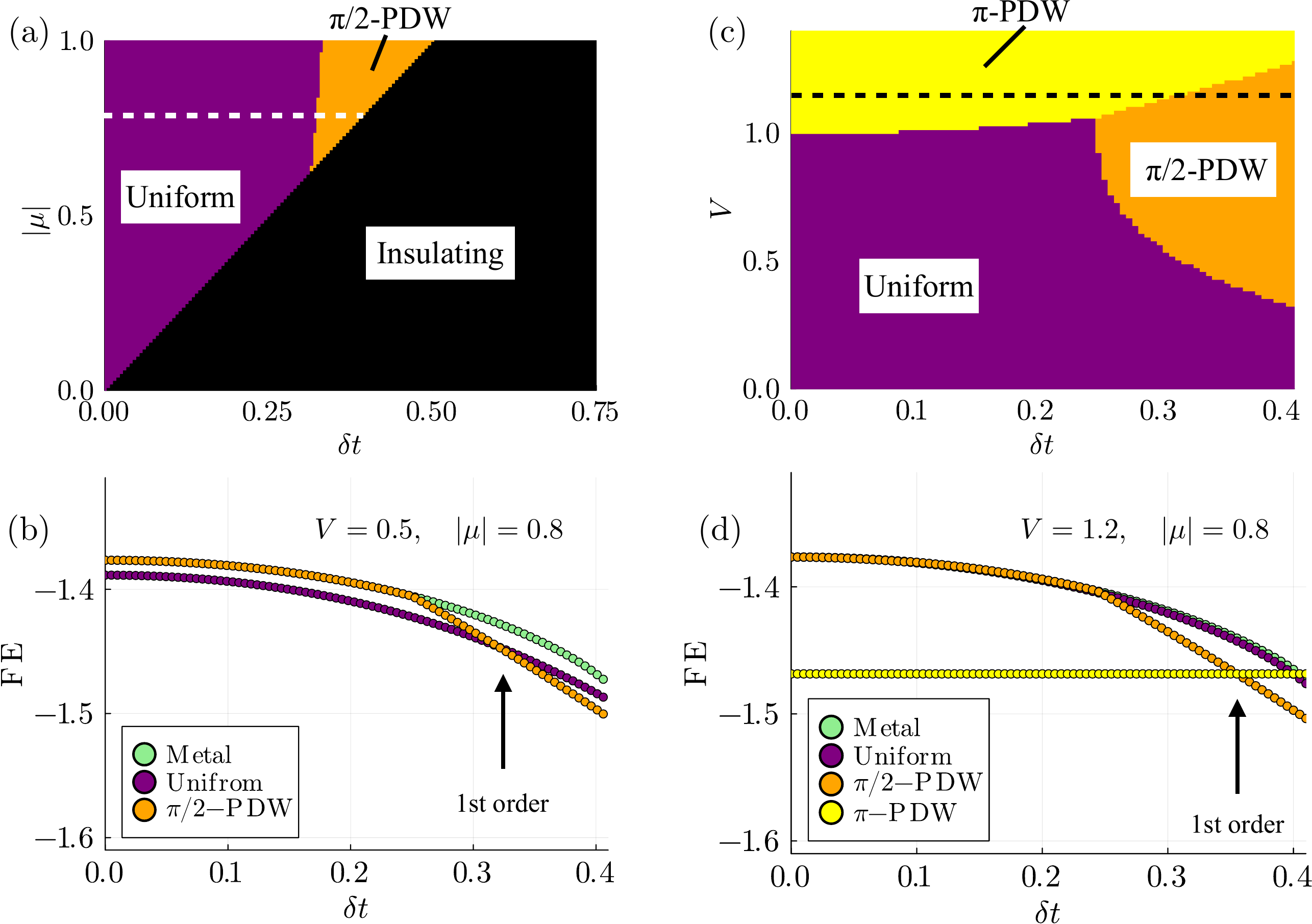}
    \caption{(a) Phase diagram based on the microscopic mean-field theory for the 1D model (\ref{Hamiltonian_1D}) in the space of $(\delta t,|\mu|)$ with $V=0.5$ and $T=0.001$.
    The black, purple, and yellow regions correspond to the insulating, uniformly superconducting, and $\pi/2$-PDW states.
    (b) $\delta t$ dependence of the free energy along the dashed line in (a).
    The free energy of the $\pi$-PDW state (not shown) is much larger than those of the other states.
    (c) Phase diagram in the space of $(\delta t,V)$ with $|\mu|=0.8$ and $T=0.001$.
    The yellow area represents the $\pi$-PDW state.
    (d) $\delta t$ dependence of the free energy along the dashed line in (c).
    }
    \label{PD_dtmu_dtV}
\end{figure*}

Here we show the phase diagram of the model (\ref{Hamiltonian_1D}) in the space of $\delta t$ and $|\mu|$ with fixed $V=0.5$ and $T=0.001$ in Fig.~\ref{PD_dtmu_dtV}(a).
The corresponding $\delta t$ dependence of the free energies for the metallic, uniformly superconducting, and $\pi/2$-PDW states with fixed $V$ and $|\mu|$ are shown in Fig.~\ref{PD_dtmu_dtV}(b).
The uniform superconducting state appears when $\delta t$ is small, and the $\pi/2$-PDW state appears when $\delta t$ goes close to the metal-insulator boundary with large $|\mu|$.
This phase diagram signifies that the doping to increase/decrease the electron density away from half-filling is indispensable for the $\pi/2$-PDW state to realize the state with a finite density of states and large $\delta t$.
The $\pi$-PDW state does not appear in the phase diagram since the effect of the pair hopping $V$ is not superior to that of the difference of the hopping strengths $\delta t$.
Figure~\ref{PD_dtmu_dtV}(b) indicates that the phase transition from the uniform superconducting state to the $\pi/2$-PDW state is of the first order since the free energies of these states intersect as $\delta t$ is changed.
Within these parameter sets, the metallic state has a higher free energy than the uniform and $\pi/2$-PDW states and thus does not appear in the phase diagram.

\subsubsection{$\delta t$-$V$ phase diagram}
Next, we change $\delta t$ and $V$ with $|\mu|=0.8$ and $T=0.001$ being fixed.
The resulting phase diagram and $\delta t$ dependence of the free energies are shown in Fig.~\ref{PD_dtmu_dtV}(c) and (d), respectively.
The uniform superconducting state appears for small $\delta t$ and $|V/U|$, 
while the $\pi/2$- and $\pi$-PDW states emerge for large $\delta t$ and moderately large $|V/U|$ and for sufficiently large $|V/U|$, respectively.
When $\delta t>0.4$, the parent state is insulating, which is not our focus.
The phase diagram shows that the $\pi/2$-PDW state requires both the large $\delta t$ close to the metal-insulator boundary and the moderately large $|V/U|$.
Conversely, the $\pi$-PDW state always appears when $|V/U|$ is sufficiently large, except for $\delta t$ close to the metal-insulating boundary ($\delta t=0.4$).
Figure~\ref{PD_dtmu_dtV}(d) indicates that the phase transition from the $\pi$-PDW state to the $\pi/2$-PDW state is of the first order, since the free energies of these states intersect with each other, as in the transition from the uniform to the $\pi/2$-PDW state in Fig.~\ref{PD_dtmu_dtV}(b).

By comparing the phase diagrams obtained from the GL theory (Fig.~\ref{PD_GL_dtmu_dtV}) and the microscopic calculations (Fig.~\ref{PD_dtmu_dtV}), we can see that the ``small $\bm{q}$" and the ``large $\bm{q}$" states in the GL theory roughly correspond to the $\pi/2$-PDW and the $\pi$-PDW states in the microscopic theory.
The spatial modulation appearing in the microscopic model with $q=\pi/2$ and $\pi$ might be too large and could be beyond the scope of the GL theory.
Nevertheless, this qualitative agreement between the GL theory prediction and the microscopic calculation justifies the interpretation of the mechanism to induce the PDW state based on the GL theory: The competition between the ``twisting" by the Lifshitz invariant and the ``aligning" by other GL terms are important to realize the small $\bm{q}$ ($q_{0}=\pi/2$ in the microscopic calculation) state, and the dominance of the drag effect by the large pair hopping $V$ is responsible for the large $\bm{q}$ ($q_{0}=\pi$) state.

\subsubsection{Temperature dependence of the free energy and the gap function for the $\pi/2$-PDW state}
\begin{figure*}
    \centering
    \includegraphics[scale=0.5]{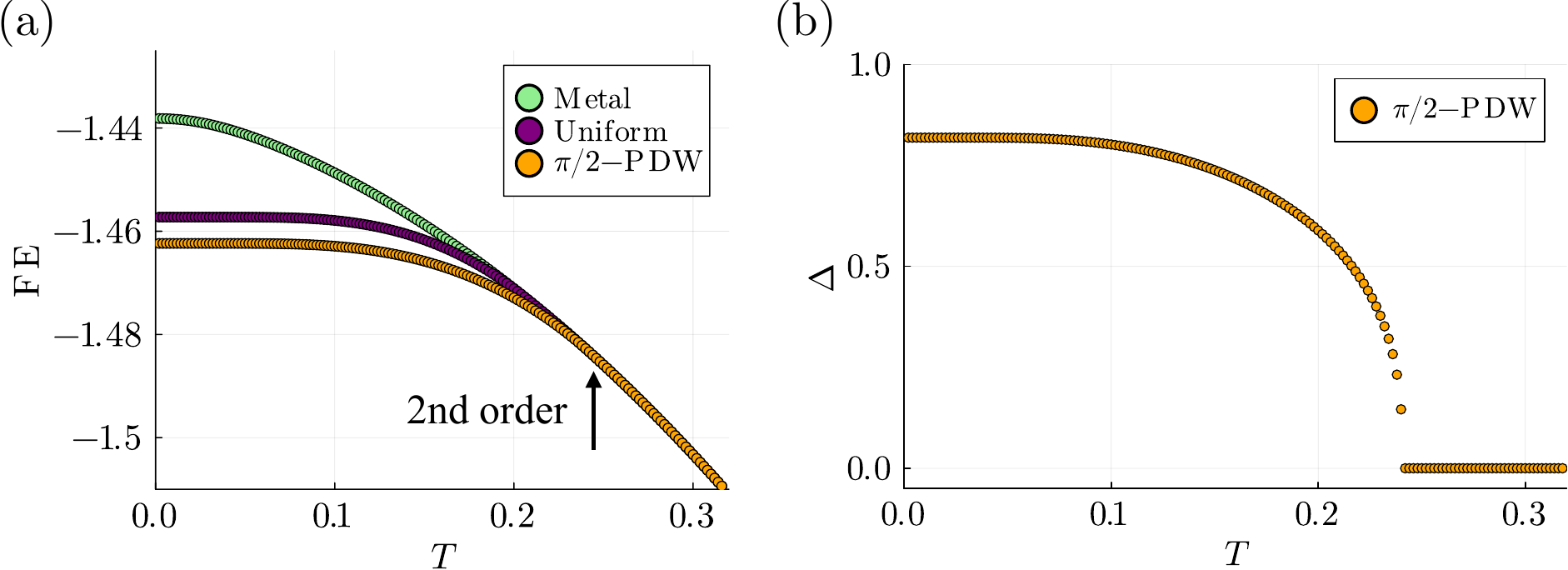}
    \caption{(a) Temperature dependence of the free energy for each phase in the 1D model (\ref{Hamiltonian_1D}) with $\delta t=0.35$, $|\mu|=0.8$, and $V=0.5$.
    The light green, purple, and orange dots correspond to the metallic, uniform, and $\pi/2$-PDW states.
    (b) Temperature dependence of the gap function for the $\pi/2$-PDW state with $\delta t=0.35$, $|\mu|=0.8$, and $V=0.5$.}
    \label{Tsweep_FE_gap}
\end{figure*}

Let us look at how the free energy and gap function of the $\pi/2$-PDW state behave when the temperature changes.
The parameters are chosen to be $\delta t=0.35$, $|\mu|=0.8$, and $V=0.5$, where the $\pi/2$-PDW state appears with $T=0.001$, as shown in Fig~\ref{PD_dtmu_dtV}.
The results are depicted in Fig.~\ref{Tsweep_FE_gap}.
In Fig.~\ref{Tsweep_FE_gap}(a), one can see that the $\pi/2$-PDW state has the lowest free energy than any other state until the system returns to the metallic state.
The $\pi$-PDW state (not shown) has the vanishing gap function for all the temperature domains in this parameter region, and its free energy overlaps with that of the metallic state.
%Hence, the $\pi$-PDW state is not depicted in Fig.~\ref{Tsweep_FE_gap}.
The phase transition from the $\pi/2$-PDW state to the metallic state is of the second order, as shown in Fig.~\ref{Tsweep_FE_gap}(b),
where the gap function of the $\pi/2$-PDW state reaches zero at around $T\approx0.24$. 

The temperature $T=0.05$ used to draw the phase diagram based on the GL theory in Fig.~\ref{PD_GL_dtmu_dtV} is not very close to the transition temperature, as shown in Fig.~\ref{Tsweep_FE_gap}(b).
Strictly speaking, this temperature regime is beyond the scope of the GL theory.
Nevertheless, the phase diagrams obtained from the GL theory are qualitatively consistent with those obtained from the microscopic calculation.
Hence, we believe the GL theory can capture the essential aspects of the PDW states.
If the temperature is near the transition temperature, the uniform and small $\bm{q}$ regions in the phase diagrams obtained from the GL theory in Fig.~\ref{PD_GL_dtmu_dtV} become narrower.

\subsection{Linear optical conductivity}

We now calculate the linear optical conductivity of the $\pi/2$- and $\pi$-PDW states to see how the collective modes of these states appear in the linear response regime.
Since we are interested in the contribution of quasiparticles and collective modes reflecting the characteristics of the superconducting state, we focus on $\sigma_{\rm QP}$ and
$\sigma_{\text{CM}}$ defined in Sec.~\ref{sec:level4-1}.
We fix the parameters $\delta t=0.35$, $|\mu|=0.8$, $U=-3.0$, and $T=0.001$, and we choose $V=0.5$ for the $\pi/2$-PDW state and $V=1.4$ for the $\pi$-PDW state.

\begin{figure*}
    \centering
    \includegraphics[scale=0.6]{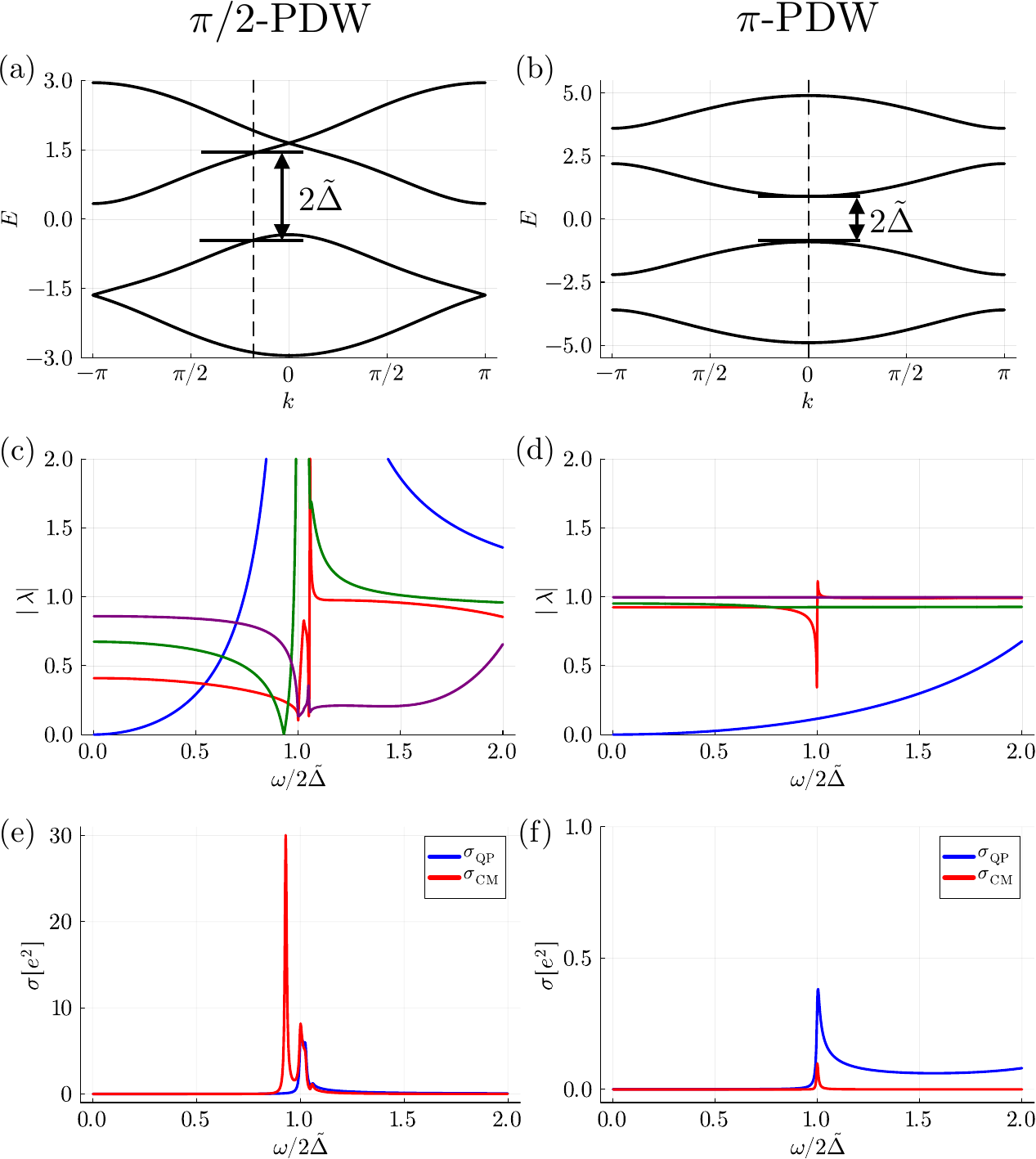}
    \caption{Band structures, absolute values of the eigenvalues of the effective interaction $\tilde{U}_{\text{eff}}$, and linear optical conductivity for the $\pi/2$- and $\pi$-PDW states in the 1D model (\ref{Hamiltonian_1D}) with $\delta t=0.35$, $|\mu|=0.8$, $U=-3.0$, and $T=0.001$. We take $V=0.5$ for the $\pi/2$-PDW state and $V=1.4$ for the $\pi$-PDW state.
    (a) Band structure of the $\pi/2$-PDW state.
    The direct gap is given by $2\tilde{\Delta}\approx1.88$ (which is different from $2\Delta_{0}\approx1.64$).
    (b) Band structure of the $\pi$-PDW state.
    The direct gap $2\tilde{\Delta}\approx1.80$ is different from $2\Delta_{0}\approx5.57$.
    [(c), (d)] Absolute values $|\lambda_{i}|$ of the eigenvalues $\lambda_{i}$ of $(1-\tilde{U}_{\rm eff}(\omega)\Pi\qty(\omega))$ of (c) the $\pi/2$- and (d) $\pi$-PDW states.
    [(e),(f)] Contributions of the quasiparticles $\sigma_{\text{QP}}$ (blue curve) and collective modes $\sigma_{\text{CM}}$ (red) of the linear optical conductivity for the (e) $\pi/2$- and (f) $\pi$-PDW states.}
    \label{1D_band_pole_OC}
\end{figure*}

To measure the direct gap $2\tilde{\Delta}$, we present the band structures of the $\pi/2$- and $\pi$-PDW states in Fig.~\ref{1D_band_pole_OC}(a) and (b), respectively.
Due to the finite center-of-mass momentum $q$, the band structure is shifted by $q$ from the original one, and the direct gap becomes different from $2|\Delta_{0}|$.
We can also plot the band structure in a back-folded way by extending the unit cell containing four sites, where the particle-hole symmetry becomes more evident but the direct gap for possible optical transitions becomes less obvious. For the details, we refer to Appendix~\ref{Apdx_C}.
For the $\pi/2$-PDW state, the direct gap $2\tilde{\Delta}$ indicated in Fig.~\ref{1D_band_pole_OC}(a) is $2\tilde{\Delta}\approx1.88$, which is different from $2|\Delta_{0}|\approx1.64$.
This is caused by the finite center-of-mass momentum $q=\pi/2$, reflected in the location of the minimum direct gap away from $k=0$ and $k=\pm\pi$.
For the $\pi$-PDW state, the direct gap $2\tilde{\Delta}$ shown in Fig.~\ref{1D_band_pole_OC}(b) is $2\tilde{\Delta}\approx1.80$, which is also different from $2|\Delta_{0}|\approx 5.57$.

We next estimate the characteristic frequencies of the collective modes by calculating the absolute values $|\lambda_{i}|$ of the eigenvalues of $(1-\tilde{U}\Pi(\omega))$, zeros of which correspond to the poles of the effective interaction $\tilde{U}_{\text{eff}}(\omega)$.
$|\lambda_{i}|$ of the $\pi/2$- and $\pi$-PDW states are shown in Fig.~\ref{1D_band_pole_OC}(c) and (d), respectively.
The zero at $\omega=0$ reached by the blue line in (c) corresponds to the NG mode, which does not contribute to the optical conductivity, as we will see later.
The situation may be different if we include the effect of the long-range Coulomb interaction or that of disorders (see, e.g.,  \cite{Cea2014} for the case of finite disorders).
The zero at $\omega\approx0.90\times2\tilde{\Delta}$ touched by the green line in (c), which corresponds to the pole of $\tilde{U}_{\text{eff}}$, significantly contributes to the linear optical conductivity, as we will see below.
The red and purple lines in (c) have two dip-like structures: one at $\omega = 2\tilde{\Delta}$ and the other at $\omega\approx1.05\times2\tilde{\Delta}$, each of which gives a non-diverging but finite contribution to the linear optical conductivity.
In the case of the $\pi$-PDW state (Fig.~\ref{1D_band_pole_OC}(d)), the zero at $\omega=0$ of the blue line corresponds to the NG mode as in (c).
The red line in (d) has a dip-like structure at $\omega=2\tilde{\Delta}$, which gives a finite contribution to the linear optical conductivity.

The linear optical conductivity of the $\pi/2$- and $\pi$-PDW states is depicted in
Fig.~\ref{1D_band_pole_OC}(e) and (f), respectively.
We show the quasiparticle contribution $\sigma_{\text{QP}}$ by the blue curves and the collective mode contribution $\sigma_{\text{CM}}$ by the red ones.
In the $\pi/2$-PDW state [Fig.~\ref{1D_band_pole_OC}(e)], the collective mode contribution $\sigma_{\text{CM}}$ has a peak at $\omega\approx0.90\times 2\tilde{\Delta}$, corresponding to the pole given by the green line in (c).
The peak becomes highest since the mode is well-defined below the gap $2\tilde{\Delta}$.
The collective mode part $\sigma_{\text{CM}}$ also has two more peaks: one at $\omega=2\tilde{\Delta}$, and the other (small one) at around $\omega/2\tilde{\Delta}\approx1.05$, corresponding to the dip-like structures in (c).
The quasiparticle contribution $\sigma_{\text{QP}}$ behaves similarly to the collective mode, except for the peak at around $\omega\approx 0.90\times 2\tilde{\Delta}$.
In the $\pi$-PDW state [Fig.~\ref{1D_band_pole_OC}(f)], on the other hand, both the contributions from the collective modes and quasiparticles have only one peak at $\omega=2\tilde{\Delta}$, and their overall magnitude is smaller than the case of the $\pi/2$-PDW state.

\begin{figure}
    \centering
    \includegraphics[scale=0.5]{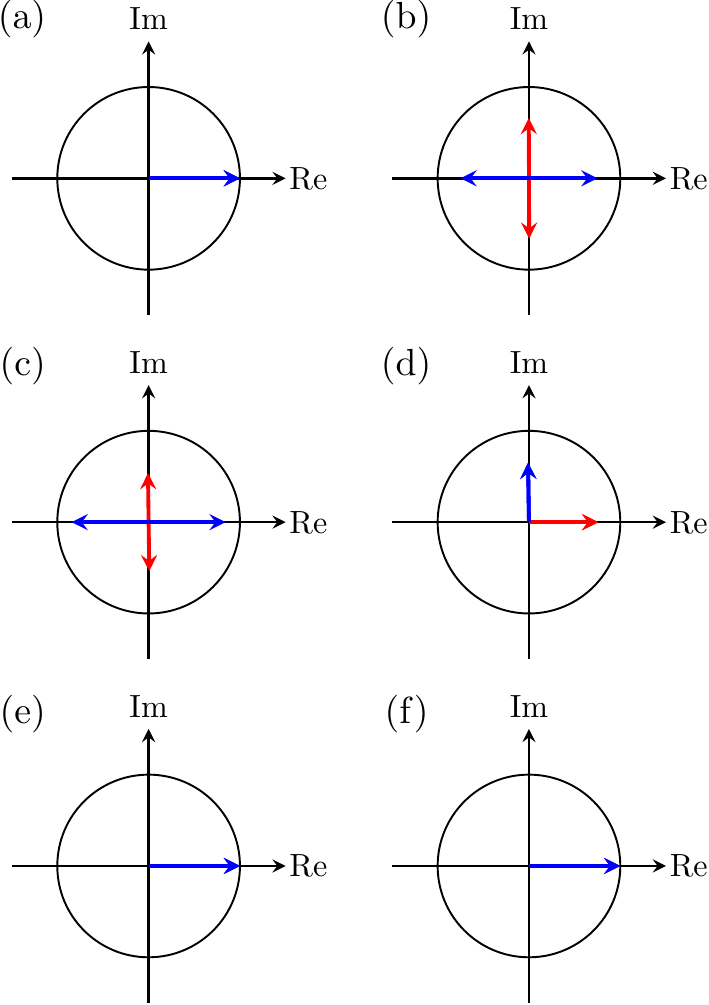}
    \caption{
    Eigenvectors of the effective interaction $\tilde U_{\rm eff}(\omega)$ with different frequencies at the peaks of the optical conductivity for the [(a)-(d)] $\pi/2$- and [(e),(f)] $\pi$-PDW states.
    The red and blue arrows represent the eigenvector components corresponding to the amplitude and phase fluctuations, respectively.
    The radius of the circle is $1/\sqrt{2}$.
    (a) $\omega=0$ (two-fold degenerate), (b) $\omega\approx 0.90\times 2\tilde{\Delta}$, (c) $\omega=2\tilde{\Delta}$, (d) $\omega\approx1.05\times2\tilde{\Delta}$ (two-fold degenerate for each),
    (e) $\omega=0$ (two-fold), and (f) $\omega=2\tilde{\Delta}$ (two-fold).}
    \label{1D_Eigenvectors}
\end{figure}

For both of the PDW states, we can check the weight of the contributions from the Higgs and Leggett modes to the linear optical conductivity by calculating the eigenvectors for the largest eigenvalue of the effective interactions $\tilde{U}_{\text{eff}}(\omega)$ at each peak.
Each eigenvector of $\tilde{U}_{\text{eff}}(\omega)$ has four complex components, two of which represent the amplitude ($\tau_{x1}$ channel) and phase ($\tau_{y1}$) fluctuations of band $1$, and the other two represent the amplitude ($\tau_{x2}$) and phase ($\tau_{y2}$) fluctuations of band $2$.
The eigenvectors at the peaks (and $\omega=0$) for both PDW states are shown in Fig.~\ref{1D_Eigenvectors}.
Note that the eigenvectors at different frequencies are not always orthogonal.
When calculating the eigenvectors, we apply a unitary transformation \cite{Aitchison1995} $\mathcal{U}=\exp(\qty(\mathrm{i}\theta_{1}\tau_{z1} + \mathrm{i}\theta_{2}\tau_{z2})/2)$ to omit the phases of the gap functions, where 
\begin{equation}
    \tau_{z\alpha} = \mqty[ A_{\alpha} & O \\ O & -A_{\alpha}]
\end{equation}
is the $z$-component generalized Pauli matrix (see also Eq.~(\ref{PauliMatrix})).
In the $\pi/2$-PDW state [Fig.~\ref{1D_Eigenvectors}(a)-(d)], the NG mode ($\omega=0$) has the weight only on the phase fluctuation (Fig.~\ref{1D_Eigenvectors}(a)), while the other peaks have contributions both from the amplitude and phase fluctuations (Fig.~\ref{1D_Eigenvectors}(b)-(d)).
This hybridization may be caused by the particle-hole asymmetric band structure of the PDW state \cite{PDW_review}.
In the $\pi$-PDW state [Fig.~\ref{1D_Eigenvectors}(e),(f)], on the other hand, both the NG mode ($\omega=0$) and the peak at $\omega=2\tilde{\Delta}$ only have contributions from the phase fluctuations.
This is unique to the $q=\pi$ PDW state: When different lattice sites have gap functions with the same magnitude and different signs, the Higgs mode contributions from each site cancel each other.
This is also mentioned in Sec.~\ref{sec:level2-2} in the context of the GL theory.
The finite response arising from the phase fluctuation in the same direction at finite $\omega$ might be related to the absence of the long-range Coulomb interaction.
If we add the long-range Coulomb interaction explicitly, the peat at $\omega=2\tilde{\Delta}$ in the $\pi$-PDW state might vanish. 

We also confirm that the same results are obtained when we expand the unit cell from two lattice sites to four lattice sites,
since the two-site translation in the $\pi/2$-PDW state and the one-site translation in the $\pi$-PDW state is equivalent to shifting the phase of the gap function by $\pi$ (or multiplying $e^{\mathrm{i}\pi}$), and the physics does not change, as mentioned in the last part of Sec.~\ref{sec:level3}.

We briefly mention the relationship between the Leggett mode and the Bardasis--Schrieffer mode \cite{Bardasis1961}.
If we properly decompose the order parameter in the irreducible representation of the point group of the system, sometimes the oscillating mode belongs to a different representation than the ground-state order parameter. In this case, the oscillation may be called the Bardasis-Schrieffer mode.
Here, we do not elaborate on the detailed distinction and simply call the mode the Leggett mode.
Recently, the Bardasis--Schrieffer mode in relation to the THz near-field optics has theoretically been discussed \cite{Sun2020}.

We also note that considering the sum rule of the optical conductivity, the spectral weight of each peak arising from the collective modes has a similar (or slightly smaller) order of the delta-function weight of the diamagnetic term that we do not consider here. 
So we expect that the peak signal can be observed in experiments.
We also note that there is a supercurrent contribution given in Sec.~\ref{LinearOpticalConductivity}, which should be considered to satisfy the sum rule.

In general, the mean-field theory gives similar results regardless of the dimension of the model, but strictly speaking, the superconducting order does not appear in 1D systems.
We demonstrate in the next section that our formalism and interpretations still work in a 2D model.

\section{\label{sec:level5}Two-dimensional model for the PDW states}
\begin{figure}
    \centering
    \includegraphics[scale=1.0]{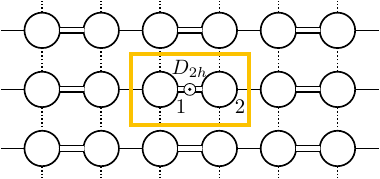}
    \caption{Schematic picture of the 2D model (\ref{Hamiltonian_2D}) that exhibits a PDW state.
    This model has two hopping amplitudes $-(t\pm\delta t)$ along the $x$ axis, while the hopping along the $y$ axis is uniformly $-t_{y}$.
    The yellow box indicates the unit cell containing two lattice sites labeled by $1$ and $2$.
    The point group of this model is $D_{2h}$, in which the Lifshitz invariant is allowed to exist in the free energy \cite{Nagashima2024}.}
    \label{Pic_model_2D}
\end{figure}

This section examines how the formulations and interpretations discussed so far also hold for a 2D system.
We first see that the $(\pi/2,0)$- and $(\pi,\pi)$-PDW state can appear in a 2D system by calculating the gap function in real space.
We then move on to the momentum space to calculate the free energy and linear optical conductivity of the $(\pi/2,0)$- and $(\pi,\pi)$-PDW states to see how the collective modes appear in the linear response regime.
To this end, we employ a model as shown in Fig.~\ref{Pic_model_2D}, where we stack multiple 1D models depicted in Fig.~\ref{Pic_model}, and the pair-hopping interaction is uniformly applied along the $x$ and $y$ axes.

\subsubsection{Real-space analysis}
We first calculate the gap function in the real space.
The mean-field Hamiltonian $\mathcal{H}_{\text{rMF2D}}$ is given by
\begin{align}
    \mathcal{H}_{\text{rMF2D}} %\notag \\
    &=\sum_{j,l}^{N}\sum_{\sigma}\left[ -\left\{t+\qty(-1)^{j}(\delta t)\right\}c^{\dagger}_{j,l,\sigma}c_{j+1,l,\sigma} \right. \notag \\
    &\quad - \left. t_{y}c^{\dagger}_{j,l,\sigma}c_{j,l+1,\sigma} + \text{H.c.} \right] -\mu\sum_{j,l}^{N}\sum_{\sigma}c^{\dagger}_{j,l,\sigma}c_{j,l,\sigma} \notag \\
    &\quad + \sum_{j,l}^{N}\qty[ \Delta_{j,l}c^{\dagger}_{j,l\up}c^{\dagger}_{j,l\down} + \text{H.c.}],
    \label{Hamiltonian_2D}
\end{align}
where
\begin{align}
    \Delta_{j,l} %\notag \\
    &= U\ev*{c_{j,l\down}c_{j,l\up}} + V\ev*{c_{j+1,l\down}c_{j+1,l\up}} + V\ev*{c_{j-1,l\down}c_{j-1,l\up}} \notag \\ 
    & \quad + V\ev*{c_{j,l+1\down}c_{j,l+1\up}} + V\ev*{c_{j,l-1\down}c_{j,l-1\up}},
\end{align}
is the site-dependent gap function.
The indices $j$ and $l$ specify the $x$ and $y$ positions, respectively.
We obtain the gap function by the self-consistent calculation as in the 1D case.
We choose $\delta t=0.45$, $t_{y}=1.0$, $\mu=-0.8$, and $U=-4.0$, and change $V$ for the $(\pi/2,0)$- and $(\pi,\pi)$-PDW state.
For the $(\pi/2,0)$-PDW state, $V=0.4$ is chosen, while for the $(\pi,\pi)$-PDW state, $V=0.8$ is chosen.
The number of the total lattice sites is $40\times40$. 
If one would study larger systems, there are several specialized numerical techniques to calculate the gap function within the mean-field approximation~\cite{gap_RealSpace2012, gap_RealSpace2017, gap_RealSpace2020}.

\begin{figure}
    \centering
    \includegraphics[scale=0.55]{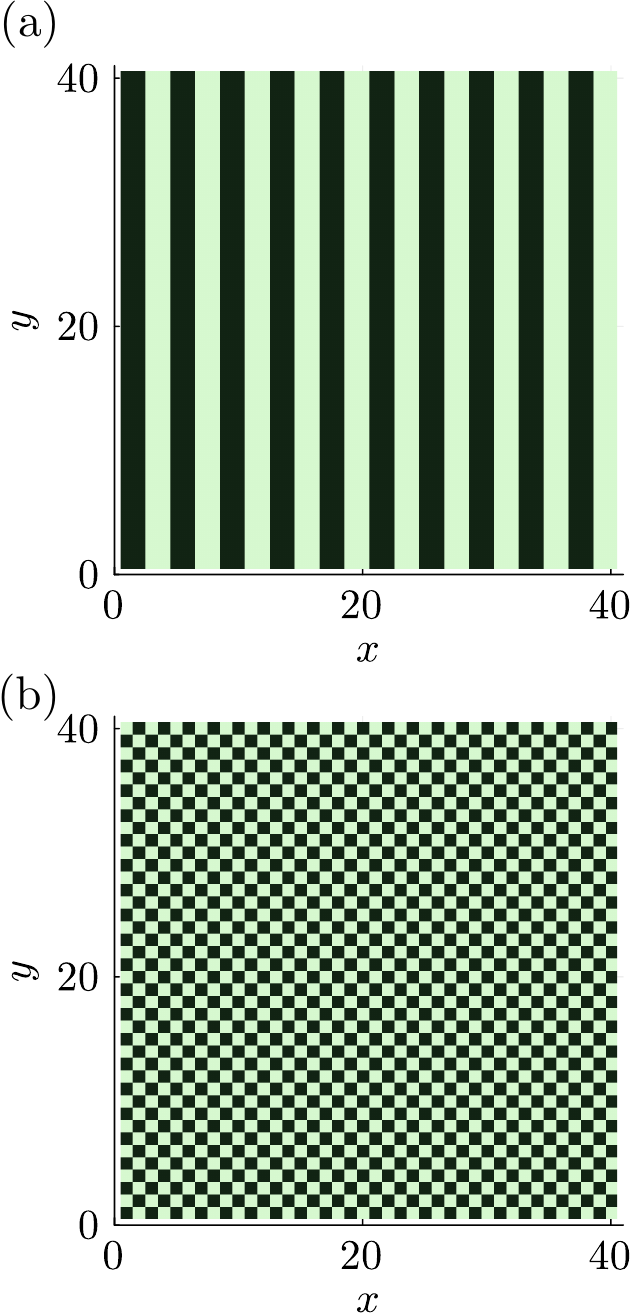}
    \caption{Gap function of the 2D model (\ref{Hamiltonian_2D}) in Fig.~\ref{Pic_model_2D} calculated in real space.
    We fix $\delta t=0.45$, $t_{y}=1.0$, $\mu=-0.8$, $U=-4.0$, and $T=0.001$, and change $V$ values for the $(\pi/2,0)$-PDW and  $(\pi,\pi)$-PDW states.
    The number of the total lattice sites is $40\times40$.
    (a) The gap function of the $(\pi/2,0)$-PDW state with $V=0.4$ in the $x$ direction.
    The light green part is $|\Delta_{0}|\approx0.79$, while the dark green part is $-|\Delta_{0}|\approx-0.79$.
    (b) The gap function of the $(\pi,\pi)$-PDW state with $V=0.8$ in the $x$ and $y$ direction.
    The light green part is $|\Delta_{0}|\approx2.1$, while the dark green part is $-|\Delta_{0}|\approx-2.1$.}
    \label{RealSpaceGap_2D}
\end{figure}

The results are shown in Fig.~\ref{RealSpaceGap_2D}.
When $V=0.4$ (Fig.~\ref{RealSpaceGap_2D}(a)), the unidirectional $(\pi/2,0)$-PDW is realized, similar to the 1D $\pi/2$-PDW state.
In this state, the PDW structure reflects the lattice structure, suggesting that the bond-order contribution surpasses that of the pair hopping.
On the other hand, when $V=0.8$ (Fig.~\ref{RealSpaceGap_2D}(b)), the $(\pi,\pi)$-PDW is realized, which does not reflect the nematic structure of the lattice as in the case of $\pi$-PDW state in one dimension.
We can conclude that the two distinct PDW states can also appear in a 2D system when the parameters are appropriately chosen.

\subsubsection{Momentum-space analysis}
We here first confirm that the $(\pi/2,0)$- and $(\pi,\pi)$-PDW states have the lowest free energy by calculating the free energy as a function of $\bm{q}$.
We can use almost the same formalism as in the case of the 1D model states in Sec.~\ref{sec:level4-1}.
The different parts are the kinetic part $\xi(\bm{k})$:
\begin{equation}
    \xi\qty(\bm{k}) = \mqty[ -\mu+g(k_{y}) & f(k_{x}) \\ f(k_{x})^{\dagger} & -\mu+g(k_{y})], 
\end{equation}
where $g(k_{y})=-2t_{y}\cos\qty(k_{y})$ and $f(k)$ is the same as the one in the 1D model, and the interaction part $\tilde{U}\qty(\bm{q})$:
\begin{equation}
    \tilde{U}\qty(\bm{q}) = \mqty[ U + 2V\cos\qty(q_{y}) & 2V\cos\qty(q_{x}) \\ 2V\cos\qty(q_{x}) & U + 2V\cos\qty(q_{y})].
\end{equation}

\begin{figure}
    \centering
    \includegraphics[scale=0.3]{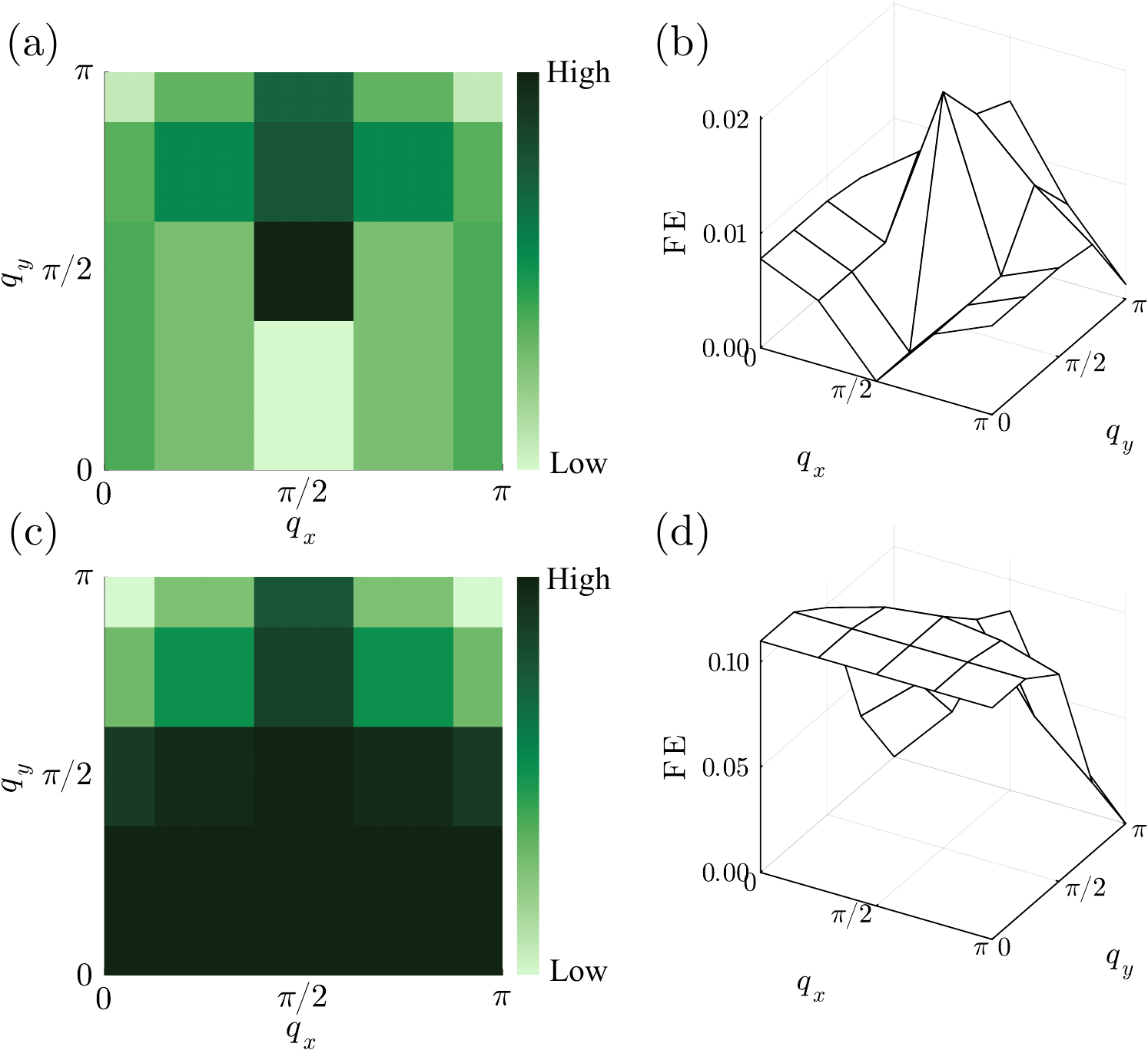}
    \caption{[(a),(c)] 2D plots of the free energy measured from the lowest value for the 2D model (\ref{Hamiltonian_2D}) with (a) $V=0.4$ and (c) $V=0.8$.
    [(b),(d)] 3D plots of the free energy with (b) $V=0.4$ and (d) $V=0.8$.
    The temperature is fixed to $T=0.05$.}
    \label{2D_FE_qdep}
\end{figure}

The $\bm{q}$-dependent free energy is shown in Fig.~\ref{2D_FE_qdep}
with $V=0.4$ and $V=0.8$ at $T=0.05$ (the other parameters are chosen to be the same as in Fig.~\ref{RealSpaceGap_2D}).
Figures~\ref{2D_FE_qdep}(a) and (c) are the 2D plots of the free energy with $V=0.4$ and $V=0.8$.
The lighter green area has a lower free energy, and the darker green area has a higher free energy.
Figures~\ref{2D_FE_qdep}(b) and (d) are the 3D plots of the free energy with $V=0.4$ and $V=0.8$ measured from the lowest free energy.
Figures~\ref{2D_FE_qdep}(a) and (b) indicate that the $(\pi/2,0)$-PDW state has the lowest free energy, while (c) and (d) indicate that the $(0,\pi)$- and $(\pi,\pi)$-PDW states have the same lowest free energy.

Although the $(0,\pi)$- and $(\pi,\pi)$-PDW states seem different, they are equivalent.
The equivalence becomes clear when we return to the real space: The gap function is $\Delta(\bm{r}) = \Delta_{0}\cos\qty(\pi x + \pi y)$.
When we move to the momentum space, we have two options to perform Fourier transformation.
One way is to take $\Delta_{0,1}=\Delta_{0}$, $\Delta_{0,2}=\Delta_{0}$, and $\bm{q}_{0}=(\pi,\pi)$, and the other way is to take $\Delta_{0,1}=\Delta_{0}$, $\Delta_{0,2}=-\Delta_{0}$, and $\bm{q}_{0}=\qty(0,\pi)$.
We can check the equivalence of these two cases by calculating the free energy or the Fourier transform to the real space.
For later calculation, we use $\bm{q}_{0}=\qty(\pi,\pi)$.

We can also see the consistency between the microscopic calculation and the GL theory in Fig.~\ref{2D_FE_qdep}.
In Fig.~\ref{2D_FE_qdep}(a) and (b), the free energy has the largest gradient from $\bm{q}=(0,0)$ to $(\pi/2,0)$, which corresponds to Eq.~(\ref{SufCondition}) (small $\bm{q}$ state).
In Fig.~\ref{2D_FE_qdep}(c) and (d), in contrast, the gradient close to $\bm{q}=(0,0)$ is zero, while the lowest free energies at $\bm{q}=\qty(0,\pi)$ and $\bm{q}=(\pi,\pi)$ are achieved, which corresponds to Eq.~(\ref{StrongCondition}) (large $\bm{q}$ state).

\begin{figure*}
    \centering
    \includegraphics[scale=0.45]{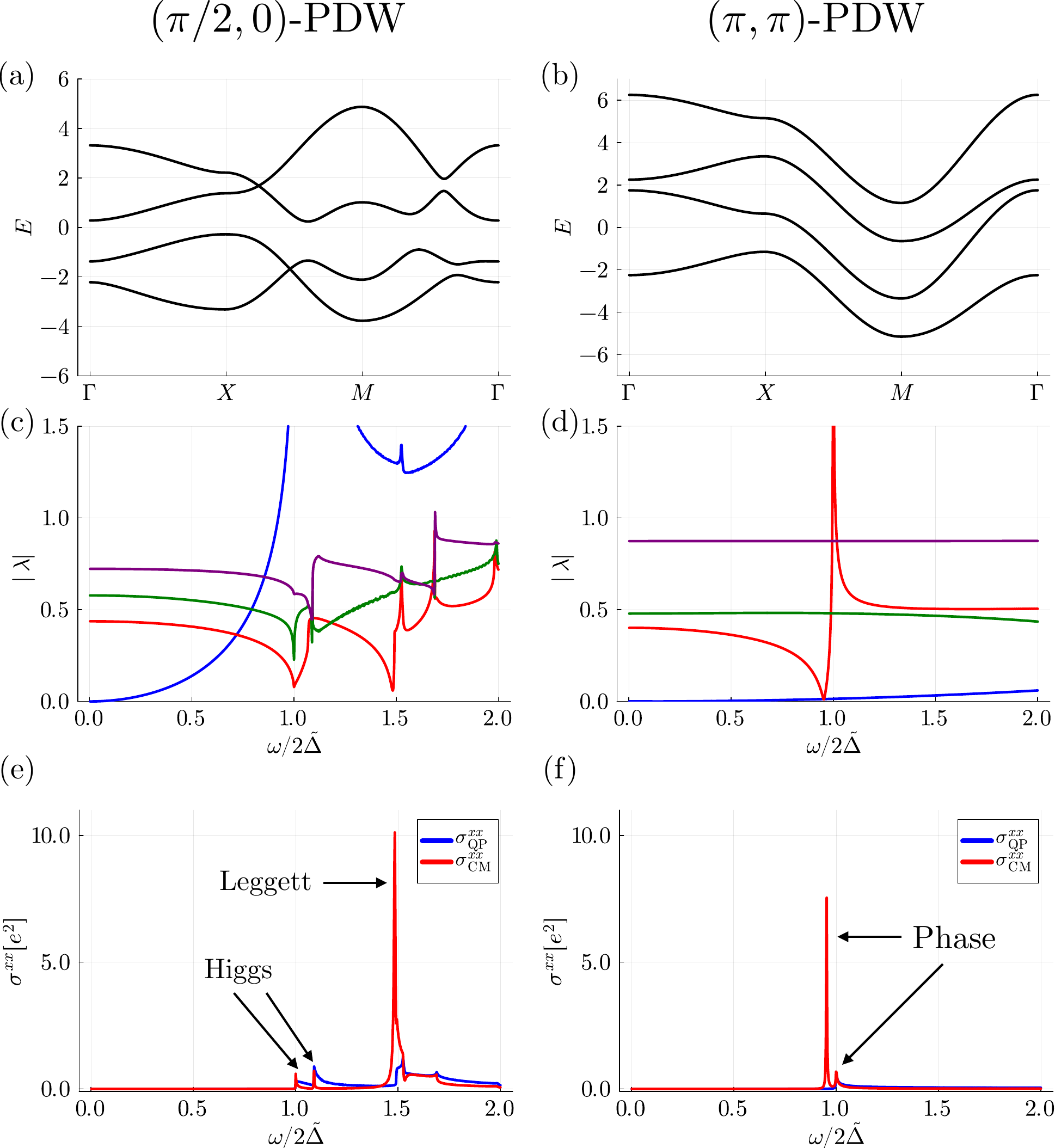}
    \caption{Band structures, absolute values of the eigenvalues of the effective interaction $\tilde{U}_{\text{eff}}$, and linear optical conductivity for the $(\pi/2,0)$- and $(\pi,\pi)$-PDW states in the 2D model (\ref{Hamiltonian_2D}) with $t_{y}=1.0$, $\delta t=0.45$, $\mu=-0.8$, $U=-4.0$, and $T=0.001$.
    We take $V=0.4$ for the $(\pi/2,0)$-PDW state and $V=0.8$ for the $(\pi,\pi)$-PDW state.
    [(a), (b)] Band structures of the (a) $(\pi/2,0)$-
    and (b) $(\pi,\pi)$-PDW states.
    [(c), (d)] Absolute values $|\lambda_{i}|$ of the eigenvalues $\lambda_{i}$ of $(1-\tilde{U}_{\rm eff}(\omega)\Pi(\omega))$ of the (c) $(\pi/2,0)$- and (d) $(\pi,\pi)$-PWD states.
    [(e),(f)] Contributions of the quasiparticles $\sigma^{xx}_{\text{QP}}$ (blue curve) and collective modes $\sigma^{xx}_{\text{CM}}$ (red) of the linear optical conductivity for the (e) $(\pi/2,0)$- and (f) $(\pi,\pi)$-PDW states.
    }
    \label{2D_OCs}
\end{figure*}

We now calculate the linear optical conductivity of the $\bm{q}_{0}=(\pi/2,0)$- and $(\pi,\pi)$-PDW states.
As in the case of the 1D model, we focus on $\sigma_{\text{QP}}$ and $\sigma_{\rm CM}$, which reflect the properties of the quasiparticles and collective modes.
We must be careful about the finite center-of-mass momentum shift in the momentum space.
In our model depicted in Fig.~\ref{Pic_model_2D}, the unit cell overlaps by the two-site translation along the horizontal axis or the one-site translation along the vertical axis.
This inequivalence means that the momentum shift in the kinetic part of the Hamiltonian is not $\bm{q}_{0}=(q_{0x},q_{0y})$, but $\bm{Q}=(q_{0x},q_{0y}/2)$ (the interaction part can use $\bm{q}_{0}$).
Paying attention to the finite momentum shift by $\bm{Q}$, the calculation can be done in the same way as in the 1D case.

We give the band structures of the $(\pi/2,0)$- and $(\pi,\pi)$-PDW states, absolute values $|\lambda_{i}|$ of the eigenvalues of $(1-\tilde{U}_{\rm eff}(\omega)\Pi(\omega))$, and the linear optical conductivity.
The parameters are fixed as $t_{y}=1.0$, $\delta t=0.45$, $\mu=-0.8$, $U=-4.0$, and $T=0.001$, and we choose $V=0.4$ for the $(\pi/2,0)$-PDW state and $V=0.8$ for the $(\pi,\pi)$-PDW state.
The results are given in Fig.~\ref{2D_OCs}.
Figures~\ref{2D_OCs}(a) and (b) depict the band structures of the $(\pi/2,0)$- and $(\pi,\pi)$-PDW states, respectively.
The direct gap $2\tilde{\Delta}$ of the $(\pi/2,0)$-PDW state in Fig.~\ref{2D_OCs}(a) is $2\tilde{\Delta}\approx1.48$, which is different from $2|\Delta_{0}|\approx1.58$.
In the $(\pi,\pi)$-PDW state in (b), they are $2\tilde{\Delta}\approx0.50$ and $2|\Delta_{0}|\approx4.2$.

We then evaluate the characteristic frequencies of the collective modes, as in the case of the 1D model in Sec.~\ref{sec:level4-4}.
The eigenvalues $|\lambda_{i}|$ for the $(\pi/2,0)$- and $(\pi,\pi)$-PDW states are shown in Fig.~\ref{2D_OCs}(c) and (d), respectively.
The zero at $\omega=0$ of the blue line in (c) corresponds to the NG mode, which is not a physical degree of freedom due to the Anderson--Higgs mechanism.
The green and purple curves have multiple dip-like structures at $\omega=2\tilde{\Delta}$, $\omega\approx 1.1\times2\tilde{\Delta}$, and $\omega\approx 1.75\times2\tilde{\Delta}$, all of which finitely contribute to the linear optical conductivity.
The red curve also has multiple dip-like structures at $\omega=2\tilde{\Delta}$ and around $1.5\times2\tilde{\Delta}$.
The former contribution is small, while the latter is much larger than the other dip-like structures in (e).
There are two zeros in (d), one of which is at $\omega=0$ corresponding to the NG mode, and the other of which is at $\omega\approx0.95\times2\tilde{\Delta}$ responsible for the sharp peak in (f).

We next consider the linear optical conductivity of the $(\pi/2,0)$- and $(\pi,\pi)$-PDW states.
Figures~\ref{2D_OCs}(e) and (f) represent the $xx$ component of the linear optical conductivity from the quasiparticle contribution $\sigma^{xx}_{\text{QP}}$ (blue curve) and the collective mode contribution $\sigma^{xx}_{\text{CM}}$ (red) of the $(\pi/2,0)$- and $(\pi,\pi)$-PDW states, respectively.
In Fig.~\ref{2D_OCs}(e), the quasiparticle $\sigma^{xx}_{\text{QP}}$ and collective mode contributions $\sigma^{xx}_{\text{CM}}$ have multiple small peaks at $\omega=2\tilde{\Delta}$, $\omega\approx 1.1\times2\tilde{\Delta}$, and $\omega\approx 1.7\times2\tilde{\Delta}$.
At each peak, the quasiparticles and collective modes have almost the same magnitude.
On the other hand, only the collective mode contribution $\sigma^{xx}_{\text{CM}}$ has a sharp peak at $\omega\approx1.5\times2\tilde{\Delta}$.
All of these peaks arise from the dip-like structures in (c).
In Fig.~\ref{2D_OCs}(f), the quasiparticle $\sigma^{xx}_{\text{QP}}$ and the collective mode contribution $\sigma^{xx}_{\text{CM}}$ have a small peak at $\omega=2\tilde{\Delta}$, but the origin of these small peaks are not apparent in (d).
The collective mode contribution $\sigma^{xx}_{\text{CM}}$ has a sharp peak at $\omega\approx0.95\times2\tilde{\Delta}$, which comes from the pole structure in (d).

\begin{figure}
    \centering
    \includegraphics[scale=0.5]{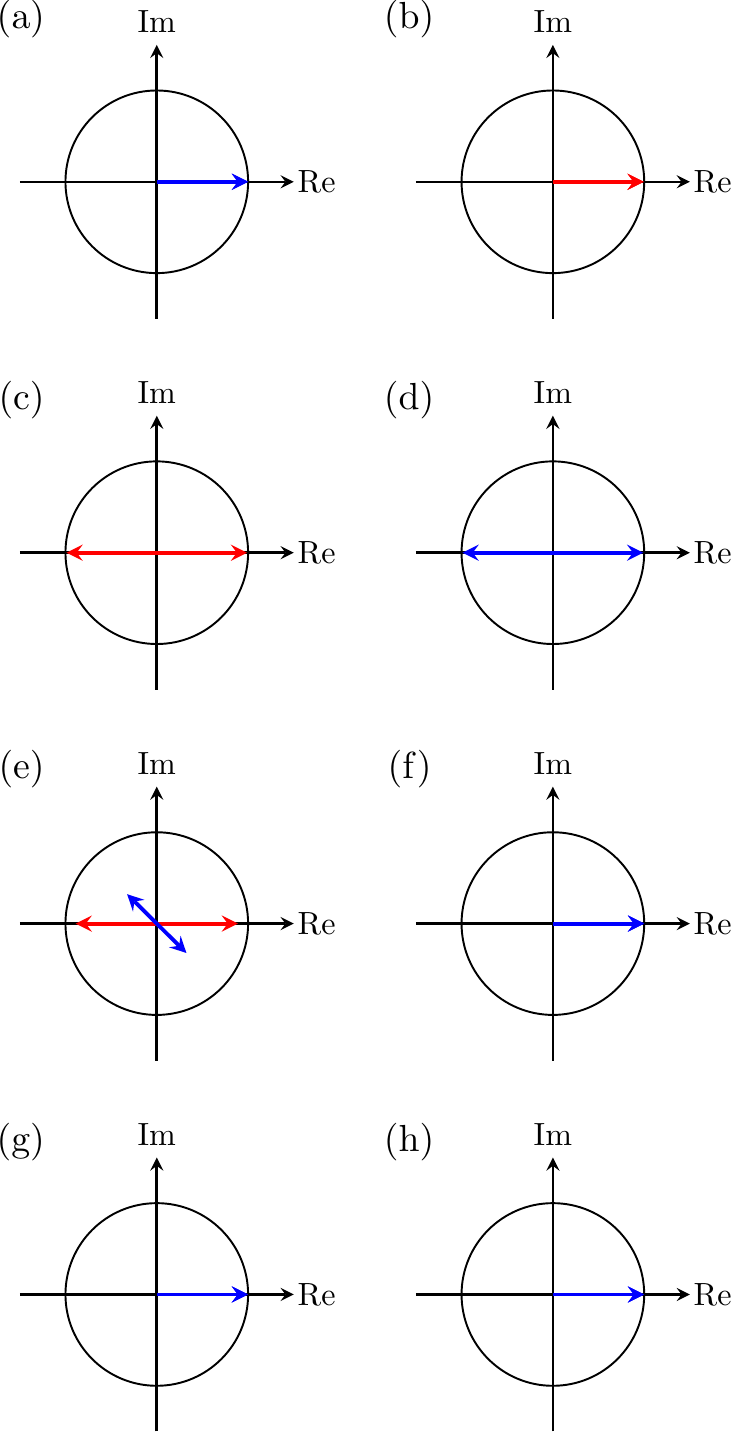}
    \caption{Eigenvectors of the effective interaction $\tilde U_{\rm eff}(\omega)$ with different frequencies at the peaks of the optical conductivity for the [(a)-(e)] $(\pi/2,0)$- and [(f)-(h)] $(\pi,\pi)$-PDW states.
    The red and blue arrows represent the eigenvector components corresponding to the amplitude and phase fluctuations, respectively.
    The radius of the circle is $1/\sqrt{2}$.
    (a) $\omega=0$ (two-fold degenerate), (b) $\omega= 2\tilde{\Delta}$ (two-fold), (c) $\omega\approx 1.1\times 2\tilde{\Delta}$, (d) $\omega\approx1.5\times2\tilde{\Delta}$, (e) $\omega\approx1.7\times2\tilde{\Delta}$,
    (f) $\omega=0$ (two-fold), (g) $\omega\approx0.95\times2\tilde{\Delta}$ (two-fold), and (h) $\omega=2\tilde{\Delta}$ (two-fold).}
    \label{2D_Eigenvectors}
\end{figure}

We further examine the weight of the contributions from the amplitude and phase fluctuations to the linear optical conductivity by checking the eigenvectors of the effective interaction $\tilde{U}_{\text{eff}}$ at the peaks (and $\omega=0$ corresponding to the NG mode).
The eigenvectors at both states' peaks (and $\omega=0$) are shown in Fig.~\ref{2D_Eigenvectors}.
As in the 1D case, we apply a unitary transformation $\mathcal{U}=\exp((\mathrm{i}\theta_{1}\tau_{z1}+\mathrm{i}\theta_{2}\tau_{z2})/2)$ to omit the phases of the gap functions.

In the $(\pi/2,0)$-PDW state (Fig.~\ref{2D_Eigenvectors}(a)-(e)), the NG mode has a contribution only from the phase fluctuations in the same direction (Fig.~\ref{2D_Eigenvectors}(a)).
In Fig.~\ref{2D_Eigenvectors}(b)((c)), one can see that the contribution is coming solely from the amplitude fluctuations in the same (opposite) direction, with which we can identify the peaks of the optical conductivity at $\omega=2\tilde\Delta$ and $\omega\approx 1.1\times 2\tilde\Delta$ in Fig.~\ref{2D_OCs}(e) originate from the Higgs mode.
In Fig.~\ref{2D_Eigenvectors}(d), the eigenvector has a weight in the phase fluctuations with the opposite signs for the different bands.
We can conclude from this fact that the highest peak in Fig.~\ref{2D_OCs}(e) at $\omega\approx1.5\times2\tilde{\Delta}$ arises from the Leggett mode.
The $(\pi/2,0)$-PDW state results signify that the Higgs and Leggett modes can be observed separately in the linear response regime, which differs from those in the 1D case.
In Fig.~\ref{2D_Eigenvectors}(e), the eigenvector has a weight both on the amplitude and phase fluctuations, implying that the shoulder structure at $\omega\approx 1.7\times 2\tilde\Delta$ in Fig.~\ref{2D_OCs}(e) corresponds to an amplitude-phase-hybridized mode.

In the $(\pi,\pi)$-PDW state [Fig.~\ref{2D_Eigenvectors}(f)-(h)], 
the eigenvectors have weights only on the phase fluctuations, suggesting that
each peak in the optical conductivity in Fig.~\ref{2D_OCs}(f) originates from certain phase modes.
This unique property of the $(\pi,\pi)$-PDW state is linked to the GL theory explained in Sec.~\ref{sec:level2-1}; Each lattice site has a gap function with the same magnitude and alternating signs, letting the Higgs mode contribution from each site cancel each other.

We note that the different components of the optical conductivity, $\sigma^{xy}$ and $\sigma^{yy}$, are all zero.
Since the time-reversal symmetry is conserved in our model, the Hall component $\sigma^{xy}$ must vanish.
Concerning $\sigma^{yy}$, the hopping amplitude along the $y$-axis is uniform in our model.
This is similar to a uniform superconductor without impurities,
where the optical conductivity totally vanishes at finite frequencies in the clean limit.

\section{\label{sec:level6} Discussions}
We have investigated how the two distinct PDW states appear in multiband superconductors with the Lifshitz invariant and how these PDW states respond linearly to external fields by calculating the linear optical conductivity.
We first utilized the macroscopic GL theory to see how a nonuniform (nonzero $\bm{q}$) state is favored when the Lifshitz invariant exists, and derived two conditions for qualitatively different finite center-of-mass momentum states.
The first one (small $\bm{q}$ state) originates from the competition between the effect of ``twisting" the phase by the Lifshitz invariant and the effect of ``aligning" the phase by other GL terms. 
If the ``phase twisting" effect is superior to the other effects, the nonzero $\bm{q}$ state is realized since the free energy $\mathcal{F}$ becomes unstable at $\bm{q}=\bm{0}$, which is manifested in the condition (\ref{SufCondition}).
The second one (large $\bm{q}$ state), on the other hand, occurs due to the dominant drag effect, and the presence/absence of the Lifshitz invariant does not matter, shown in the condition~(\ref{StrongCondition}).

We constructed the microscopic 1D two-band model to realize the nonzero $\bm{q}$ state corresponding to the PDW state.
We compare the results of the GL theory and the microscopic calculations and see a qualitative agreement between them.
The results in equilibrium can be used to explore the optical properties of the PDW states, where we calculate the optical conductivity of the $\pi/2$- and $\pi$-PDW states.
In the $\pi/2$-PDW state, the Higgs and Leggett modes are hybridized, while in the $\pi$-PDW state, the phase modes contribute to the linear optical conductivity, presenting a significant difference between the two distinct PDW states.

To demonstrate the validity of our theory in higher dimensions, we also constructed the 2D two-band model to calculate the gap function, free energy, and optical conductivity.
We obtained two distinct $(\pi/2,0)$- and $(\pi,\pi)$-PDW states, which qualitatively correspond to the small and large $\bm{q}$ states in the GL theory.
Their linear optical responses were drastically different: In the $(\pi/2,0)$-PDW state, the Higgs, Leggett, and their hybridized modes arose in the linear response, but in the $(\pi,\pi)$-PDW state phase modes mainly contributed to the linear response.
These results indicate that the PDW state is an appropriate platform for exploring the superconducting collective modes in the linear response regime.

The macroscopic GL theory agreed fairly well with the microscopic theories.
Our results demonstrate the effectiveness of the GL theory in describing the conditions for the two distinct PDW states, although the value of $\bm{q}$ realized in the microscopic theory might be beyond the scope of the GL theory.
The conditions (\ref{SufCondition}) and (\ref{StrongCondition}) give clear physical pictures within the GL framework to realize the PDW states.
Moreover, these conditions give us guidelines to explore real materials exhibiting PDW states.
The former one (\ref{SufCondition}) suggests that materials with a large Lifshitz invariant are suitable for the small $\bm q$ PDW state, while the latter one (\ref{StrongCondition}) implies that materials with a large effective mass, such as heavy fermion superconductors \cite{HeavyFermionSC_review1984} and flat band superconductors \cite{Tian2023, Khasanov2024}, would be preferred for the large $\bm q$ PDW state.

We finally comment on possible experimental observations of the collective modes in the linear response regime (note that the Leggett mode has been suggested to be observed in nonlinear responses \cite{Blumberg2007, Balatsuky2000, Ota2011, Lin2012, Marciani2013, Bittner2015, Krull2016, Cea2016, Giorgianni2019, Murotani2019, Fiore2022, Yuan2024}).
Interesting candidate materials would be cuprates \cite{Cuprate_expr_Fujita2012, Cuprate_expr_Edkins2019, Cuprate_expr_Du2020}, iron-based superconductors \cite{Iron_expr_Liu2023,Iron_expr_Zhao2023}, Kagome superconductors \cite{Kagome_expr_Chen2021, Kagome_expr_Zhao2021}, and $\text{UTe}_{2}$ \cite{UTe2_expr_Aishwarya2023, UTe2_expr_Gu2023}.
We can see peaks from the quasiparticles and collective modes, such as Higgs and Leggett modes, when we observe the linear optical conductivity.
Still, it may be difficult to distinguish the contributions from the quasiparticles and collective modes unless the superconducting gap function is $s$-wave and the collective mode is inside the gap.
Nevertheless, we can pick up the collective modes' contributions in real materials by considering the effect of impurities.
On the other hand, the Higgs mode contribution is found to be larger than that of quasiparticles in the presence of nonmagnetic impurities in the dirty regime \cite{Silaev2019, Tsuji_and_Nomura2020, Seibold2021}.
Therefore, by tracking the impurity dependence of the peak structures, one could extract the contributions from the collective Higgs and Leggett modes.

\section{\label{sec:level7} Acknowledgement}
We thank K. Takasan, S. Imai, T. Nakamoto, Y. Nagai, T. Kitamura, M. Sigrist, and H. Tsunetsugu for fruitful discussions. This work is supported by JST FOREST (Grant No. JPMJFR2131) and JSPS KAKENHI (Grant No. JP24H00191).

\onecolumngrid
\appendix
\section{Details of the microscopic calculations in momentum space}
\label{Apdx.A}
In this Appendix, we microscopically derive the GL parameters, free energy, gap function, and linear optical conductivity with the path integral in imaginary time $\tau$.
We formulate in one dimension keeping the model in Fig.~\ref{Pic_model} in mind for simplicity, but extending the dimension higher than one is straightforward.
We begin with the Hamiltonian,
%$\mathcal{H}$:
\begin{align}
    \mathcal{H} &= \mathcal{H}_{\text{kin}} + \mathcal{H}_{\text{int}}, \notag \\
    \mathcal{H}_{\text{int}} &= \sum_{k\alpha\alpha'\sigma}\xi_{\alpha\alpha'}(k)c^{\dagger}_{k\alpha\alpha'\sigma}c_{k\alpha'\sigma}, \notag \\
    \mathcal{H}_{\text{int}} &= \sum_{kk'\alpha\alpha'}\sum_{q=q_{0}}\frac{1}{2}\tilde{U}_{\alpha\alpha'}(q)c^{\dagger}_{k+q,\alpha\up}c^{\dagger}_{-k+q,\alpha\down}c_{-k'+q,\alpha'}c_{k'+q,\alpha'\up},
    \label{Hamiltonian_k_space}
\end{align}
where the interaction $\tilde{U}(q)$ is
\begin{equation}
    \tilde{U}(q) = \mqty[ U & 2V\cos(q) \\ 2V\cos(q) & U].
    \label{Interaction_k_space}
\end{equation}
Considering the real space formulation in Sec.~\ref{sec:level3}, why we have this interaction becomes clear.
The PDW gap function $\Delta(x)=\Delta_{0}\cos\qty(q_{0}x+\theta)$ is decomposed into two parts in real space.
\begin{align}
    \Delta(x) = \Delta_{0}\cos\qty(q_{0}x+\theta) &= \frac{\Delta_{0}}{2}e^{\mathrm{i}\qty(q_{0}x+\theta)} + \frac{\Delta_{0}}{2}e^{-\mathrm{i}\qty(q_{0}x+\theta)} \notag \\
    &= \frac{\Delta_{q_{0}}}{2}e^{\mathrm{i}q_{0}x} + \frac{\Delta_{-q_{0}}}{2}e^{-\mathrm{i}q_{0}x},
\end{align}
where $\Delta_{q_{0}} = \Delta_{0}e^{\mathrm{i}\theta}$, $\Delta_{-q_{0}} = \Delta_{0}e^{-\mathrm{i}\theta}$, and $\Delta_{q_{0}}^{*}=\Delta_{-q_{0}}$.
Since the Fourier transform of the operator $c^{\dagger}_{2j-1,\sigma}$ and $c^{\dagger}_{2j,\sigma}$ is given by
\begin{equation}
    c_{2j-1,\sigma} = \frac{1}{\sqrt{N}}\sum_{k}c_{k,1,\sigma}e^{\mathrm{i}k\cdot\qty(j-1/2)}, \quad c_{2j,\sigma} = \frac{1}{\sqrt{N}}\sum_{k}c_{k,2,\sigma}e^{\mathrm{i}k\cdot j},
\end{equation}
where the second index of $c_{k,\alpha,\sigma}$ is the band index $\alpha=1,2$, and by using Eq.~(\ref{Gap_realspace_MF}) the gap function at the site $(2j-1)$ is written as
\begin{align}
    \frac{\Delta_{q_{0}}}{2}e^{\mathrm{i}q_{0}\qty(2j-1)} + \frac{\Delta_{-q_{0}}}{2}e^{-\mathrm{i}q_{0}\qty(2j-1)} &= U\ev*{c_{2j-1,\down}c_{2j-1,\up}} + V\ev*{c_{2j,\down}c_{2j,\up}} + V\ev*{c_{2j-2,\down}c_{2j-2,\up}} \notag \\
    &= U \frac{1}{N}\sum_{kk'}\ev*{c_{k,1,\down}c_{k',1,\up}e^{\mathrm{i}(k+k')\cdot(j-1/2)}} \notag \\
    &\quad + V\frac{1}{N}\sum_{kk'}\ev*{c_{k,2,\down}c_{k',2,\up}e^{\mathrm{i}(k+k')\cdot(j-1/2)}e^{\mathrm{i}(k+k')}} \notag \\
    &\quad + V\frac{1}{N}\sum_{kk'}\ev*{c_{k,2,\down}c_{k',2,\up}e^{\mathrm{i}(k+k')\cdot(j-1/2)}e^{-\mathrm{i}(k+k')}},
\end{align}
the gap function transforms as follows.
    \begin{align}
    \text{(LHS)} &= \sum_{j}e^{-\mathrm{i}q\cdot(2j-1)}\qty(\frac{\Delta_{q_{0}}}{2}e^{\mathrm{i}q_{0}\cdot(2j-1)} + \frac{\Delta_{-q_{0}}}{2}e^{-\mathrm{i}q_{0}\cdot(2j-1)}) \notag \\
    &= \frac{\Delta_{q_{0}}}{2}\delta_{q,q_{0}} + \frac{\Delta_{-q_{0}}}{2}\delta_{q,-q_{0}} \notag \\
    \text{(RHS)} &= \frac{1}{N}\sum_{j}\sum_{kk'}U\ev*{c_{k,1,\down}c_{k,'1,\up}}e^{\mathrm{i}(k+k'-2q)\cdot(j-1/2)} \notag \\
    & \quad + \frac{1}{N}\sum_{j}\sum_{kk'}V\ev*{c_{k,2,\down}c_{k',2,\up}}e^{\mathrm{i}(k+k'-2q)\cdot(j-1/2)}e^{\mathrm{i}(k+k')/2} \notag \\
    & \quad + \frac{1}{N}\sum_{j}\sum_{kk'}V\ev*{c_{k,2,\down}c_{k',2,\up}}e^{\mathrm{i}(k+k'-2q)\cdot(j-1/2)}e^{-\mathrm{i}(k+k')/2} \notag \\
    &= \sum_{k}\qty( U\ev*{c_{k+q,1,\down}c_{-k+q,1,\up}} + 2V\cos{(q)}\ev*{c_{k+q,2,\down}c_{-k+q,2,\up}} ) \notag \\
    &= \sum_{k}\frac{1}{2}\qty( U\ev*{c_{k+q,1,\down}c_{-k+q,1,\up}} + 2V\cos{(q)}\ev*{c_{k+q,2,\down}c_{-k+q,2,\up}} ) \notag \\
    & \quad + \sum_{k}\frac{1}{2}\qty( U\ev*{c_{k-q,1,\down}c_{-k-q,1,\up}} + 2V\cos{(q)}\ev*{c_{k-q,2,\down}c_{-k-q,2,\up}} )
\end{align}
The gap function at the site $2j$ transforms in the same way.
Hence, the interaction in a momentum space is put in Eq.~(\ref{Interaction_k_space}).
As expected, $q=0$ is applied to the uniform case, while $q=\pm q_{0}$ for the PDW case (in our model $q_{0}=\pi/2$ or $\pi$).
We can also check that the Hamiltonian Eq.~(\ref{Hamiltonian_k_space}) returns to the uniform one when $q_{0}\to 0$.

We shift the momentum $k$ to $k\pm q_{0}$ in $\mathcal{H}_{\text{kin}}$ by using the Brillouin zone periodicity to use the Hubbard--Stratonovich transformation:
\begin{equation}
    \mathcal{H}_{\text{kin}} = \sum_{k\alpha\alpha'\sigma}\sum_{q=\pm q_{0}}\frac{1}{2}\xi_{\alpha\alpha'}\qty(k+q)c^{\dagger}_{k+q,\alpha\sigma}c_{k+q,\alpha'\sigma}.
\end{equation}
We also include the electromagnetic contribution up to the first order $\mathcal{H}_{\text{EM}}$ arising from the kinetic part with the shifted momentum to consider the linear optical conductivity later:
\begin{equation}
    \mathcal{H}_{\text{EM}} = \sum_{q=\pm q_{0}}\sum_{k\alpha\alpha'\sigma}\qty(\partial_{k}\xi_{\alpha\alpha'})(k+q)\qty(-eA)c^{\dagger}_{k+q,\alpha\sigma}c_{k+q,\alpha'\sigma}.
\end{equation}
Then, we use the path integral approach in imaginary time $\tau$.
The partition function of the whole system $\mathcal{Z}$ is stated as $\mathcal{Z}=\int \mathcal{D}(c^{\dagger}c)e^{-S[c^{\dagger},c]}$ with Euclidean action
\begin{equation}
    S[c^{\dagger},c] = \int_{0}^{\beta}d\tau\qty( \sum_{k\alpha\sigma}c^{\dagger}_{k\alpha\sigma}\partial_{\tau}c_{k\alpha\sigma} + \mathcal{H} ) = \int_{0}^{\beta}d\tau\qty( \sum_{k\alpha\sigma}\sum_{q=\pm q_{0}}\frac{1}{2}c^{\dagger}_{k+q,\alpha\sigma}\partial_{\tau}c_{k+q,\alpha\sigma} + \mathcal{H} ),
\end{equation}
where the volume of the system is set to be one for simplicity.
Hubbard--Stratonovich transformation splits the interaction part of the Hamiltonian $\mathcal{H}_{\text{int}}$ into the auxiliary bosonic fields $\Delta_{\pm q_{0}\alpha}$, $\Delta_{\pm q_{0}\alpha}^{*}$, and fermions (electrons).
\begin{align}
    &\exp\qty( \sum_{q=\pm q{0}}\int_{0}^{\beta}d\tau \sum_{\alpha\alpha'}\sum_{kk'}\frac{1}{2}\tilde{U}_{\alpha\alpha'}\qty(q)c^{\dagger}_{k+q,\alpha\up}c^{\dagger}_{-k+q,\alpha'\down}c_{-k'+q,\alpha\down}c_{k'+q,\alpha'\up} ) \notag \\
    &= \prod_{q=\pm q_{0}}\int \mathcal{D}\qty(\Delta_{q}^{*},\Delta_{q})\exp\left(-\int_{0}^{\beta}d\tau\left[ \sum_{\alpha\alpha'}2\frac{\Delta_{q,\alpha}^{*}}{2}\tilde{U}_{\alpha\alpha'}^{-1}\qty(q)\frac{\Delta_{q,\alpha'}}{2} \right.\right. \notag \\
    &\quad\quad\quad\quad\quad \left.\left.+ \sum_{\alpha,k}\left(\frac{\Delta_{q,\alpha}^{*}}{2}c_{k+q,\alpha\up}c_{-k+q,\alpha\down} + \frac{\Delta_{q,\alpha}}{2}c^{\dagger}_{-k+q,\alpha\down}c^{\dagger}_{k+q,\alpha\up}\right)\right]\right)
\end{align}
To simplify the notation, We take a basis 
\begin{equation}
    \Psi_{q,k} = \qty[ C^{\dagger}_{q\up}(k)\ C_{q\down}(-k)\ C_{-q\down}(-k)\ C^{\dagger}_{-q\up}(k)]^{\text{T}},
\end{equation}
where $C^{\dagger}_{q\sigma}(k)=\qty[c^{\dagger}_{k+q,1\sigma}\  c^{\dagger}_{k+q,2\sigma}]$.
This gives a simple formulation of the action with Green's function:
\begin{align}
    &S[c^{\dagger},c,\Delta^{*}_{q_{0}}, \Delta_{q_{0}},\Delta^{*}_{-q_{0}}, \Delta_{-q_{0}}] \notag \\
    &=-2\sum_{q=\pm q_{0}}\int_{0}^{\beta}d\tau\sum_{\alpha\alpha'}\frac{\Delta_{q,\alpha}^{*}}{2}\tilde{U}^{-1}_{\alpha\alpha'}\frac{\Delta_{q,\alpha'}}{2} - \int_{0}^{\beta}d\tau\Psi^{\dagger}_{q_{0},k}(\tau)\tilde{G}_{q_{0}}(k,\tau)\Psi_{q_{0},k}(\tau),
\end{align}
where
\begin{align}
    \tilde{G}^{-1}_{q_{0}}(k,\tau) &=\mqty[ { \begin{array}{cc} -\frac{1}{2}\partial_{\tau} - \frac{1}{2}\xi\qty(k+q_{0}) & -\Delta_{q_{0}}(\tau)/2 \\ -\Delta_{-q_{0}}(\tau)/2 & - \frac{1}{2}\partial_{\tau} + \frac{1}{2}\xi\qty(k-q_{0}) \end{array} } & \mbox{\huge O} \\ \mbox{\huge O} & { \begin{array}{cc} -\frac{1}{2}\partial_{\tau} + \frac{1}{2}\xi\qty(k+q_{0}) & -\Delta_{q_{0}}(\tau)/2 \\ -\Delta_{-q_{0}}(\tau)/2 & - \frac{1}{2}\partial_{\tau} - \frac{1}{2}\xi\qty(k-q_{0}) \end{array} } ] \notag \\
    & \quad + eA(\tau)\mqty[ {\begin{array}{cc} \frac{1}{2}\partial_{k}\xi\qty(k+q_{0}) & O \\ O & \frac{1}{2}\partial_{k}\xi\qty(k-q_{0}) \end{array} } & \mbox{\huge O} \\ \mbox{\huge O} & {\begin{array}{cc} \frac{1}{2}\partial_{k}\xi\qty(k-q_{0}) & O \\ O & \frac{1}{2}\partial_{k}\xi\qty(k+q_{0}) \end{array} } ],
\end{align}
We here used the hermiticity of the kinetic part $\xi(k)$ and the property of the gap function $\Delta_{-q_{0}}=\Delta_{q_{0}}^{*}$.

We move on to the frequency space by using $\Psi_{k,q_{0}}(\tau) = (1/\beta)\sum_{n}\Psi_{k,q_{0}}(\mathrm{i}\omega_{n})e^{-\mathrm{i}\omega_{n}\tau}$ and obtain with the fluctuation $\delta\Delta_{\pm q_{0}}$
\begin{align}
    S\qty[c^{\dagger},c, \Delta^{*}_{q_{0}}, \Delta_{q_{0}}, \Delta^{*}_{-q_{0}}, \Delta_{-q_{0}} ] =& -2\sum_{q=\pm q_{0}}\beta\sum_{\alpha\alpha'}\frac{\Delta^{*}_{q,0\alpha}}{2}\tilde{U}^{-1}_{\alpha\alpha'}(q)\frac{\Delta_{q,0\alpha'}}{2} \notag \\
    &-2\sum_{q=\pm q_{0}}\sum_{\alpha\alpha'}\sum_{n}\frac{\delta\Delta^{*}_{q,\alpha}(\mathrm{i}\omega_{n})}{2}\tilde{U}^{-1}_{\alpha\alpha'}(q)\frac{\delta\Delta_{q,\alpha'}(\mathrm{i}\omega_{n})}{2} \notag \\
    &-\frac{1}{\beta}\sum_{k}\sum_{m,n}\Psi^{\dagger}_{q_{0},k}(\mathrm{i}\omega_{m})\tilde{G}_{q_{0}}(\mathrm{i}\omega_{m},\mathrm{i}\omega_{n};k)\Psi_{q_{0},k}(\mathrm{i}\omega_{n}).
\end{align}
The Green function $\tilde{G}_{q_{0}}(\mathrm{i}\omega_{m},\mathrm{i}\omega_{n};k)$ is defined as
\begin{align}
&\tilde{G}^{-1}_{q_{0}}(\mathrm{i}\omega_{m},\mathrm{i}\omega_{n};k) \notag \\
&= \mqty[ { \begin{array}{cc} \qty( \mathrm{i}\omega_{n} - \xi\qty(k+q_{0}) )/2 & -\Delta_{q_{0},0}/2 \\ -\Delta_{-q_{0},0}/2 & \qty( \mathrm{i}\omega_{n} + \xi\qty(k-q_{0}) )/2 \end{array} } & \mbox{\huge O} \\ \mbox{\huge O} & { \begin{array}{cc} \qty( \mathrm{i}\omega_{n} + \xi\qty(k+q_{0}) )/2 & -\Delta_{q_{0},0}/2 \\ -\Delta_{-q_{0},0}/2 & \qty( \mathrm{i}\omega_{n} - \xi\qty(k-q_{0}) )/2 \end{array} } ]\beta\delta_{m,n} \notag \\
&\quad + \mqty[ { \begin{array}{cc} O & -\delta\Delta_{q_{0}}(\mathrm{i}\omega_{m}-\mathrm{i}\omega_{n})/2 \\ -\delta\Delta_{-q_{0}}(\mathrm{i}\omega_{m}-\mathrm{i}\omega_{n})/2 & O \end{array} } & \mbox{\huge O} \\ \mbox{\huge O} & { \begin{array}{cc} O & -\delta\Delta_{q_{0}}(\mathrm{i}\omega_{m}-\mathrm{i}\omega_{n})/2 \\ -\delta\Delta_{-q_{0}}(\mathrm{i}\omega_{m}-\mathrm{i}\omega_{n})/2 & O \end{array}} ]\beta \notag \\
&\quad + \mqty[ {\begin{array}{cc} \partial_{k}\xi\qty(k+q_{0})/2 & O \\ O & \partial_{k}\xi\qty(k-q_{0})/2 \end{array} } & \mbox{\huge O} \\ \mbox{\huge O} & {\begin{array}{cc} \partial_{k}\xi\qty(k-q_{0})/2 & O \\ O & \partial_{k}\xi\qty(k+q_{0})/2 \end{array} } ] eA(\mathrm{i}\omega_{m} - \mathrm{i}\omega_{n})\beta \notag \\
&= \frac{\beta\delta_{m,n}}{2}\mqty[ G^{-1}_{+,0}(\mathrm{i}\omega_{n};k) & O \\ O & G^{-1}_{-,0}(\mathrm{i}\omega_{n};k)] \notag \\
&\quad\quad -\frac{\beta}{2} \mqty[ \delta\Delta(\mathrm{i}\omega_{m}-\mathrm{i}\omega_{n}) & O \\ O & \delta\Delta(\mathrm{i}\omega_{m}-\mathrm{i}\omega_{n}) ]  + \frac{\beta}{2}\mqty[ v_{q_{0}}(k) & O \\ O & v_{-q_{0}}(k) ]eA(\mathrm{i}\omega_{m} - \mathrm{i}\omega_{n}).
\end{align}
Here, we also defined 
\begin{equation}
    G^{-1}_{\pm,0}\qty(\mathrm{i}\omega_{n};k) := \mqty[ \mathrm{i}\omega_{n} \mp \xi(k+ q_{0}) & -\Delta_{q_{0},0} \\ -\Delta_{-q_{0},0} & \mathrm{i}\omega_{n} \pm \xi(k- q_{0})], \quad 
    v_{\pm q_{0}} := \mqty[ \partial_{k}\xi(k\pm q_{0}) & O \\ O & \partial_{k}\xi(k\mp q_{0})].
\end{equation}
Other quantities in frequency space are defined as
\begin{equation}
    \delta\Delta_{\pm q_{0}}\qty(\mathrm{i}\omega_{m}-\mathrm{i}\omega_{n}) = \frac{1}{\beta}\int_{0}^{\beta}d\tau \delta\Delta_{\pm q_{0}}e^{\mathrm{i}(\omega_{m}-\omega_{n})\tau}, \quad A\qty(\mathrm{i}\omega_{m}-\mathrm{i}\omega_{n}) = \frac{1}{\beta}\int_{0}^{\beta}d\tau A(\tau)e^{\mathrm{i}(\omega_{m}-\omega_{n})\tau}.
\end{equation}
The fermionic path integral with the constraint $\Delta^{*}_{q_{0}}=\Delta_{-q_{0}}$ gives us the expression below.
\begin{align}
S[\Delta_{q_{0}},\Delta_{-q_{0}}] =& -\beta\sum_{\alpha\alpha'}\Delta_{-q_{0},0\alpha}\tilde{U}^{-1}_{\alpha\alpha'}(q_{0})\Delta_{q_{0},0\alpha'} - \sum_{\alpha\alpha'}\sum_{m}\delta\Delta_{-q_{0},\alpha}(\mathrm{i}\omega_{m})\tilde{U}^{-1}_{\alpha\alpha'}(q_{0})\delta\Delta_{q_{0},\alpha'}(\mathrm{i}\omega_{m}) \notag \\
&-\frac{1}{2\beta}\sum_{k}\sum_{m,n}\text{Tr}\ln{\qty[ -\tilde{G}^{-1}_{q_{0}}\qty(\mathrm{i}\omega_{m},\mathrm{i}\omega_{n};k)] },
\end{align}
where we used the property $\tilde{U}^{-1}_{\alpha\alpha'}(q_{0}) = \tilde{U}^{-1}_{\alpha'\alpha}(q_{0}) = \tilde{U}^{-1}_{\alpha\alpha'}(-q_{0})$, and the coefficient $1/2$ in front of the trace term appears from the fact that we expand the basis ($c_{k}\to c_{k\pm q_{0}}$).
By choosing the reference state and extracting the corresponding Green's function $\tilde{G}_{q_{0},0}$ with the self-energy $\Sigma$, we arrive at
\begin{align}
S[\Delta_{q_{0}},\Delta_{-q_{0}}] =& -\beta\sum_{\alpha\alpha'}\Delta_{-q_{0},0\alpha}\tilde{U}^{-1}_{\alpha\alpha'}(q_{0})\Delta_{q_{0},0\alpha'} - \sum_{\alpha\alpha'}\sum_{m}\delta\Delta_{-q_{0},\alpha}(\mathrm{i}\omega_{m})\tilde{U}^{-1}_{\alpha\alpha'}(q_{0})\delta\Delta_{q_{0},\alpha'}(\mathrm{i}\omega_{m}) \notag \\
& -\frac{1}{2\beta}\sum_{k}\sum_{m,n}\text{Tr}\ln{\qty[ -\tilde{G}^{-1}_{q_{0},0}] } -\frac{1}{2\beta}\sum_{k}\sum_{m,n}\text{Tr}\qty[ \sum_{L=1}^{\infty}\frac{\qty(\tilde{G}_{q_{0},0}\Sigma)^{L}}{L}]. 
\end{align}
Note that we assumed a $s$-wave pairing in the interaction, and the matrix of the gap part $\Delta_{\pm q_{0},0}$ is always diagonal.\\

\subsection{GL effective action and parameters}
%\textit{GL effective action and parameters}.
We derive the GL effective action neglecting the fluctuation and the vector potential.
Green's function of the reference state and the self-energy are written as
\begin{equation}
    \tilde{G}^{-1}_{q_{0},0}\qty(\mathrm{i}\omega_{m},\mathrm{i}\omega_{n};k) = \frac{\beta\delta_{mn}}{2}\mqty[\mathcal{G}^{-1}_{+,0}(\mathrm{i}\omega_{n};k) & O \\ O & \mathcal{G}^{-1}_{-,0}(\mathrm{i}\omega_{n};k) ], \quad 
    \Sigma = \frac{\beta\delta_{mn}}{2}\mqty[ \Delta_{0} & O \\ O & \Delta_{0}],
\end{equation}
where
\begin{equation}
    \mathcal{G}^{-1}_{\pm,0}\qty(\mathrm{i}\omega_{n};k) := \mqty[ \mathrm{i}\omega_{n} \mp \xi(k+q_{0}) & O \\ O & \mathrm{i}\omega_{n} \pm \xi(k-q_{0})], \quad 
    \Delta_{0} = \mqty[ O & \Delta_{q_{0},0} \\ \Delta_{-q_{0},0} & O].
\end{equation}
We focus on up to the second-order terms.
The component of the infinite sum needed to calculate them is just the $L=2$, namely
\begin{align}
    &\text{Tr}\qty[ \frac{1}{2}\tilde{G}_{q_{0},0}\Sigma\tilde{G}_{q_{0},0}\Sigma] \notag \\
    &=\frac{\delta_{mn}}{2}\text{Tr}\qty(\mqty[\mathcal{G}_{+,0} & O \\ O & \mathcal{G}_{-,0} ]\mqty[ \Delta_{0} & O \\ O & \Delta_{0}]\mqty[\mathcal{G}_{+,0} & O \\ O & \mathcal{G}_{-,0} ]\mqty[ \Delta_{0} & O \\ O & \Delta_{0}]) \notag \\
    &= \frac{\delta_{mn}}{2}\text{Tr}\qty(\mqty[ \mathcal{G}_{+,0}\Delta_{0}\mathcal{G}_{+,0}\Delta_{0} & O \\ O & \mathcal{G}_{-,0}\Delta_{0}\mathcal{G}_{-,0}\Delta_{0}]) \notag \\
    &= \frac{\delta_{m,n}}{2}\text{Tr}\qty( \mqty[ \begin{array}{cc} g_{\text{e}+}\Delta_{q_{0},0}g_{\text{h}-}\Delta_{-q_{0},0} & O \\ O & g_{\text{h}-}\Delta_{-q_{0},0}g_{\text{e}+}\Delta_{q_{0},0} \end{array} & \mbox{\huge O} \\ \mbox{\huge O} & \begin{array}{cc} g_{\text{h}+}\Delta_{q_{0},0}g_{\text{e}-}\Delta_{-q_{0},0} & O \\ O & g_{\text{e}-}\Delta_{-q_{0},0}g_{\text{h}+}\Delta_{q_{0},0} \end{array} ]) \notag \\
    &= \delta_{m,n}\text{Tr} \qty( g_{\text{e}+}\Delta_{q_{0},0}g_{\text{h}-}\Delta_{-q_{0},0} + g_{\text{h}+}\Delta_{q_{0},0}g_{\text{e}-}\Delta_{-q_{0},0}).
\end{align}
Here, the normal state Green's functions of electron and hole part are introduced by
\begin{align}
g_{\text{e}+} &= \qty[ \mathrm{i}\omega_{n} - \xi\qty(k + q_{0}) ]^{-1}, \\
g_{\text{e}-} &= \qty[ \mathrm{i}\omega_{n} - \xi\qty(k - q_{0}) ]^{-1}, \\
g_{\text{h}+} &= \qty[ \mathrm{i}\omega_{n} + \xi\qty(k + q_{0}) ]^{-1}, \\
g_{\text{h}-} &= \qty[ \mathrm{i}\omega_{n} + \xi\qty(k - q_{0}) ]^{-1}.
\end{align}
Expansion of the normal state Green's function about $q_{0}$ connects our momentum space formalism and the effective action in real space. We take the terms of $q_{0}$ up to the second order in the trace as
\begin{align}
&\text{Tr}\qty( g_{\text{e}+}\Delta_{q_{0},0}g_{\text{h}-}\Delta_{-q_{0},0} )\notag \\
&= \sum_{\alpha\alpha'}\Delta_{-q_{0},0\alpha}\qty(g_{\text{e}+,\alpha\alpha'}g_{\text{h}-,\alpha'\alpha})\Delta_{q_{0},0\alpha'} \notag \\
&= \sum_{\alpha\alpha'}\Big(\Delta_{-q_{0},0\alpha}\qty(g_{\text{e},\alpha\alpha'}g_{\text{h},\alpha'\alpha})\Delta_{q_{0},0\alpha'}  \notag \\
&\quad + \Delta_{-q_{0},0\alpha} \qty[ -g_{\text{e},\alpha\alpha'}\qty(\partial_{k}g_{\text{h},\alpha'\alpha})+ \qty(\partial_{k}g_{\text{e},\alpha\alpha'})g_{\text{h},\alpha'\alpha} ]\Delta_{q_{0},0\alpha'}q_{0} \notag \\
&\quad  + \Delta_{-q_{0},0\alpha} \qty[ g_{\text{e},\alpha\alpha'}\qty(\frac{1}{2}\partial^{2}_{k}g_{\text{h},\alpha'\alpha}) + \qty(\frac{1}{2}\partial^{2}_{k}g_{\text{e},\alpha\alpha'})g_{\text{h},\alpha'\alpha} - \qty(\partial_{k}g_{\text{e},\alpha\alpha'})\qty(\partial_{k}g_{\text{h},\alpha'\alpha}) ]\Delta_{q_{0},0\alpha'}q_{0}^{2}\Big) ,
\end{align}
and $\text{Tr}\qty( g_{\text{h}+}\Delta_{q_{0},0}g_{\text{e}-}\Delta_{-q_{0},0} )$ gives the same one with $q_{0}\to -q_{0}$.
We here define
\begin{equation}
g_{\text{e}} = \qty[ \mathrm{i}\omega_{n} - \xi\qty(k)]^{-1} \quad \text{and} \quad g_{\text{h}} = \qty[ \mathrm{i}\omega_{n} +\xi\qty(k)]^{-1},
\end{equation}
and reach the expression
\begin{align}
S_{\text{eff}} =& -\beta\sum_{\alpha\alpha'}\Delta_{-q_{0},0\alpha}\tilde{U}^{-1}_{\alpha\alpha'}(q_{0})\Delta_{q_{0},0\alpha'} - \frac{1}{2\beta}\sum_{k}\sum_{m,n}\text{Tr}\ln{ \qty[ -\tilde{G}^{-1}_{q_{0},0}]} \notag \\
&-\beta\sum_{\alpha\alpha'}\sum_{k}\frac{1}{2\beta}\sum_{n}\Delta_{-q_{0},0\alpha}\qty(g_{\text{e},\alpha\alpha'}g_{\text{h},\alpha'\alpha} + g_{\text{h},\alpha\alpha'}g_{\text{e},\alpha'\alpha})\Delta_{q_{0},0\alpha'} \notag \\
&-\beta\sum_{\alpha\alpha'}\sum_{k}\frac{1}{2\beta}\sum_{n}\Delta_{-q_{0},0\alpha}\Big[ \qty(\partial_{k}g_{\text{e},\alpha\alpha'})g_{\text{h},\alpha'\alpha} + \qty(\partial_{k}g_{\text{h},\alpha\alpha'})g_{\text{e},\alpha'\alpha}  \Big]\Delta_{q_{0},0\alpha'}\qty(2q_{0}) \notag \\
&-\beta\sum_{\alpha\alpha'}\sum_{k}\frac{1}{2\beta}\sum_{n}\Delta_{-q_{0},0\alpha}\Big[ \qty(\frac{1}{2}\partial_{k}^{2}g_{\text{e},\alpha\alpha'})g_{\text{h},\alpha'\alpha} 
 + \qty(\frac{1}{2}\partial_{k}^{2}g_{\text{h},\alpha\alpha'})g_{\text{e},\alpha'\alpha} \Big]\qty(2q_{0})^{2},
 \label{GL_action}
\end{align}
where we used the integration by part to simplify the expression.
This expression Eq.~(\ref{GL_action}) clearly shows that that the ``order parameters" in momentum space are $\Delta_{\pm q_{0},0\alpha}$ and the cosine term does not appear in the effective action.

We can acquire the GL parameters in the GL free energy $\mathcal{F}$ by using $\mathcal{F} = S_{\text{eff}}/\beta$.
Paying attention to the following property
\begin{align}
    -g_{\text{h},\alpha'\alpha}\qty(-\mathrm{i}\omega_{n};-k) &= \qty(\qty[\mathrm{i}\omega_{n} - \xi(-k)]^{-1})_{\alpha'\alpha} \notag \\
    &= \qty(\qty[\mathrm{i}\omega_{n} - \xi^{\text{T}}(k)]^{-1})_{\alpha'\alpha} \notag \\
    &= \qty(\qty[\mathrm{i}\omega_{n} - \xi(k)]^{-1})_{\alpha\alpha'} \notag \\
    &= g_{\text{e},\alpha\alpha'}\qty(\mathrm{i}\omega_{n};k),
\end{align}
the GL parameters arising from the kinetic parts can be written down immediately from Eq.~(\ref{GL_action}) as
\begin{align}
    \epsilon_{\text{kin},\alpha\alpha'} &= -\frac{1}{2\beta}\sum_{n}\sum_{k}g_{\text{e},\alpha\alpha'}\qty(-\mathrm{i}\omega_{n}; -k)g_{\text{e},\alpha\alpha'}\qty(\mathrm{i}\omega_{n}; k), \\
    \mathrm{i}d_{\text{I},\text{kin},\alpha\alpha'} &= \frac{1}{2\beta}\sum_{n}\sum_{k}g_{\text{e},\alpha\alpha'}\qty(-\mathrm{i}\omega_{n}; -k)\partial_{k}g_{\text{e},\alpha\alpha'}\qty(\mathrm{i}\omega_{n}; k), \\
    \frac{1}{2m^{*}_{\text{kin},\alpha}} &= -\frac{1}{2\beta}\sum_{n}\sum_{k}g_{\text{e},\alpha\alpha}\qty(-\mathrm{i}\omega_{n}; -k)\frac{1}{2}\partial_{k}^{2}g_{\text{e},\alpha\alpha}\qty(\mathrm{i}\omega_{n}; k) \quad (\alpha=\alpha'), \\
    \eta_{\text{kin},\alpha\alpha'} &= -\frac{1}{2\beta}\sum_{n}\sum_{k}g_{\text{e},\alpha\alpha'}\qty(-\mathrm{i}\omega_{n}; -k)\frac{1}{2}\partial_{k}^{2}g_{\text{e},\alpha\alpha'}\qty(\mathrm{i}\omega_{n}; k) \quad (\alpha\neq\alpha').
\end{align}
We need to take into account the contribution from the supercurrent part $-\sum_{\alpha\alpha'}\Delta_{-q_{0},0\alpha}\tilde{U}^{-1}_{\alpha\alpha'}\Delta_{q_{0},0\alpha'}$.
We use our 1D model depicted in Fig.~\ref{Pic_model} ($\alpha,\alpha'=1,2$) for simplicity.
Since the interaction $\tilde{U}(q_{0})$ is given by
\begin{equation}
    \tilde{U}\qty(q_{0}) = \mqty[ U & 2V\cos(q_{0}) \\ 2V\cos(q_{0}) & U], \notag
\end{equation}
its inverse is easily written down as
\begin{equation}
    \tilde{U}^{-1}\qty(q_{0}) = \frac{1}{U^{2} - 4V^{2}\cos^{2}(q_{0})}\mqty[ U & -2V\cos(q_{0}) \\ -2V\cos(q_{0}) & U].
\end{equation}
We can obtain the GL parameters from the supercurrent part by expanding each component about $q_{0}$.
The results without the index are as follows.
\begin{align}
    \epsilon_{\text{SC}} &= \frac{2V}{U^{2}-4V^{2}}, \\
    \mathrm{i}d_{\text{I},\text{SC}} &= 0, \\
    \frac{1}{2m^{*}_{\text{SC}}} &= \frac{4UV^{2}}{\qty(U^{2}-4V^{2})^{2}}, \\
    \eta_{\text{SC}} &= -\frac{U^{2}+4V^{2}}{\qty(U^{2}-4V^{2})^{2}}V.
\end{align}
The important point is that the strength of the Lifshitz invariant stays intact by the supercurrent part because the interaction $\tilde{U}(q_{0})$ is an even function of $q_{0}$.\\

\subsection{Free energy without fluctuation}
%\textit{Free energy without fluctuation}.
The free energy $\mathcal{F}$ without fluctuation is necessary to determine which state has a lower free energy and is more stable than others.
We neglect the fluctuations and the vector potential as before, and the reference state Green's function is
\begin{align}
&\tilde{G}^{-1}_{q_{0},0}\qty(\mathrm{i}\omega_{m},\mathrm{i}\omega_{n};k) \notag \\
&= \mqty[ { \begin{array}{cc} \qty( \mathrm{i}\omega_{n} - \xi\qty(k+q_{0}) )/2 & -\Delta_{q_{0},0}/2 \\ -\Delta_{-q_{0},0}/2 & \qty( \mathrm{i}\omega_{n} + \xi\qty(k-q_{0}) )/2 \end{array} } & \mbox{\huge O} \\ \mbox{\huge O} & { \begin{array}{cc} \qty( \mathrm{i}\omega_{n} + \xi\qty(k+q_{0}) )/2 & -\Delta_{q_{0},0}/2 \\ -\Delta_{-q_{0},0}/2 & \qty( \mathrm{i}\omega_{n} - \xi\qty(k-q_{0}) )/2 \end{array} } ]\beta\delta_{m,n} \notag \\
&= \frac{\beta\delta_{mn}}{2}\mqty[ G^{-1}_{+,0}\qty(\mathrm{i}\omega_{n};k) & O \\ O & G^{-1}_{-,0}\qty(\mathrm{i}\omega_{n};k)].
\end{align}
The effective action reads
\begin{equation}
S_{\text{eff}} = -\beta\sum_{\alpha\alpha'}\Delta_{-q_{0},0\alpha}\tilde{U}^{-1}_{\alpha\alpha'}(q_{0})\Delta_{q_{0},0\alpha'} - \frac{1}{2\beta}\sum_{k}\sum_{m,n}\text{Tr}\ln{\qty[ -\tilde{G}^{-1}_{q_{0},0}\qty(\mathrm{i}\omega_{m},\mathrm{i}\omega_{n};k)]}.
\end{equation}
The second term of the effective action can be rigorously calculated by using the identity $\text{Tr ln}=\text{ln det}$:
\begin{align}
&\frac{1}{2\beta}\sum_{k}\sum_{m,n}\text{Tr}\ln{\qty[ -\tilde{G}^{-1}_{q_{0},0}\qty(\mathrm{i}\omega_{m},\mathrm{i}\omega_{n};k)]} \notag \\
=& \frac{1}{2\beta}\sum_{k}\sum_{m,n}\ln{}\text{det}\qty( \mqty[ G^{-1}_{+,0}\qty(\mathrm{i}\omega_{n};k)\beta\delta_{m,n}/2 & O \\ O & G^{-1}_{-,0}\qty(\mathrm{i}\omega_{n};k)\beta\delta_{m,n}/2] ) \notag \\
=& \frac{1}{2}\sum_{k}\sum_{n}\ln{ \qty( \prod_{\alpha=1}^{2n_{0}}\qty[ \qty(\mathrm{i}\omega_{n} - E_{+, \alpha}(k))\frac{\beta}{2} ] \cdot \prod_{\alpha=1}^{2n_{0}}\qty[ \qty(\mathrm{i}\omega_{n} - E_{-, \alpha}(k))\frac{\beta}{2} ] )} \notag \\
=& \sum_{k}\frac{1}{2}\sum_{n}\sum_{\alpha=1}^{2n_{0}}\ln{ \qty[ \beta(\mathrm{i}\omega_{n}) - \beta E_{+, \alpha}(k)] } + \sum_{k}\frac{1}{2}\sum_{n}\sum_{\alpha=1}^{2n_{0}}\ln{ \qty[ \beta(\mathrm{i}\omega_{n}) - \beta E_{-, \alpha}(k)] } + \text{const.} \notag \\
=& \sum_{k}\frac{1}{2}\sum_{\alpha=1}^{n_{0}}\qty[ \beta E_{+, \alpha}(k) + 2\ln{ \qty( 1 + e^{-\beta E_{+, \alpha}(k)}) } ]  + \sum_{k}\frac{1}{2}\sum_{\alpha=1}^{n_{0}}\qty[ \beta E_{-, \alpha}(k) + 2\ln{ \qty( 1 + e^{-\beta E_{-, \alpha}(k)}) } ] + \mathrm{const.} \notag \\
=& \sum_{k}\sum_{\alpha=1}^{n_{0}}\qty[ \beta E_{+, \alpha}(k) + 2\ln{ \qty( 1 + e^{-\beta E_{+, \alpha}(k)}) } ] + \text{const.}
\end{align}
We can use the following identity
\begin{align}
\beta E_{+,\alpha}(k) + 2\ln\qty(1 + e^{-\beta E_{+,\alpha}(k)}) &= \beta E_{+,\alpha}(k) + 2\ln\qty(e^{-\beta E_{+,\alpha}(k)}\qty( 1 + e^{\beta E_{+,\alpha}(k)}) ) \notag \\
&= -\beta E_{+,\alpha}(k) + 2\ln\qty(1+e^{\beta E_{+,\alpha}(k)}),
\end{align}
for the positive $E_{+,\alpha}(k)$, so the range of the summation of $\alpha$ is $1$ to $n_{0}$.

The free energy $\mathcal{F}$ is immediately written down by the relation $\mathcal{F} = S_{\text{eff}}/\beta$ with neglecting the unimportant constant term:
\begin{equation}
\mathcal{F} = -\sum_{\alpha\alpha'}\Delta_{-q_{0},0\alpha}\tilde{U}^{-1}_{\alpha\alpha'}(q_{0})\Delta_{q_{0},0\alpha} - \sum_{k}\sum_{\alpha=1}^{n_{0}}\qty[ E_{+, \alpha}(k) + \frac{2}{\beta}\ln{ \qty( 1+e^{-\beta E_{+, \alpha}(k)}) }],
\end{equation}
with $E_{+, \alpha}(k)$ for $\alpha=1,\cdots 2n_{0}$ being the $\alpha$-th largest eigenvalue of the matrix
\begin{align}
\mqty[ \xi\qty(k+q_{0}) & \Delta_{q_{0},0} \\ \Delta_{-q_{0},0} & -\xi\qty(k-q_{0}) ].
\end{align}

\subsection{Gap equation}
%\textit{Gap equation}.
Here, we derive the self-consistent gap equation from the action.
We neglect the fluctuations and the vector potential again and concentrate on the reference state with the gap functions.
The reference state Green's function is the same as in the previous free energy calculation:
\begin{equation}
\tilde{G}^{-1}_{q_{0},0}\qty(\mathrm{i}\omega_{m},\mathrm{i}\omega_{n};k) = \frac{\beta\delta_{mn}}{2}\mqty[ G^{-1}_{+,0}\qty(\mathrm{i}\omega_{n};k) & O \\ O & G^{-1}_{-,0}\qty(\mathrm{i}\omega_{n};k)],
\end{equation}
and the effective action is the same:
\begin{equation}
S_{\text{eff}} = -\beta\sum_{\alpha\alpha'}\Delta_{-q_{0},0\alpha}\tilde{U}^{-1}_{\alpha\alpha'}(q_{0})\Delta_{q_{0},0\alpha'} - \frac{1}{2\beta}\sum_{k}\sum_{m,n}\text{Tr}\ln{\qty[ -\tilde{G}^{-1}_{q_{0},0}\qty(\mathrm{i}\omega_{m},\mathrm{i}\omega_{n};k)]}.
\end{equation}
Since the gap equation for $\Delta_{q_{0},0\alpha'}$ is derived by the minimization of the effective action for $\Delta_{-q_{0},0\alpha}$, we take a functional derivative of $S_{\text{eff}}$ about $\Delta_{-q_{0},0\alpha}$ and attain
\begin{equation}
\frac{\delta S_{\text{eff}}}{\delta \Delta_{-q_{0},0\alpha}} = -\beta\sum_{\alpha'}\tilde{U}^{-1}_{\alpha\alpha'}\Delta_{q_{0},0\alpha'} - \frac{1}{2\beta}\sum_{k}\sum_{m,n}\frac{\delta}{\delta \Delta_{-q_{0},0\alpha}}\text{Tr}\ln{ \qty[ -\tilde{G}^{-1}_{q_{0},0}(\mathrm{i}\omega_{m},\mathrm{i}\omega_{n};k)] }.
\end{equation}
The functional derivative of the trace term follows
\begin{align}
\frac{\delta}{\delta \Delta_{-q_{0},0\alpha}}\text{Tr}\ln{ \qty[ -\tilde{G}^{-1}_{q_{0},0}] } &= \text{Tr}\qty[ \qty(-\tilde{G}^{-1}_{q_{0},0})^{-1}\frac{\delta}{\delta \Delta_{-q_{0},0\alpha}}\qty( -\tilde{G}^{-1}_{q_{0},0} )] \notag \\
&= \text{Tr}\qty[ \tilde{G}_{q_{0},0}\frac{\delta}{\delta \Delta_{-q_{0},0\alpha}}\tilde{G}^{-1}_{q_{0},0} ],
\end{align}
and
\begin{align}
\frac{\delta}{\delta \Delta_{-q_{0},0\alpha}}\tilde{G}^{-1}_{q_{0},0} &= \mqty[ {\begin{array}{cc} O & O \\ -A_{\alpha}/2 & O \end{array} } & \mbox{\huge O} \\ \mbox{\huge O} & {\begin{array}{cc} O & O \\ -A_{\alpha}/2 & O \end{array} } ]\beta\delta_{m,n} \notag \\
&= \mqty[ -\qty(\tau_{x,\alpha} - \mathrm{i}\tau_{y,\alpha} )/4 & O \\ O & -\qty(\tau_{x,\alpha} - \mathrm{i}\tau_{y,\alpha} )/4 ]\beta\delta_{m,n}.
\end{align}
Bearing in mind that the reference state Green's function has the same block form, we can obtain
\begin{equation}
    \frac{\delta}{\delta\Delta_{-q_{0},0\alpha}}\text{Tr }\text{ln}\qty[-\tilde{G}^{-1}_{q_{0},0}] = -\text{Tr}\qty[ \qty(G_{+,0}\qty(\mathrm{i}\omega_{n};k) + G_{-,0}\qty(\mathrm{i}\omega_{n};k) )\frac{1}{2}\qty(\tau_{x,\alpha} - \mathrm{i}\tau_{y,\alpha})]\delta_{mn}.
\end{equation}
Using the Brillouin zone periodicity, we reach the gap equation as follows.
\begin{equation}
    \Delta_{q_{0},0\alpha} = \frac{1}{\beta}\sum_{n}\sum_{k}\sum_{\alpha'}\tilde{U}_{\alpha\alpha'}\text{Tr}\qty[ \frac{1}{2}\qty(\tau_{x,\alpha'}-\mathrm{i}\tau_{y,\alpha'})G_{+,0}\qty(\mathrm{i}\omega_{n};k)].
\end{equation}
This is the same formula as in the uniform case except for the finite momentum shift.\\

\subsection{Linear optical conductivity}
%\textit{Linear optical conductivity.} 
Now we turn to the linear optical conductivity.
We first consider the linear optical conductivity arising from the fluctuations that reflect the contributions from quasiparticles and collective modes.
As we have seen in previous calculations, the formulations are the same as those in the uniform case except for the finite momentum shift.
The result in the PDW case should go back to the uniform case when we take $q_{0}\to 0$.
Keeping this in mind, we start from the Green's function and the self-energy below.
\begin{align}
\tilde{G}^{-1}_{q_{0},0} &= \mqty[ { \begin{array}{cc} \qty( \mathrm{i}\omega_{n} - \xi\qty(k+q_{0}) )/2 & -\Delta_{q_{0},0}/2 \\ -\Delta_{-q_{0},0}/2 & \qty( \mathrm{i}\omega_{n} + \xi\qty(k-q_{0}) )/2 \end{array} } & \mbox{\huge O} \\ \mbox{\huge O} & { \begin{array}{cc} \qty( \mathrm{i}\omega_{n} + \xi\qty(k+q_{0}) )/2 & -\Delta_{q_{0},0}/2 \\ -\Delta_{-q_{0},0}/2 & \qty( \mathrm{i}\omega_{n} - \xi\qty(k-q_{0}) )/2 \end{array} } ]\beta\delta_{m,n} \notag \\
&= \frac{\beta\delta_{mn}}{2}\mqty[ G^{-1}_{+, 0}(\mathrm{i}\omega_{n};k) & O \\ O & G^{-1}_{-, 0}(\mathrm{i}\omega_{n};k) ], \notag \\
\Sigma &= \frac{1}{2}\mqty[ { \begin{array}{cc} O & \delta\Delta_{q_{0}}(\mathrm{i}\omega_{m}-\mathrm{i}\omega_{n}) \\ \delta\Delta_{-q_{0}}(\mathrm{i}\omega_{m}-\mathrm{i}\omega_{n}) & O \end{array} } & \mbox{\huge O} \\ \mbox{\huge O} & { \begin{array}{cc} O & \delta\Delta_{q_{0}}(\mathrm{i}\omega_{m}-\mathrm{i}\omega_{n}) \\ \delta\Delta_{-q_{0}}(\mathrm{i}\omega_{m}-\mathrm{i}\omega_{n}) & O \end{array}} ]\beta \notag \\
&\quad -\frac{1}{2}eA(\mathrm{i}\omega_{m} - \mathrm{i}\omega_{n})\mqty[ {\begin{array}{cc} \partial_{k}\xi\qty(k+q_{0}) & O \\ O & \partial_{k}\xi\qty(k-q_{0}) \end{array} } & \mbox{\huge O} \\ \mbox{\huge O} & {\begin{array}{cc} \partial_{k}\xi\qty(k+q_{0}) & O \\ O & \partial_{k}\xi\qty(k-q_{0}) \end{array} } ] \beta \notag \\
&= \frac{\beta}{2}\mqty[ \delta\Delta\qty(\mathrm{i}\omega_{m}-\mathrm{i}\omega_{n}) & O \\ O &  \delta\Delta\qty(\mathrm{i}\omega_{m}-\mathrm{i}\omega_{n}) ] - \frac{\beta}{2}eA\qty(\mathrm{i}\omega_{m}-\mathrm{i}\omega_{n})\mqty[ v_{q_{0}}(k) & O \\ O & v_{-q_{0}}(k)],
\end{align}
where
\begin{align}
G^{-1}_{+,0}(\mathrm{i}\omega_{n};k) &= \mqty[ \mathrm{i}\omega_{n} - \xi\qty(k+q_{0}) & -\Delta_{q_{0},0} \\ -\Delta_{-q_{0},0} & \mathrm{i}\omega_{n} + \xi\qty(k-q_{0}) ], \\
G^{-1}_{-,0}(\mathrm{i}\omega_{n};k) &= \mqty[ \mathrm{i}\omega_{n} + \xi\qty(k+q_{0}) & -\Delta_{q_{0},0} \\ -\Delta_{-q_{0},0} & \mathrm{i}\omega_{n} - \xi\qty(k-q_{0}) ], \\
\delta\Delta\qty(\mathrm{i}\omega_{m}-\mathrm{i}\omega_{n}) &= \mqty[ O & \delta\Delta_{q_{0}}\qty(\mathrm{i}\omega_{m}-\mathrm{i}\omega_{n}) \\ \delta\Delta_{-q_{0}}\qty(\mathrm{i}\omega_{m}-\mathrm{i}\omega_{n}) & O ] = \sum_{\mu,\alpha}\Delta_{\mu\alpha}\qty(\mathrm{i}\omega_{m}-\mathrm{i}\omega_{n})\tau_{\mu\alpha}, \\
v_{q_{0}}(k) &= \mqty[ \partial_{k}\xi\qty(k+q_{0}) & O \\ O & \partial_{k}\xi\qty(k-q_{0}) ],
\end{align}
and here we decompose the fluctuation
\begin{align}
\delta\Delta_{q_{0},\alpha}(\mathrm{i}\omega_{m}) &= \Delta_{x,\alpha}(\mathrm{i}\omega_{m}) - \mathrm{i}\Delta_{y,\alpha}(\mathrm{i}\omega_{m}) \\
\delta\Delta_{-q_{0},\alpha}(\mathrm{i}\omega_{m}) &= \qty( \delta\Delta_{q_{0},\alpha}(\mathrm{i}\omega_{m}) )^{*} = \Delta_{x,\alpha}(-\mathrm{i}\omega_{m}) + \mathrm{i}\Delta_{y,\alpha}(-\mathrm{i}\omega_{m})
\end{align}
(we omit $\delta$ and $q_{0}$ after the decomposition for simplicity).
Then, the nontrace fluctuation term can be expressed as
\begin{align}
&\sum_{\alpha\alpha'}\sum_{m}4\frac{\delta\Delta_{-q_{0},\alpha}(\mathrm{i}\omega_{m})}{2}\tilde{U}^{-1}_{\alpha\alpha'}(q_{0})\frac{\delta\Delta_{q_{0},\alpha'}(\mathrm{i}\omega_{m})}{2} \notag \\
=&\sum_{\alpha\alpha'}\sum_{m}\qty[ \Delta_{x,\alpha}(-\mathrm{i}\omega_{m}) + \mathrm{i}\Delta_{y,\alpha}(-\mathrm{i}\omega_{m}) ]\tilde{U}^{-1}_{\alpha\alpha'}(q_{0})\qty[ \Delta_{x,\alpha'}(\mathrm{i}\omega_{m}) - \mathrm{i}\Delta_{y,\alpha'}(\mathrm{i}\omega_{m}) ] \notag \\
=& \sum_{\alpha\alpha'}\sum_{\mu\mu'}\sum_{m}\Delta_{\mu\alpha}\qty(-\mathrm{i}\omega_{m})\tilde{U}^{-1}_{\alpha\alpha'}(q_{0})\Delta_{\mu'\alpha'}\qty(\mathrm{i}\omega_{m}),
\end{align}
where the imaginary part disappears when taking a summation.
Since we are interested in linear optical conductivity, the term we need to treat in the trace is $L=2$ again.
This reads
\begin{align}
\frac{1}{2\beta}\sum_{k}\sum_{m,n}\text{Tr}\qty[ \frac{1}{2}\tilde{G}_{q_{0},0}\Sigma\tilde{G}_{q_{0},0}\Sigma ]=& \frac{1}{2\beta}\sum_{k}\sum_{m,n}\text{Tr}\qty[ G_{+,0}\delta\Delta G_{+,0}\delta\Delta] \notag \\
+& \frac{1}{2\beta}\sum_{k}\sum_{m,n}\text{Tr}\qty[ G_{+,0}\delta\Delta G_{+,0}\qty(-eA)v_{q_{0}}] \notag \\
+& \frac{1}{2\beta}\sum_{k}\sum_{m,n}\text{Tr}\qty[ G_{+,0}\qty(-eA)v_{q_{0}} G_{+,0}\delta\Delta ] \notag \\
+& \frac{1}{2\beta}\sum_{k}\sum_{m,n}\text{Tr}\qty[ G_{+,0}\qty(-eA)v_{q_{0}}  G_{+,0}\qty(-eA)v_{q_{0}}] \notag \\
+& \frac{1}{2\beta}\sum_{k}\sum_{m,n}\text{Tr}\qty[ G_{-,0}\delta\Delta G_{-,0}\delta\Delta] \notag \\
+& \frac{1}{2\beta}\sum_{k}\sum_{m,n}\text{Tr}\qty[ G_{-,0}\delta\Delta G_{-,0}\qty(-eA)v_{-q_{0}}] \notag \\
+& \frac{1}{2\beta}\sum_{k}\sum_{m,n}\text{Tr}\qty[ G_{-,0}\qty(-eA)v_{-q_{0}} G_{-,0}\delta\Delta ] \notag \\
+& \frac{1}{2\beta}\sum_{k}\sum_{m,n}\text{Tr}\qty[ G_{-,0}\qty(-eA)v_{-q_{0}}  G_{-,0}\qty(-eA)v_{-q_{0}}].
\label{OC_expansion_terms}
\end{align}
By shifting the momentum, we can simultaneously obtain
\begin{equation}
G_{-,0}(k) \to G_{+,0}(k), \quad v_{-q_{0}}(k) \to -v_{-q_{0}}(k) = v^{\text{T}}_{q_{0}}(-k).
\end{equation}
The example of its application is as follows.
\begin{align}
\sum_{k}\sum_{m,n}\text{Tr}\qty[ G_{-,0}(k)\delta\Delta G_{-,0}(k)v_{-q_{0}}(k) ] &= \sum_{k}\sum_{m,n}\text{Tr}\qty[ G_{+,0}(k)\qty(\delta\Delta)^{\text{T}} G_{+,0}(k)v^{\text{T}}_{q_{0}}(-k)] \notag \\
&= \sum_{k}\sum_{m,n}\text{Tr}\qty[ G_{+,0}(-k)\qty(\delta\Delta)^{\text{T}} G_{+,0}(-k)v^{\text{T}}_{q_{0}}(k) ] \notag \\
&= \sum_{k}\sum_{m,n}\text{Tr}\qty[ G^{\text{T}}_{+,0}(k)\qty(\delta\Delta)^{\text{T}} G^{\text{T}}_{+,0}(k)v^{\text{T}}_{q_{0}}(k) ] \notag \\
&= \sum_{k}\sum_{m,n}\text{Tr}\qty[ G_{+,0}(k)\delta\Delta G_{+,0}(k)v_{q_{0}}(k) ],
\end{align}
where we exchange $\omega_{m}$ and $\omega_{n}$ that leads to $\delta\Delta\to \delta\Delta^{*} = \qty(\delta\Delta)^{\text{T}}$ ($G_{\pm,0}$ does not change by exchanging the frequency) in the first line.
Using this property in the last four terms of Eq.~(\ref{OC_expansion_terms}), we obtain
\begin{align}
\sum_{k}\text{Tr}\qty[ G_{-,0}\delta\Delta G_{-,0}\delta\Delta] &= \sum_{k}\text{Tr}\qty[ G_{+,0}\delta\Delta G_{+,0}\delta\Delta] \notag \\
\sum_{k}\text{Tr}\qty[ G_{-,0}\delta\Delta G_{-,0}\qty(-eA)v_{q_{0}}] &= \sum_{k}\text{Tr}\qty[ G_{+,0}\delta\Delta G_{+,0}\qty(-eA)v_{q_{0}}] \notag \\
\sum_{k}\text{Tr}\qty[ G_{-,0}\qty(-eA)v_{q_{0}} G_{-,0}\delta\Delta] &= \sum_{k}\text{Tr}\qty[ G_{+,0}\qty(-eA)v_{q_{0}} G_{+,0}\delta\Delta] \notag \\
\sum_{k}\text{Tr}\qty[ G_{-,0}\qty(-eA)v_{q_{0}} G_{-,0}\qty(-eA)v_{q_{0}}] &= \sum_{k}\text{Tr}\qty[ G_{+,0}\qty(-eA)v_{q_{0}} G_{+,0}\qty(-eA)v_{q_{0}}].
\end{align}
Hence it follows
\begin{align}
\frac{1}{2\beta}\sum_{k}\sum_{m,n}\text{Tr}\qty[ \frac{1}{2}\tilde{G}_{q_{0},0}\Sigma\tilde{G}_{q_{0},0}\Sigma ]=&\frac{1}{\beta}\sum_{k}\sum_{m,n}\text{Tr}\qty[ G_{+,0}\delta\Delta G_{+,0}\delta\Delta] \notag \\
+& \frac{1}{\beta}\sum_{k}\sum_{m,n}\text{Tr}\qty[ G_{+,0}\delta\Delta G_{+,0}\qty(-eA)v_{q_{0}}] \notag \\
+& \frac{1}{\beta}\sum_{k}\sum_{m,n}\text{Tr}\qty[ G_{+,0}\qty(-eA)v_{q_{0}} G_{+,0}\delta\Delta ] \notag \\
+& \frac{1}{\beta}\sum_{k}\sum_{m,n}\text{Tr}\qty[ G_{+,0}\qty(-eA)v_{q_{0}}  G_{+,0}\qty(-eA)v_{q_{0}}].
\end{align}
This result says that the formalism is the same as in the case of uniform Hamiltonian except for the finite momentum shift $q_{0}$.
Thus, we can immediately arrive at the expression of the linear optical conductivity within the random phase approximation because the formulation with the BCS Hamiltonian is well-known~\cite{Nagashima2024, Kamatani2022}:
\begin{align}
\sigma_{\text{QP}}(\omega) &= \frac{\mathrm{i}e^{2}}{\omega}\Phi(\omega), \\
\sigma_{\text{CM}}(\omega) &= \frac{\mathrm{i}e^{2}}{2\omega}Q^{\text{T}}(\omega)\tilde{U}_{\text{eff}}(q_{0}; \omega)Q(-\omega),
\end{align}
where
\begin{align}
\Phi(\mathrm{i}\Omega) &= \frac{1}{\beta}\sum_{n}\sum_{k}\text{Tr}\qty[ v_{q_{0}}(k)G_{+,0}(\mathrm{i}\omega_{n}+\mathrm{i}\Omega; k)v_{q_{0}}(k)G_{+,0}(\mathrm{i}\omega_{n};k) ], \\
\qty[ Q(\mathrm{i}\Omega)]_{\mu\alpha} &= \frac{1}{\beta}\sum_{n}\sum_{k}\text{Tr}\qty[ v_{q_{0}}(k)G_{+,0}(\mathrm{i}\omega_{n}+\mathrm{i}\Omega;k)\tau_{\mu\alpha}G_{+,0}(\mathrm{i}\omega_{n};k)], \\
\qty[\tilde{U}_{\text{eff}}(q_{0}; \mathrm{i}\Omega)]_{\mu\alpha,\mu'\alpha'} &= \qty( \delta_{\mu\mu'}\delta_{\alpha\alpha'} - \tilde{U}_{\alpha\alpha'}(q_{0})\qty[\Pi(\mathrm{i}\Omega)]_{\mu\alpha,\mu'\alpha'} )^{-1}\tilde{U}_{\alpha\alpha'}(q_{0}), \\
\qty[\Pi(\mathrm{i}\Omega)]_{\mu\alpha,\mu'\alpha'} &= \frac{1}{2\beta}\sum_{n}\sum_{k}\text{Tr}\qty[ \tau_{\mu\alpha}G_{+,0}(\mathrm{i}\omega + \mathrm{i}\Omega ;k)\tau_{\mu'\alpha'}G_{+,0}(\mathrm{i}\omega_{n};k) ].
\end{align}

We finally consider the linear optical conductivity from the supercurrent part of the effective action $S_{\text{eff},\text{SC}}=-\beta\sum_{\alpha\alpha'}\Delta_{-q_{0},0\alpha}\tilde{U}^{-1}_{\alpha\alpha'}(q_{0})\Delta_{q_{0},0\alpha'}$ when we use our 1D model depicted in Fig.~\ref{Pic_model}.
The interaction, including the effect of vector potential arising from the pair-hopping, is
\begin{equation}
    \tilde{U}(q,A) = \mqty[ U & 2V\cos\qty(q-2eA) \\ 2V\cos\qty(q-2eA) & U].
\end{equation}
The factor $2$ appearing with $eA$ comes from the pair-hopping term because the pair-hopping interaction has two creation/annihilation operators on adjacent sites.
We concentrate on three cases: uniform ($q=0$), $\pi/2$-PDW ($q=\pi/2$), and $\pi$-PDW ($q=\pi$), although we can derive the general formula of the linear optical conductivity from the supercurrent part.

\textit{The uniform case}.
The inverse of the interaction $\tilde{U}^{-1}(q,A)$ is
\begin{equation}
    \tilde{U}^{-1}(q=0,A) = \frac{1}{U^{2} - 4V^{2}\cos^{2}\qty(2eA)}\mqty[ U & -2V\cos(2eA) \\ -2V\cos(2eA) & U].
\end{equation}
By expanding the interaction about $A$ with the assumption that $A$ is sufficiently small and taking only the second-order terms of $A$, we obtain
\begin{equation}
    \frac{(2e)^{2}A^{2}}{\qty(U^{2} - 4V^{2})^{2}}\mqty[ -4UV^{2} & V(U^{2} + 4V^{2}) \\ V(U^{2} + 4V^{2}) & -4UV^{2}].
\end{equation}
Then the current of this part $j_{\text{SC},q=0}$ is calculated from $j_{\text{SC}}=-(\delta S_{\text{eff},\text{SC}}/\delta A)$ with $S_{\text{eff},\text{SC}} = -\beta\sum_{\alpha\alpha'}\Delta_{-q_{0},0\alpha}\tilde{U}^{-1}_{\alpha\alpha'}(q_{0})\Delta_{q_{0},0\alpha'}$
\begin{align}
    j_{\text{SC},q=0} &= \frac{2(2e)^{2}A}{(U^{2}-4V^{2})^{2}}\mqty[ \Delta_{0} & \Delta_{0}]\mqty[ -4UV^{2} & V(U^{2}+4V^{2}) \\ V(U^{2}+4V^{2}) & -4UV^{2}]\mqty[ \Delta_{0} \\ \Delta_{0}] \notag \\
    &= -\frac{(2e)^{2}4V\Delta_{0}^{2}}{(U+2V)^{2}}A.
\end{align}
We thus acquire the linear optical conductivity arising from the supercurrent part $\sigma_{\text{gap},q=0}$ by using $j(\omega)=\sigma(\omega)E(\omega)$ and $E(\omega)=\mathrm{i}\omega A(\omega)$:
\begin{equation}
    \sigma_{\text{SC},q=0}(\omega) = \frac{(2e)^{2}}{\mathrm{i}\omega}\frac{4V\Delta_{0}^{2}}{(U+2V)^{2}}.
\end{equation}

\textit{The $\pi/2$-PDW case}.
The procedure is the same as in the uniform case.
The interaction is
\begin{equation}
    \tilde{U}^{-1}\qty(q=\frac{\pi}{2},A) = \frac{1}{U^{2} - 4V^{2}\sin^{2}\qty(2eA)}\mqty[ U & -2V\sin(2eA) \\ -2V\sin(2eA) & U].
\end{equation}
Its inverse is written up to the second-order terms of $A^{2}$ as
\begin{equation}
    \tilde{U}^{-1}\qty(q_{0}=\frac{\pi}{2},A) \approx \qty(\frac{1}{U^{2}} + \frac{4V^{2}}{U^{4}}\qty(2eA)^{2})\mqty[ U & -2V\qty(2eA) \\ -2V\qty(2eA) & U].
\end{equation}
Supposing that the gap functions on the site $1$ and $2$ satisfy $|\Delta_{\pi/2,01}|=|\Delta_{\pi/2,02}|=\Delta_{0}$, we finally obtain
\begin{equation}
    \sigma_{\text{SC},q=\pi/2} = \frac{(2e)^{2}}{\mathrm{i}\omega}\frac{16V^{2}\Delta_{0}^{2}}{U^{3}}.
\end{equation}

\textit{The $\pi$-PDW case}.
We can arrive at the result at once by changing the sign of $V$ with the assumption $\Delta_{\pi/2,01}=\Delta_{\pi/2,02}=\Delta_{0}$:
\begin{equation}
    \sigma_{\text{SC},q=\pi} = -\frac{(2e)^{2}}{\mathrm{i}\omega}\frac{4V\Delta_{0}^{2}}{(U+2V)^{2}}.
\end{equation}

\section{Absence of hybridization between the Higgs and Leggett modes in the GL theory}
\label{Apdx.B}
In this Appendix, we show that the Higgs and Leggett modes do not hybridize with each other based on the macroscopic GL theory.
To understand this, let us neglect the spatial dependence of fluctuations of order parameters and begin with the following free energy $\mathcal{F}$ with the amplitude fluctuation $H_{i}$ ($i=1,2$) and a relative phase fluctuation $\phi_{\text{L}}$:
\begin{align}
    \mathcal{F} &= \sum_{i=1,2}\qty[ \qty(a+\frac{q^{2}}{2m^{*}}) \qty(\psi_{0}+H_{i})^{2} + \frac{b}{2}\qty(\psi_{0} + H_{i})^{4} ] \notag \\
&\quad + 2\qty(\psi_{0} + H_{1})\qty(\psi_{0} + H_{2})\qty(\epsilon + \eta q^{2})\cos\qty(\phi + \phi_{\text{L}}) \notag \\
&\quad + 4\qty(\psi_{0} + H_{1})\qty(\psi_{0} + H_{2}) d_{\text{I}}q\sin\qty(\phi + \phi_{\text{L}}).
\end{align}
Here, $\sin\phi$ and $\cos\phi$ are defined as
\begin{align}
\sin\phi &= - \frac{2d_{\text{I}}q}{\sqrt{ \qty(\epsilon + \eta q^{2})^{2} + 4\qty(d_{\text{I}}q)^{2} }}, \label{sin} \\
\cos\phi &= -\frac{\epsilon + \eta q^{2}}{\sqrt{ \qty(\epsilon + \eta q^{2})^{2} + 4\qty(d_{\text{I}}q)^{2} }} \label{cos},
\end{align}
where we put $a_{1}=a_{2}=a$, $b_{1}=b_{2}=b$, and $m^{*}_{1}=m^{*}_{2}=m^{*}$.
We also assume that the vectors $\bm{d}_{I}$ and $\bm{q}$ are in the same direction for simplicity.
To minimize the free energy around the ground state, we expand $\mathcal{F}$ in terms of $H_{i}$ and $\phi_{\text{L}}$, and collect terms up to the second order in the fluctuation:
\begin{align}
\mathcal{F} &\approx 2\qty(a + \frac{q^{2}}{2m^{*}}) \psi_{0}^{2} + b\psi_{0}^{4} + 2\qty(\epsilon + \eta q^{2})\psi_{0}^{2}\cos\phi + 4d_{\text{I}}q\psi_{0}\sin\phi \notag \\
&\quad + \qty[ 2\qty(a + \frac{q^{2}}{2m^{*}})\psi_{0} + 2b\psi_{0}^{3} + 2\qty(\epsilon + \eta q^{2})\psi_{0}\cos\phi + 4d_{\text{I}}q\psi_{0}\sin\phi ]\qty(H_{1} + H_{2}) \notag \\
&\quad + \qty[ -2\qty(\epsilon + \eta q^{2})\psi_{0}^{2}\sin\phi + 4d_{\text{I}}q\psi_{0}^{2}\cos\phi ]\phi_{\text{L}} \notag \\
&\quad+ \qty( a + 3b\psi_{0}^{2} + \frac{q^{2}}{2m^{*}})\qty(H_{1}^{2} + H_{2}^{2}) \notag \\
&\quad + 2\qty(\epsilon + \eta q^{2})\qty[ \cos\phi\qty( H_{1}H_{2} - \frac{1}{2}\psi_{0}^{2}\phi_{\text{L}}^{2}) - \psi_{0}\sin\phi\qty(H_{1} + H_{2})\phi_{\text{L}} ] \notag \\
&\quad + 4d_{\text{I}}q\qty[ \sin\phi\qty(H_{1}H_{2} - \frac{1}{2}\psi_{0}^{2}\phi_{\text{L}}^{2}) + \psi_{0}\cos\phi\qty(H_{1}+H_{2})\phi_{\text{L}} ].
\label{Minimization}
\end{align}
The third line of Eq.~(\ref{Minimization}) vanishes because of the definitions of $\sin\phi$ and $\cos\phi$.
The coefficient of $\qty(H_{1} + H_{2})$ in the second line of Eq.~(\ref{Minimization}) gives the minimization condition of the free energy around the ground state.
This reads
\begin{equation}
\psi_{0} = 0 \quad \text{or} \quad \psi_{0}^{2} = -\frac{1}{b}\qty[ a + \frac{q^{2}}{2m^{*}} - \sqrt{ \qty(\epsilon + \eta q^{2})^{2} + 4(d_{\text{I}}q)^{2} } ].
\end{equation}
The second condition gives the superconducting ground state.
With this,
%Keeping the second condition in mind, 
$\mathcal{F}$ can be rewritten as
\begin{align}
\mathcal{F} = \text{const.} &+ \qty( a + 3b\psi_{0}^{2} + \frac{q^{2}}{2m^{*}})\qty(H_{1}^{2} + H_{2}^{2}) \notag \\
& + 2\qty(\epsilon + \eta q^{2})\qty[ \cos\phi\qty( H_{1}H_{2} - \frac{1}{2}\psi_{0}^{2}\phi_{\text{L}}^{2}) - \psi_{0}\sin\phi\qty(H_{1} + H_{2})\phi_{\text{L}} ] \notag \\
& + 4d_{\text{I}}q\qty[ \sin\phi\qty(H_{1}H_{2} - \frac{1}{2}\psi_{0}^{2}\phi_{\text{L}}^{2}) + \psi_{0}\cos\phi\qty(H_{1}+H_{2})\phi_{\text{L}} ].
\label{minimizedF}
\end{align}
Because of the constraint conditions (\ref{sin}) and (\ref{cos}), the coupling between the Higgs and Leggett modes in Eq.~(\ref{minimizedF}) (which is proportional to $(H_{1}+H_{2})\phi_{\text{L}}$) vanishes:
\begin{align}
&-2\qty(\epsilon + \eta q^{2})\psi_{0}\sin\phi\qty(H_{1}+H_{2})\phi_{\text{L}} + 4d_{\text{I}}q\cos\phi\qty(H_{1}+H_{2})\phi_{\text{L}} \notag \\
&= \frac{2\qty(\epsilon + \eta q^{2})\cdot 2d_{\text{I}}q }{\sqrt{ (\epsilon + \eta q^{2})^{2} + 4\qty(d_{\text{I}}q)^{2} } }\psi_{0}\qty(H_{1}+H_{2})\phi_{\text{L}} - \frac{2\qty(\epsilon + \eta q^{2})\cdot 2d_{\text{I}}q }{\sqrt{ \qty(\epsilon + \eta q^{2})^{2} + 4\qty(d_{\text{I}}q)^{2} } }\psi_{0}\qty(H_{1}+H_{2})\phi_{\text{L}} \notag \\
&= 0.
\end{align}
Thus, we obtain the following expression for $\mathcal{F}$:
\begin{align}
\mathcal{F} = \text{const.} &+ \qty( a + 3b\psi_{0}^{2} + \frac{q^{2}}{2m^{*}})\qty(H_{1}^{2} + H_{2}^{2})
%\notag \\
%&
- \sqrt{ (\epsilon + \eta q^{2})^{2} + 4(d_{\text{I}}q)^{2} }\qty( H_{1}H_{2} - \frac{1}{2}\psi_{0}^{2}\phi_{\text{L}}^{2} ).
\label{appendix: F}
\end{align}
There is no coupling term between the amplitude and phase fluctuations ($\propto H_{i}\phi_{\text{L}}$).

\section{The ``backfolded" band structure of the PDW states}
\label{Apdx_C}
\begin{figure}
    \centering
    \includegraphics[scale=0.7]{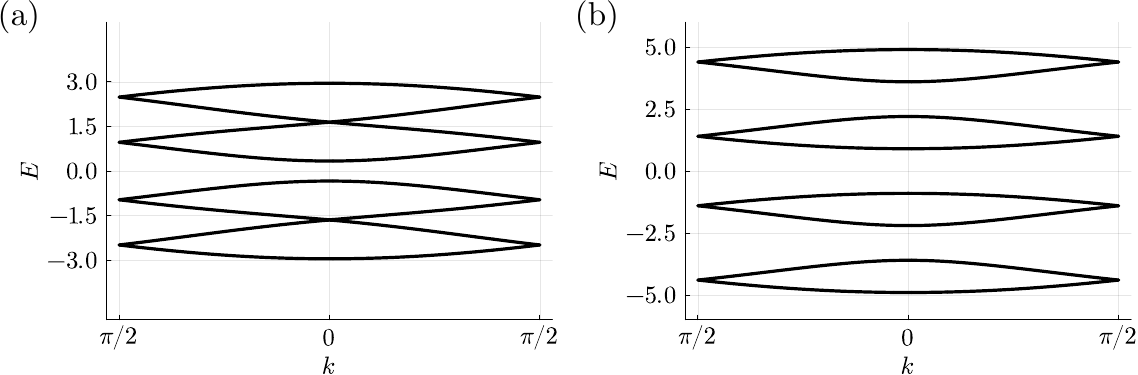}
    \caption{The back-folded band structures of the (a) $\pi/2$- and (b) $\pi$-PDW states.
    The parameters are chosen as $\delta t=0.35$, $|\mu|=0.8$, $U=-3.0$, $T=0.001$, and $V=0.5$ for the $\pi/2$-PDW state (a), and $V=1.4$ for the $\pi$-PDW state (b).
    }
    \label{fig: backfolded}
\end{figure}

In Sec. IV E, we have plotted the band structures of the PDW states with the finite center-of-mass momentum $\bm{q}$.
The particle-hole symmetry is not violated (as discussed in Sec.~\ref{sec:level2-3}), since it is modified to include the effect of $\bm{q}$.
This form is useful to pick up the direct gap for possible optical transitions.
Here, we show the back-folded band structure with the doubled unit cell, which becomes particle-hole symmetric for the $\pi/2$- and $\pi$-PDW states in one dimension.
The parameters are chosen as $\delta t=0.35$, $|\mu|=0.8$, $U=-3.0$, $T=0.001$, and $V=0.5$ for the $\pi/2$-PDW state, and $V=1.4$ for the $\pi$-PDW state.
We extend the unit cell to include four sites, back-fold the Brillouin zone, and plot the band structure corresponding to the eigenvalues of the BdG Hamiltonian (Eq.~(\ref{BdG})).
The resulting band structures are particle-hole symmetric (see Fig.~\ref{fig: backfolded}), as we expected.

\twocolumngrid

\bibliography{main.bib}

\end{document}